\documentclass[pra,reprint,superscriptaddress]{revtex4-2}

\usepackage{amssymb}
\usepackage{physics}
\usepackage{dsfont}
\usepackage{mathtools}
\usepackage{natbib}
\usepackage{graphicx}
\usepackage[T1]{fontenc}
\usepackage{lmodern}
\usepackage{hyperref}

\DeclarePairedDelimiter\floor{\lfloor}{\rfloor}
\newcommand{\san}[1]{\mathsf{#1}}

\graphicspath{{figures/}}

\begin{document}
\title{Quantum repeaters based on concatenated bosonic and discrete-variable quantum codes}
\author{Filip Rozp\k{e}dek}
\thanks{frozpedek@uchicago.edu}
\affiliation{Pritzker School of Molecular Engineering, University of Chicago, Chicago, IL 60637, USA}
\author{Kyungjoo Noh}
\affiliation{AWS Center for Quantum Computing, Pasadena, CA 91125, USA}
\author{Qian Xu}
\affiliation{Pritzker School of Molecular Engineering, University of Chicago, Chicago, IL 60637, USA}
\author{Saikat Guha}
\affiliation{James C. Wyant College of Optical Sciences, University of Arizona, Tucson, AZ 85721, USA}
\author{Liang Jiang}
\thanks{liang.jiang@uchicago.edu}
\affiliation{Pritzker School of Molecular Engineering, University of Chicago, Chicago, IL 60637, USA}

\begin{abstract}
We propose an architecture of quantum-error-correction-based quantum repeaters that combines techniques used in discrete- and continuous-variable quantum information. Specifically, we propose to encode the transmitted qubits in a concatenated code consisting of two levels. On the first level we use a continuous-variable GKP code encoding the qubit in a single bosonic mode. On the second level we use a small discrete-variable code. Such an architecture has two important features. Firstly, errors on each of the two levels are corrected in repeaters of two different types. This enables for achieving performance needed in practical scenarios with a reduced cost with respect to an architecture for which all repeaters are the same. Secondly, the use of continuous-variable GKP code on the lower level generates additional analog information which enhances the error-correcting capabilities of the second-level code such that long-distance communication becomes possible with encodings consisting of only four or seven optical modes.
\end{abstract}

\maketitle


\section{Introduction}

Quantum cryptographic and quantum computing tasks offer qualitative advantages over their classical counterparts. However, in order to implement these tasks, it is essential to be able to transmit quantum information across long distances. There have been significant efforts in recent years in designing future large-scale quantum networks that could offer such a functionality by overcoming the exponential signal decay with distance in the optical fibre through the use of quantum repeaters~\cite{briegel1998quantum, Munro_15}. Multiple different types of quantum repeaters have been proposed, utilising different techniques to overcome losses and operational errors of the devices.

The original repeater proposals utilise heralded entanglement generation between repeater stations~\cite{briegel1998quantum, Munro_15}. These elementary links can then be connected into end-to-end entanglement using Bell-measurements at the repeaters. The entanglement rate of these schemes is significantly limited by the communication time between repeaters, where the communication is needed to herald success of both elementary link generation and probabilistic entanglement distillation used for correcting operational errors. These limitations can be overcome using one-way quantum repeaters based on forward error correction~\cite{Munro_15, munro2012quantum, muralidharan2014ultrafast, glaudell2016serialized, namiki2016role, ewert2017ultrafast, ewert2016ultrafast, muralidharan2018one, miatto2018hamiltonians, lee2019fundamental, borregaard2019one}. Such repeaters are not limited by two-way communication, as a stream of qubits, encoded in a loss-tolerant code, is sent over a multi-hop channel. A repeater station uses quantum decoding and re-encoding operations to near-deterministically correct errors (loss and operational) and forwards the encoded state to the next station. Most of the repeater schemes belonging to the above two categories need quantum memories, which could be substituted by all-photonic entangled states~\cite{Azuma_15, pant2017rate}. However, there also exist one-way schemes which do not require any storage of quantum information and where all the operations performed inside the repeaters involve only optical elements~\cite{ewert2017ultrafast, lee2019fundamental}. Yet, the use of a few matter qubits in such repeaters could enable for more efficient generation and error correction of the photonic encoded states~\cite{borregaard2019one}. 

The significant rate improvement of these error-correction-based schemes comes at the cost of large physical resource overhead. Specifically, in order to overcome losses over a $1000$ km path, most one-way and all-photonic architectures would require the ability to generate, transfer, store and operate on hundreds or thousands of highly entangled qubits within each repeater~\cite{muralidharan2014ultrafast, ewert2017ultrafast, lee2019fundamental, borregaard2019one, Azuma_15, pant2017rate}.

So far, all existing one-way quantum repeaters only considered quantum error correction based on two-level or multi-level encoding to correct excitation loss errors, without taking advantage of the bosonic nature of the quantum channel. Here we propose a new type of quantum repeater architecture based on concatenated quantum error correction, with continuous-variable (CV) bosonic encoding at the lower level (inner code) and discrete-variable (DV) encoding at the higher level (outer code). The specific bosonic code that we consider here is the single-mode Gottesman-Kitaev-Preskill (GKP) code~\cite{gottesman2001encoding} which has been demonstrated to perform well against photon loss errors given a suitable encoding strategy~\cite{albert2018performance, noh2018quantum}. We note that while implementation of GKP encodings is challenging, there have been experimental demonstrations of approximate GKP states in trapped ions~\cite{fluhmann2019encoding, de2020error} and the superconducting microwave cavity~\cite{campagne2019stabilized}. While various architectures for quantum computing based on GKP encodings have been proposed~\cite{menicucci2014fault, bourassa2020blueprint, fukui2018high, fukui2019high, vuillot2019quantum, noh2020fault}, no corresponding quantum communication protocol has yet been considered. Similarly as in the proposed quantum computing architectures with GKP code, we propose to concatenate the GKP code with a higher-level multi-qubit code to boost its performance. We show here that if sufficiently high-quality GKP states can be prepared and operated on, then long-distance quantum communication can be achieved by using only few qubits in the higher-level multi-qubit encoding. Moreover, our repeater architecture is also cost-efficient. This is because we find that in order to maintain high performance it is not necessary for all quantum repeaters to be able to perform error correction on both encoding levels. Specifically, it is sufficient for the more powerful but at the same time more costly repeaters correcting errors on both levels to be placed only sporadically, with majority of repeaters correcting only the lower level errors as shown in FIG.~\ref{fig}. This enables significant reduction of the required resources as the repeaters correcting only the lower level errors need to operate only on single GKP data modes at a time.

\section{Results}

\subsection{GKP repeater chain architecture}
\label{sec:GKPrep}

In this section we describe a simple repeater architecture in which quantum information is encoded in the GKP code and the GKP repeaters placed along the channel are used to correct errors arising from the communication through a lossy channel. We will see that GKP encoding alone together with a specific considered decoding strategy is not sufficient to achieve long-distance quantum communication, which motivates the introduction of the concatenated-coded scheme, described in Section~\ref{sec:concatenatedrep}. Firstly however, we provide some basic information about the fundamental principles behind the GKP error correction.

\subsubsection{GKP-qubit error correction}

Similarly to quadrature amplitude modulation encoding used in classical communication~\cite{rouphael2009rf}, we may use the quantum GKP encoding to correct loss errors. The basic idea behind the GKP code~\cite{gottesman2001encoding} is that while the exact value of two conjugate continuous observables $\hat{q}$ and $\hat{p}$ cannot be measured simultaneously, the two operators 
\begin{equation}
\hat{S}_q = \exp(i2\sqrt{\pi}\hat{q}), \quad \quad \hat{S}_p = \exp(-i2\sqrt{\pi}\hat{p}),
\end{equation}
which are periodic functions of $\hat{q}$ and $\hat{p}$, commute with each other and therefore can be measured simultaneously. The GKP code thus encodes a qubit in a two-dimensional subspace of an infinite-dimensional oscillator space. This subspace is stabilized by these two operators and the GKP state can be visualized as an infinite, periodic grid structure in the $(q,p)$ phase space. For the GKP code based on a square lattice, which we will consider here, the standard basis states are given as:
\begin{equation}
\begin{aligned}
\ket{0_{\text{GKP}}} &= \sum_{n \in \mathbb{Z}} \ket{q = 2n\sqrt{\pi}}, \\
\ket{1_{\text{GKP}}} &= \sum_{n \in \mathbb{Z}} \ket{q = (2n+1)\sqrt{\pi}}.
\end{aligned}
\label{eq:GKPStandardBasis}
\end{equation}
Similarly the GKP $X$ basis logical states are:
\begin{equation}
\begin{aligned}
\ket{+_{\text{GKP}}} &= \sum_{n \in \mathbb{Z}} \ket{p = 2n\sqrt{\pi}}, \\
\ket{-_{\text{GKP}}} &= \sum_{n \in \mathbb{Z}} \ket{p = (2n+1)\sqrt{\pi}}.
\end{aligned}
\label{eq:GKPXBasis}
\end{equation}

We can see that the grid corresponding to the basis state $\ket{0}$ ($\ket{+}$) is shifted by $\sqrt{\pi}$ along the $\hat{q}$ ($\hat{p}$) quadrature with respect to $\ket{1}$ ($\ket{-}$). Hence, by measuring the two stabilisers, which amounts to measuring both $\hat{q}$ and $\hat{p}$ quadratures of the GKP state modulo $\sqrt{\pi}$, we can detect and correct any small shifts (of size smaller than $\sqrt{\pi}/2$) in both quadratures, in a way that does not reveal the encoded logical information.

The two GKP stabilisers can in fact be measured using additional GKP ancilla modes through a Steane error-correction process. Application of a two-mode operation between the GKP data mode and the GKP ancilla can transfer the information about the noise from the data qubit onto the ancilla in such a way that the logical information is not revealed. Hence measuring the ancilla and applying a feedback displacement based on the measurement outcome enables GKP quantum error correction. More detailed information about this procedure can be found in Appendix~\ref{sec:GKPErrCorr}.

We note here that an ideal GKP state corresponds to a superposition of infinitely many infinitely squeezed states hence requiring infinite energy. Such states are unphysical and realistic GKP states have finite amount of squeezing, see Section~\ref{sec:realGKP}. This means that the information obtained from the measurement on the GKP ancilla is effectively noisy and therefore the feedback displacement will not bring the data state exactly to the logical space but will leave some residual displacement reflecting the finite amount of squeezing of the GKP ancilla. Here we consider a specific strategy of rescaling the measured GKP syndrome by a real number $c \in (0,1]$ before applying the feedback displacement~\cite{noh2019encoding, fukui2019high, yamasaki2020polylog}. The value of $c$ depends on the relation between the channel noise and the amount of GKP squeezing and is chosen such that the variance of the residual displacement after the feedback correction can be minimised, see Appendix~\ref{sec:GKPrescale} for details.

In practice, communication channels are corrupted by loss, not a shift in phase space. Nevertheless, it is known that the GKP code also works well against loss errors. This is because the sender can phase-insensitively amplify the GKP states (with a gain determined by the expected loss) such that the action of the effective channel results in random shift errors which the GKP code is designed to correct for~\cite{kim1996quantum,sabapathy2011robustness,ivan2011operator,noh2018quantum}. This strategy is described in more detail in Appendix~\ref{sec:GKPagainstLoss}.
 
\subsubsection{Repeater model}

In the considered architecture quantum information encoded in the GKP qubits is sent through the repeater chain as follows. After Alice performs the encoding operation, she applies the phase-insensitive amplification and sends the GKP qubit through the lossy channel towards the first GKP repeater. The repeater performs GKP correction first in $\hat{q}$ and then in $\hat{p}$ quadrature. After that it again applies the phase-insensitive amplification and sends the state to the next repeater. In this way the encoded qubit can effectively be transmitted to Bob. In our model we consider two sources of imperfections apart from loss in the communication channel. Firstly we assume a finite photon in- and out-coupling efficiency $\eta_0$ which quantifies the efficiency of transferring the photon from the fibre to the repeater and back into the fibre. Hence the total transmissivity of the lossy channel between two neighbouring repeaters separated by the distance $L$ is:
\begin{equation}
\eta = \eta_0 e^{-L/L_0},
\label{eq:eta}
\end{equation}
where $L_0$ is the attenuation length of the channel. Here we assume transmission at telecom frequency at which $L_0 = 22$ km. The second imperfection we consider is the finite amount of GKP squeezing. Under finite squeezing the GKP grid does not consist of delta functions but of Gaussian peaks with an overlaying envelope function such that the peaks in the Wigner function decay to zero height in the limit of large quadrature values. The standard deviation of these finitely squeezed Gaussian peaks is given by $\sigma_{\text{GKP}}$ and the amount of squeezing can also be quantified by comparing $\sigma_{\text{GKP}}$ to the standard deviation of a Gaussian peak of a coherent state given by $1/\sqrt{2}$. Hence squeezing expressed in dB can be defined as:
\begin{equation}
s = -10 \log_{10}(2\sigma_{\text{GKP}}^2).
\end{equation}
In our analysis, similarly to~\cite{noh2020fault}, we consider a conservative error model which allows us to describe a finitely squeezed GKP qubit as an ideal GKP state subjected to a random displacement according to a probability distribution parameterised by $\sigma_{\text{GKP}}$, see Section~\ref{sec:realGKP} for details. Since both finite GKP squeezing and the channel noise lead now to random displacement errors, we can reliably approximate the repeater performance by considering perfect error correction using infinitely squeezed ancillas, which however is now performed on the data qubits subjected to an effective communication channel. This effective channel now includes not only the noise coming from $\eta$ defined in Eq.~\eqref{eq:eta} but also from non-zero $\sigma_{\text{GKP}}$. That is we consider an approximation in which the noise from finite squeezing can be effectively incorporated into the channel and combined with the noise due to photon loss. This approximation enables us to construct a simple analytical model through which we can efficiently evaluate repeater performance, including optimisation over repeater spacing. We validate this analytical model against a numerical Monte-Carlo simulation, see Appendix~\ref{sec:approxAncIntoChannel} and Appendix~\ref{sec:GKrep} for more information about the model.

\subsubsection{Performance of the GKP repeater chain}

We quantify the performance of our scheme by calculating the achievable secret-key rate in bits per optical mode $r'$. This is a fundamental information-theoretic quantity that plays a key role in the studies of quantum communication~\cite{takeoka2014fundamental,pirandola2017fundamental} and the units of bits per mode are also often referred to as bits per channel use or bits per channel use per mode. We consider a six-state quantum key distribution (QKD) protocol~\cite{bruss1998optimal} supplemented with the two-way post-processing scheme called advantage distillation~\cite{renner2008security}. This scheme enables Alice and Bob to filter out a large fraction of erroneous rounds thus significantly increasing the achievable key rate in the high-noise regime. Specifically, we consider the advantage distillation protocol of~\cite{watanabe2007key} which for all noise regimes allows us to generate more key than with standard one-way post-processing. Moreover, since in the GKP error correction we independently correct errors in the $\hat{q}$ and $\hat{p}$ quadratures, the probability of a $Y$ flip is quadratically suppressed. This is because a logical $Y$-error can only happen if there is both a logical $X$ and $Z$ error. This asymmetry leads to the fact that the quantum bit error rate (QBER) will be much larger in the $Y$ basis than in the $X$ and $Z$ basis. Therefore we can make use of the result of~\cite{murta2020key} where it is shown that if advantage distillation is used, we will obtain the highest secret-key rate by using the basis with the highest QBER for key generation. See Appendix~\ref{sec:seckeyRateAD} for more details on the discussed QKD protocol and Appendix~\ref{sec:GKrep} for details on evaluating QBER for the GKP repeater chain.

We list the results in the top left table in FIG.~\ref{fig:table}. Specifically, we list the achievable distances over which secret-key rate in bits per optical mode stays above $r'=0.01$. We choose this specific value as a threshold as it allows us for an easy comparison of our scheme with the PLOB bound~\cite{pirandola2017fundamental}, which corresponds to the two-way assisted capacity of the pure-loss channel. This quantity describes the ultimate limit of repeater-less quantum communication. For perfect devices and as a function of the communication distance $L_{\text{tot}}$ it is given by $K(L_{\text{tot}}) = -\log_2(1-\exp(-L_{\text{tot}}/L_0))$. While it drops below $K(L_{\text{tot}}) = 0.01$ after $L_{\text{tot}}=109$ km, it stays positive for all distances $L_{\text{tot}}$. However, the amount of key that can be generated through such direct transmission becomes negligible for large distances. On the other hand, the secret-key rate of our repeater schemes starts dropping rapidly to zero at certain distance $L_{\text{tot}}$ such that the distance at which its value is given by $r'=0.01$ is close to the distance at which it falls to zero. This is due to the fact that the effective channel modelling the transmission through our repeater schemes is the Pauli channel, see Section~\ref{sec:MonteCarlo} and Appendix~\ref{sec:seckeyRateAD}. These features can also be seen in FIG.~\ref{fig:secfrac} for our concatenated-coded schemes. Hence the threshold value of 0.01 provides a good reference that allows us to investigate the communication distances for which the amount of generated key is non-negligible.

For each set of parameters we optimise the repeater separation such that the generated secret-key rate can be maximised, with the restriction that the minimum repeater separation is 250 m. We find that for all the parameter configurations, for which the achievable distance is larger than 100 km, the optimal repeater spacing that maximises secret-key rate at that achievable distance is always the minimum separation of 250 m. The main conclusion drawn from the obtained data is that in order to achieve communication over distances of 1000 km and larger, close to perfect photon coupling efficiency is needed with unrealistically high amount of GKP squeezing. This means that for the GKP encoding/decoding strategy based on phase-insensitive amplification, GKP code alone is not sufficient for achieving practical long-distance quantum communication. This motivates us to introduce the second level of encoding.

\subsection{Repeater architecture based on concatenated GKP and discrete-variable codes}
\label{sec:concatenatedrep}

Random displacements with components along the $q$ and $p$ axes that are larger in magnitude than $\sqrt{\pi}/2$ are not correctable by the GKP code alone. Therefore we consider a second level of encoding, either with a [[4,1,2]] code~\cite{grassl1997codes} which encodes one logical qubit in $4$ modes (i.e. $4$ GKP qubits) or with [[7,1,3]] Steane code~\cite{steane1996error}, which encodes one logical qubit in $7$ modes (i.e., $7$ GKP qubits). This higher-level encoding enables us to correct logical GKP errors and hence effectively to correct displacements with magnitude larger than $\sqrt{\pi}/2$.

\subsubsection{Concatenated-coded repeater architecture}

We consider a hybrid repeater architecture in a linear chain, with $N$ type-A (outer code) repeater nodes and $mN$ type-B (inner code) repeater nodes (where $m$ is an integer we optimise) such that there are $m$ type-B nodes (also referred to as GKP nodes) between consecutive type-A nodes (also referred to as multi-qubit nodes). Distance between consecutive repeater nodes (regardless of their type) is taken to be a constant we optimize. A type-A node waits to receive $4$ ($7$) modes---of a $4$-GKP-qubit-encoded ($7$-GKP-qubit-encoded) single logical qubit in a [[4,1,2]] code ([[7,1,3]] Steane code), corrupted by noise---performs a GKP error correction described in Section~\ref{sec:GKPrep} on all the modes followed by the outer-code error correction, and transmits in sequence the 4 (7) GKP modes of the $4$-mode-entangled ($7$-mode-entangled) state to the next hop (a type-B node). A type-B node simply applies GKP error correction as in the GKP repeater chain to each received mode, and sends it to the next node (which could be type-A or B).

We note that in order to maximise repeater performance, repeater architecture utilising only the more powerful type-A nodes would in most cases be sufficient. However, this would necessitate a dense placement of these multi-qubit nodes which require more quantum memory and processing, and are more resource expensive than type-B nodes. Therefore we will show that in order to optimise the performance-cost trade-off, it is beneficial to consider the hybrid architecture consisting of both types of nodes. We depict our repeater architecture in FIG.~\ref{fig}. Further details of our concatenated-coded repeater scheme are described in Appendix~\ref{sec:2ndLevel} and Appendix~\ref{sec:multiqubitrep}. 

\begin{figure*}
\includegraphics[width = 2\columnwidth]{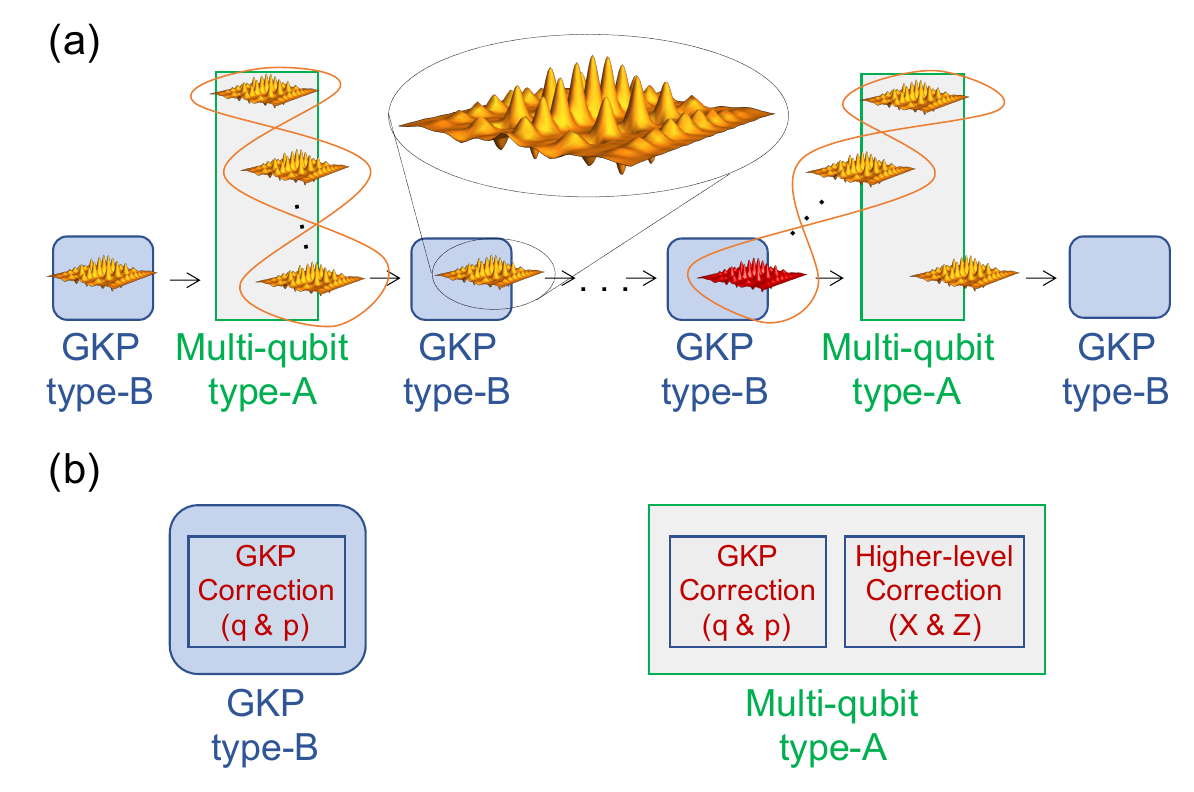}
\caption{\textbf{The proposed hybrid concatenated-coded repeater architecture.} (a) The Wigner function of the single-mode imperfect GKP state is depicted in the enlarged inset. We propose to use two levels of encoding and on the second level multiple single-mode GKP qubits are entangled together (marked with an orange ribbon) to encode a single logical qubit. The repeater architecture makes use of two types of repeaters. The first type are type-B repeaters (marked as blue) which can correct small displacement errors on the single GKP qubits that are sequentially transmitted between those stations. A displacement larger than the critical value cannot be corrected by the type-B repeaters and results in a logical error at the GKP level after the GKP correction (marked by the red GKP qubit). Therefore we sporadically introduce more powerful (and costly) type-A repeaters (marked as green) which store all the subsequently arriving GKP qubits from a given second-level encoded block. By jointly operating on all such qubits, the type-A repeaters can efficiently correct logical errors from the failed GKP corrections at the lower level. (b) High-level depiction of the operations performed in each repeater type. Type-B repeater corrects small random displacement errors in both $\hat{q}$ and $\hat{p}$ quadratures by measuring the stabilisers of the GKP code. Type-A repeater additionally corrects higher-level $X$ and $Z$ errors corresponding to the logical errors on the GKP level. These errors are corrected by measuring the $Z$ and $X$ stabilisers of the outer code. A detailed description of these operations for the type-B repeaters is provided in Appendix~\ref{sec:GKrep} and for the type-A repeaters in Appendix~\ref{sec:multiqubitrep}.}
\label{fig}
\end{figure*}

\subsubsection{GKP analog information}

The feature of our repeater scheme that enables us to significantly boost its performance with respect to other one-way repeater architectures based on error correction, is the use of analog continuous information from the GKP corrections at both type-B and type-A nodes in order to enhance the correcting capabilities of the outer code~\cite{fukui2017analog}. Specifically, measuring the GKP syndrome amounts to measuring each of the quadratures modulo $\sqrt{\pi}$, such that the syndrome is a number from a continuous interval $[-\sqrt{\pi}/2, \sqrt{\pi}/2)$. If the measured value is close to the boundary of this interval, there is a higher probability that the correction back to the logical GKP space will result in a logical error. This observation can be made mathematically rigorous, that is for a given measured syndrome value an error likelihood during correction can be established. This additional syndrome information, when passed from the type-B GKP repeater nodes to the type-A (multi-qubit) repeater nodes, enables the latter to correct more errors that are otherwise not correctable by the [[4,1,2]] and [[7,1,3]] codes. Specifically, for Pauli errors the [[4,1,2]] outer code is only an error-detection code that cannot correct any errors while the [[7,1,3]] outer code can normally correct only single-qubit errors. However, as shown in~\cite{fukui2017analog}, by utilising the continuous GKP syndrome information the [[4,1,2]] outer code can be transformed into an error-correction code which can correct most of single-qubit errors.  We also find that with the analog information the [[7,1,3]] outer code can correct most of both single- and two-qubit errors. See Appendix~\ref{sec:GKPErrCorr}, Appendix~\ref{sec:2ndLevel} and Appendix~\ref{sec:multiqubitrep} for mathematical details on calculating error likelihood from the analog information for our schemes.

\begin{figure}
\includegraphics[width = \columnwidth]{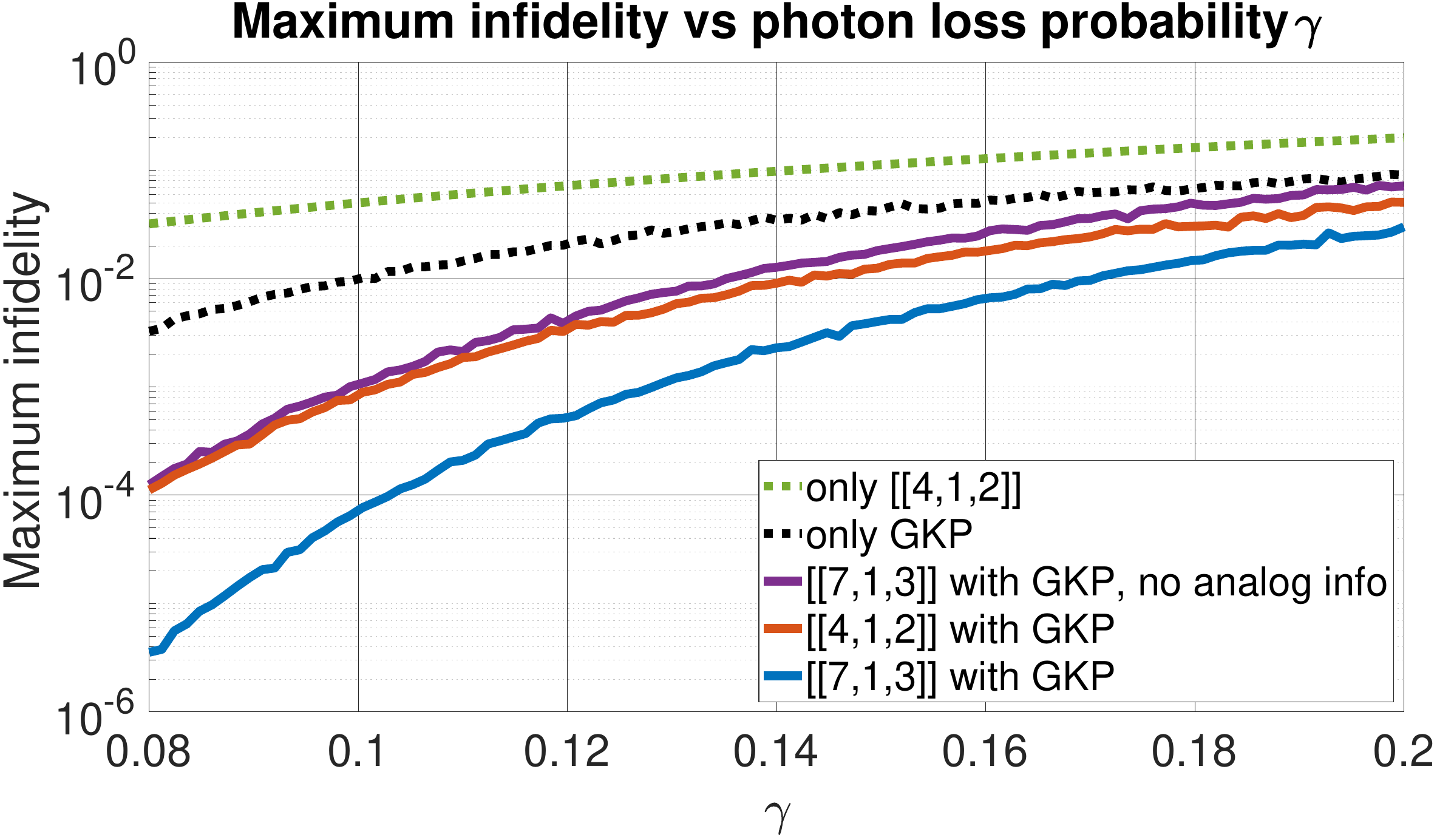}
\caption{\textbf{Maximum infidelity (maximised over all input states) versus photon loss probability $\gamma$ for the different considered encoding schemes.} We consider a scenario where Alice performs the encoding and sends the state to Bob, who performs error correction. For all plotted schemes apart from the ``only [[4,1,2]]'' scheme it is the GKP error correction, followed by the second-level error correction for the concatenated-coded schemes. We assume correction using infinitely squeezed ancilla modes, so that after correction the state will be in the code space, either with errors corrected or with a logical error. Additionally, we plot the maximum infidelity of $5\gamma^2$ for the purely discrete-variable scheme based on the [[4,1,2]] code (dotted green) as proposed in~\cite{leung1997approximate}. The curves for all the schemes apart from the ``only [[4,1,2]]'' scenario have been obtained from the simulated data and the relative error on the maximum infidelity is around $7\%$ for all the data points, see Section~\ref{sec:MonteCarlo} for details.}
\label{fig:analog}
\end{figure}

We illustrate the benefit of the analog GKP information in FIG.~\ref{fig:analog} where we consider a simple scenario in which Alice performs perfect encoding, applies phase-insensitive amplification to all the GKP qubits and then transmits the encoded state through the pure-loss channel with photon loss probability $\gamma = 1-\eta$. Bob then firstly performs a round of perfect (i.e. using infinitely squeezed ancilla modes) GKP correction on all the GKP qubits followed by the perfect outer-code correction. We then plot the maximum infidelity versus the loss probability $\gamma$ for the single-mode GKP encoding and the two-level-coded scheme. The maximum infidelity is given by one minus fidelity between the input state of Alice and the output state after transmission and correction of Bob. The specific input state is the state that minimises the fidelity or equivalently maximises the infidelity, see Appendix~\ref{sec:infidelity} for more details. For the case when the [[7,1,3]] outer code is used we plot separately the scenarios in which we do and do not make use of the additional analog information from the GKP correction round. We see that making use of this information provides a significant performance boost for our two-level-coded scheme. We also see that the concatenated-coded schemes improve the performance with respect to just GKP encoding. Furthermore, we see that for low loss, if we use the [[7,1,3]] outer code but do not make use of the analog information, the performance becomes similar to that of the scheme based on the [[4,1,2]] outer code, as both architectures can then correct only single-qubit errors. For larger losses the scheme based on the [[4,1,2]] outer code performs even better, because less qubits are used resulting in a smaller probability of an uncorrectable two-qubit error after GKP correction. The fact that these two schemes achieve a similar performance further justifies the capability of the analog information to transform the error-detecting code into an error-correcting code originally observed in~\cite{fukui2017analog}. 

Additionally we also compare the performance of our GKP-based schemes with a purely discrete-variable qubit scheme based on the [[4,1,2]] code~\cite{leung1997approximate}. Specifically, it has been shown that while this code is only an error-detection code against Pauli errors, it can be used for approximate error correction against a qubit amplitude damping channel, which corresponds to the pure-loss channel restricted to the vacuum and single-photon subspace. We see in FIG.~\ref{fig:analog} that making use of the full infinite-dimensional space with the GKP-based encodings that convert the action of the pure-loss channel into a random displacement channel allows for better performance than using only a qubit space of four optical modes against the amplitude damping channel. We note that we consider this additional strategy based on the purely discrete-variable encoding only in FIG.~\ref{fig:analog}. Therefore in the following sections whenever we refer to the schemes based on the [[4,1,2]] code and the [[7,1,3]] code, we always refer to the concatenated-coded schemes with the GKP encoding at the lower level.

\subsubsection{Performance of the concatenated-coded repeater architecture}

We again assess the performance of our repeater scheme for the task of generating shared secret key using the six-state QKD protocol with advantage distillation. We note that the two considered outer codes also correct the $X$ and $Z$ errors independently, similarly to the GKP code. This means that the quadratic suppression of $Y$-errors also applies to the concatenated-coded scheme. Therefore we can continue to make use of the result of~\cite{murta2020key} and maximise the key by extracting it in $Y$ basis. We note that Alice and Bob extract secret keys from the logical qubits. Hence, secret-key rate in bits per mode is calculated by dividing secret-key rates in bits per logical qubit, by $4$ for the case of the [[4,1,2]] outer code and by 7 for the case of the [[7,1,3]] outer code. We again refer the reader to Appendix~\ref{sec:seckeyRateAD} for the discussion of the considered QKD protocol. 

We perform Monte-Carlo simulation for the evolution of errors in the $(\hat{q},\hat{p})$ quadratures in our repeater scheme. From the simulation we estimate the quantum bit error rate (QBER) and calculate the expected asymptotic secret-key rate. We run the simulation for different placements of the type-A and type-B repeater nodes. Specifically, we assume at least one type-A station per 10 km. We then consider denser configurations with more type-A stations and for each of these cases we vary $m$, the number of type-B stations placed between neighbouring type-A stations. We consider all such configurations for which, similarly as in the case of the GKP repeater chain, the minimum separation between the neighbouring stations is 250 m, that is the sum of the number of type-A and type-B stations per 10 km cannot exceed 40. We describe the details of the simulation in Section~\ref{sec:MonteCarlo}.

\begin{figure*}
\includegraphics[width = 2.05\columnwidth]{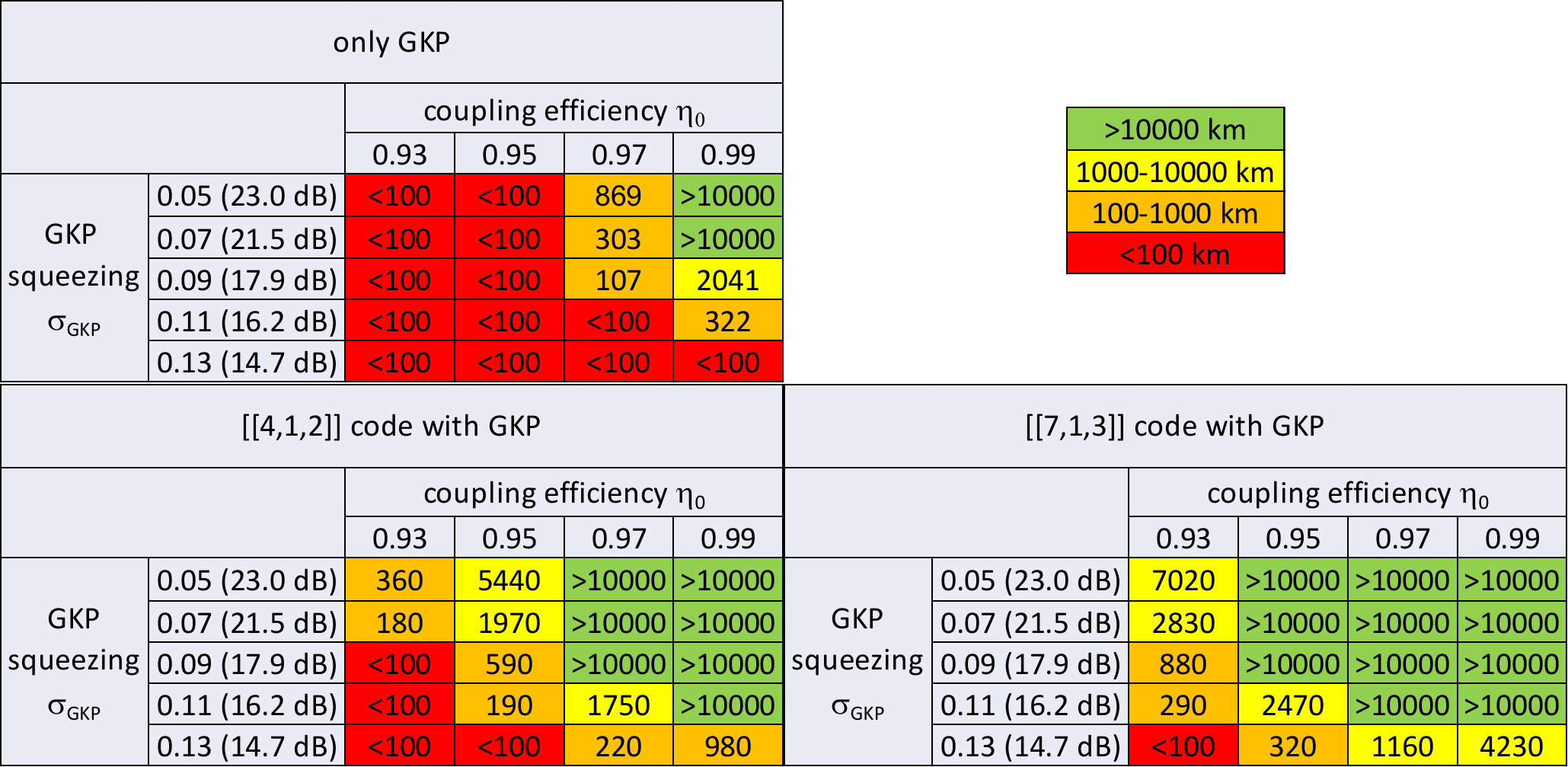}
\caption{\textbf{Achievable distances (km) over which secret-key rate in bits per mode stays above $r'=0.01$ for the three considered repeater architectures.} We see that using two levels of encoding allows for achieving larger distances with more relaxed hardware parameters than just a single level of GKP encoding alone. For comparison, the corresponding distance over which the capacity of the pure-loss channel drops to $K(L_{\text{tot}}) = 0.01$ when perfect devices and the same attenuation length are considered is $L_{\text{tot}} = 109$ km. This is the distance over which we could maintain the secret-key rate per optical mode larger than 0.01 if we did not have access to quantum repeaters but could perform direct communication using optimal encoding and decoding strategy with perfect hardware. For the ``only GKP'' case, the data were obtained using 10 iterations of the binary search method over distance in the interval $[0,10000]$ km resulting in approximately 10 km accuracy, where for each of the considered distances the secret-key rate was evaluated using the analytical model described in Appendix~\ref{sec:GKrep}. For the concatenated-coded schemes we performed numerical Monte-Carlo simulations such that the effective error on the achievable distances is around 10\%, see Section~\ref{sec:MonteCarlo} for details. }
\label{fig:table}
\end{figure*}

In the first step we consider only the repeater performance, that is we look for the repeater placement configuration that maximises the achievable secret-key rate. We look for the achievable distances with the concatenated-coded schemes, for which the achievable secret-key rate in bits per mode is larger than $r'=0.01$. The results are presented in the bottom two tables in FIG.~\ref{fig:table}. We see that the achievable distances are much larger and can be attained with more relaxed parameters than for the GKP repeater scheme. Specifically, for $\eta_0 = 0.97$, the architecture based on the [[4,1,2]] code ([[7,1,3]] code) enables to achieve secret-key rate per optical mode larger 0.01 for total distances larger than 1000 km already with 16.2 dB (14.7 dB) of squeezing. For the [[7,1,3]] code achieving such secret-key rate for total distance close to 1000 km is also possible with much lower photon coupling efficiency of $\eta_0 = 0.93$ if 17.9 dB of GKP squeezing is considered.

All the values from the tables in FIG.~\ref{fig:table}, can also be compared against the PLOB bound~\cite{pirandola2017fundamental} introduced in Section~\ref{sec:GKPrep} and describing the limits of direct transmission. Since its value drops below $K(L_{\text{tot}})=0.01$ after $L_{\text{tot}}=109$ km, we see that for most considered parameter regimes the concatenated-coded schemes easily overcome the optimal direct transmission. The relaxed hardware requirements, large achievable distances and performance against the PLOB bound show that our concatenated-coded schemes are promising architectures for long-distance quantum communication.

\subsubsection{Performance versus cost trade-off of the concatenated-coded repeater architecture}

Clearly the more costly denser placement of type-A repeaters results in better performance. Therefore in the second step we also take the repeater cost into account. Specifically, for each end-to-end distance we aim to minimise the normalised cost function defined as
\begin{equation}
C' = \frac{\text{Resources used per km}}{\text{achieved secret-key rate per mode}}.
\label{eq:costfunction}
\end{equation}
The natural way of counting the resources will clearly depend on the physical implementation of our scheme. Here we count the resources by considering the needed number of GKP storage modes and the storage duration needed for these modes in both types of repeater nodes, see Section~\ref{sec:Cost} for mathematical definition of the cost function. Specifically we consider not only the storage of the data modes but also of the ancilla modes needed for error correction. We consider a discretization of all the operations into time steps, where one time step is needed either for preparing a GKP ancilla state or for performing all the two-mode Gaussian operations between a single data mode and the ancilla mode for the purpose of the inner or outer code stabiliser measurement. We describe the details of our scheduling procedure in Appendix~\ref{sec:ScheduleAndCost}. For each of the needed storage modes we count the number of time steps that this mode must be able to store the state for without losing or decohering it. Then we sum the number of these time steps for all the needed storage modes inside each repeater type to obtain the total cost of placing a given repeater. This way of estimating repeater cost applies e.g. to an architecture in which the repeaters would consist of coupled cavities, where each cavity is effectively used as a quantum memory for a single GKP mode during the correction operations. We discuss possible implementations in more detail in Section~\ref{sec:discussion} as well as in Section~\ref{sec:Cost}.

For the above discussed strategy of estimating the resources, the exact cost values are explicitly stated in Section~\ref{sec:Cost} and derived in Appendix~\ref{sec:ScheduleAndCost}. Here we just note that in our architecture we add additional operations inside type-A repeaters which aim at decreasing the noise effect of finite GKP squeezing. This includes performing additional GKP corrections between multi-qubit stabiliser measurements and repeating the measurement of the second-level syndrome for better measurement reliability. As a result we find that for the proposed scheduling of the operations the type-A repeater for the [[4,1,2]] code costs around 17 times more than the type-B repeater, while the type-A repeater for the [[7,1,3]] code costs around 78 times more than the type-B repeater. The numerator in the cost function in Eq.~\eqref{eq:costfunction} is just the sum of the costs of Alice's encoding station and all the repeaters in the given configuration (including Bob's decoding station which also performs quantum error correction and can be treated as a type-A repeater, see Appendix~\ref{sec:seckeyRateAD} for more details) over the total communication distance $L_{\text{tot}}$, divided by this distance in km, see Section~\ref{sec:Cost} for more detail. We note that while we do not specify the scheduling of operations at Alice's encoding station, higher-level encoding from GKP qubits can be achieved by performing the same type of operations as performed inside repeaters. These include CV two-qubit Clifford gates and additional GKP corrections to limit the accumulation of errors due to finite squeezing in GKP modes. We have verified that such a procedure enables reliable higher-level encoding where the probability of a logical error on any of the GKP data qubits during this procedure is smaller than the corresponding probability of error due to performing operations with finitely squeezed GKP ancilla modes during error-correction inside the type-A repeaters. The complexity of such an encoding does not exceed the corresponding complexity of operations performed inside the type-A repeater and therefore in the cost function we assign to the encoding the same cost as to the type-A repeater. In our analysis the cost function is minimised independently for each distance over all the repeater placement configurations.

\begin{figure}
\includegraphics[width = \columnwidth]{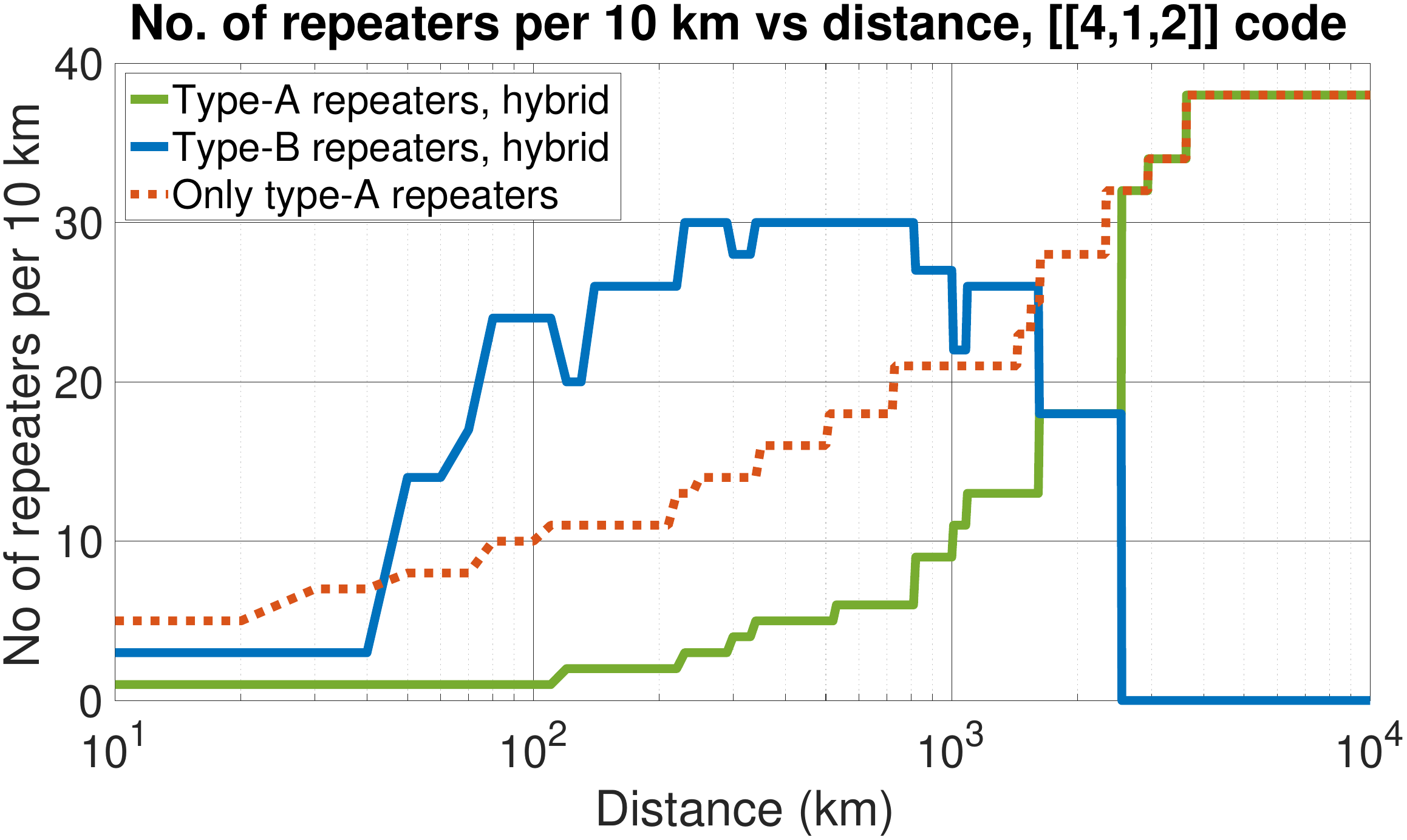}\\
\vspace{0.25cm}%
\includegraphics[width = \columnwidth]{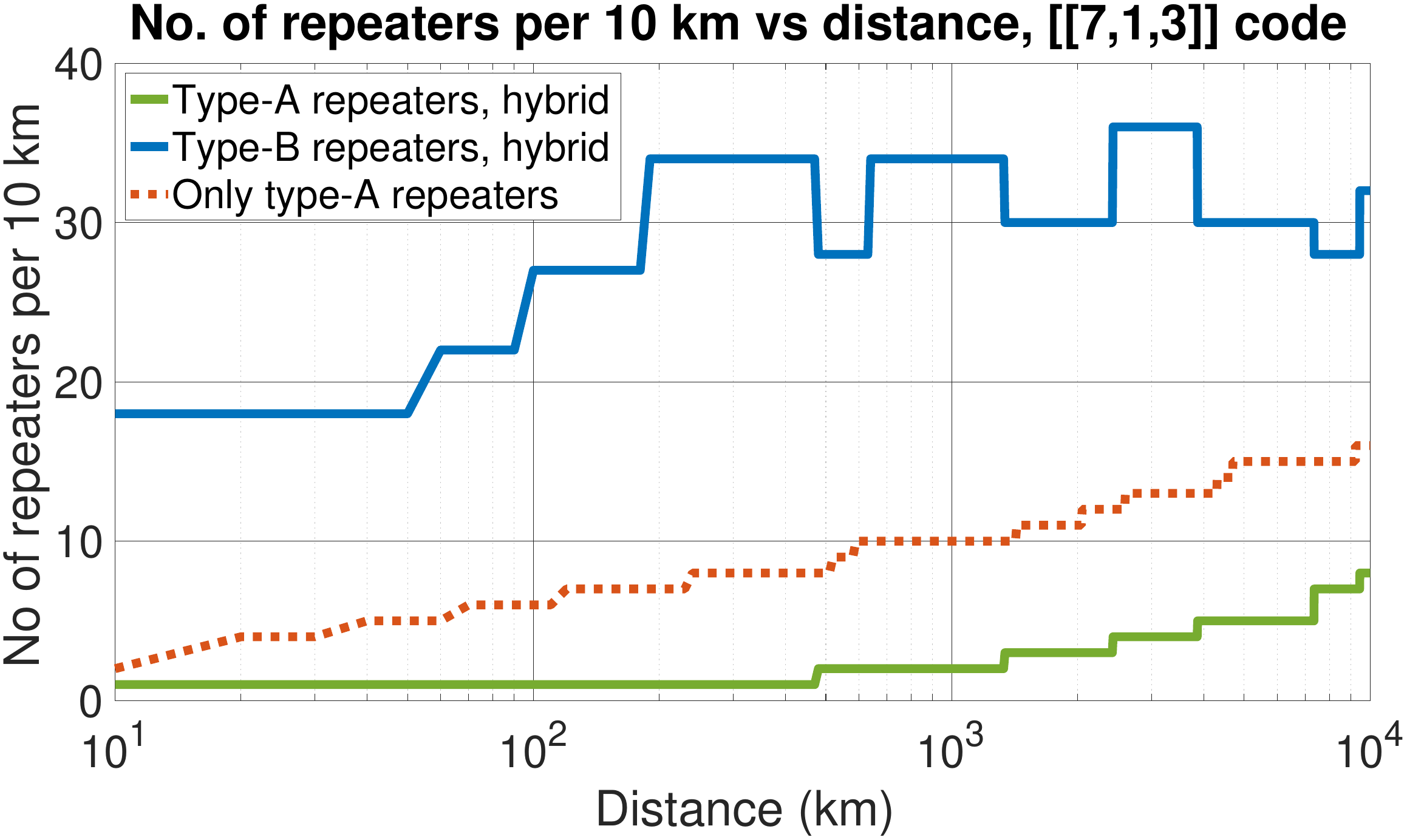}
\caption{\textbf{Optimal repeater configuration which minimises the cost function for the concatenated-coded schemes.} The figure shows the optimal repeater placement for the architectures based on the [[4,1,2]] code and [[7,1,3]] code versus communication distance. We show the optimal number of type-A repeaters (solid green), type-B repeaters (solid blue) as well as the number of stations in the architecture that uses only type-A repeaters (dotted red). The considered parameters are $\eta_0 = 0.97$ and $\sigma_{\text{GKP}} = 0.09$. We see that using the hybrid scheme allows for reducing the density of the more costly type-A repeaters, with respect to the scheme that uses only type-A repeaters. We also see that since the [[7,1,3]] code type-A repeaters are powerful but costly, we need much less of them than the [[4,1,2]] code type-A repeaters. The type-B repeaters are cheap and so already for shorter distances it is beneficial to place them densely. For the scheme based on the [[4,1,2]] code we observe that for distances larger than around 2500 km the architecture utilising only type-A repeaters becomes optimal. The effect of the simulation error is described in Section~\ref{sec:MonteCarlo}.}
\label{fig:stations}
\end{figure}

Here we perform the cost function analysis for the scenario with $\eta_0 = 0.97$ and with 17.9 dB of squeezing corresponding to $\sigma_{\text{GKP}}=0.09$. In FIG.~\ref{fig:stations} we depict the optimal repeater placement configuration for each distance for the concatenated-coded architectures. We plot the optimal number of repeaters per 10 km for the hybrid architecture and for comparison for the architecture that uses only type-A repeaters. We see that the hybrid architecture enables us to use less of the expensive type-A repeaters thanks to the help of the cheaper type-B repeaters. Since type-B repeaters are cheap, we see that already for shorter distances it is optimal to place them densely. Moreover, we see that since the [[7,1,3]] code repeaters are more powerful than [[4,1,2]] code repeaters, we need less of the former ones in the first architecture than we need of the latter ones in the second one. We also observe that for the hybrid architectures the optimal number of type-A repeaters increases monotonically with distance and the stepwise increase of this number may result in a stepwise decrease of the optimal number of type-B stations.

\begin{figure}
\includegraphics[width = \columnwidth]{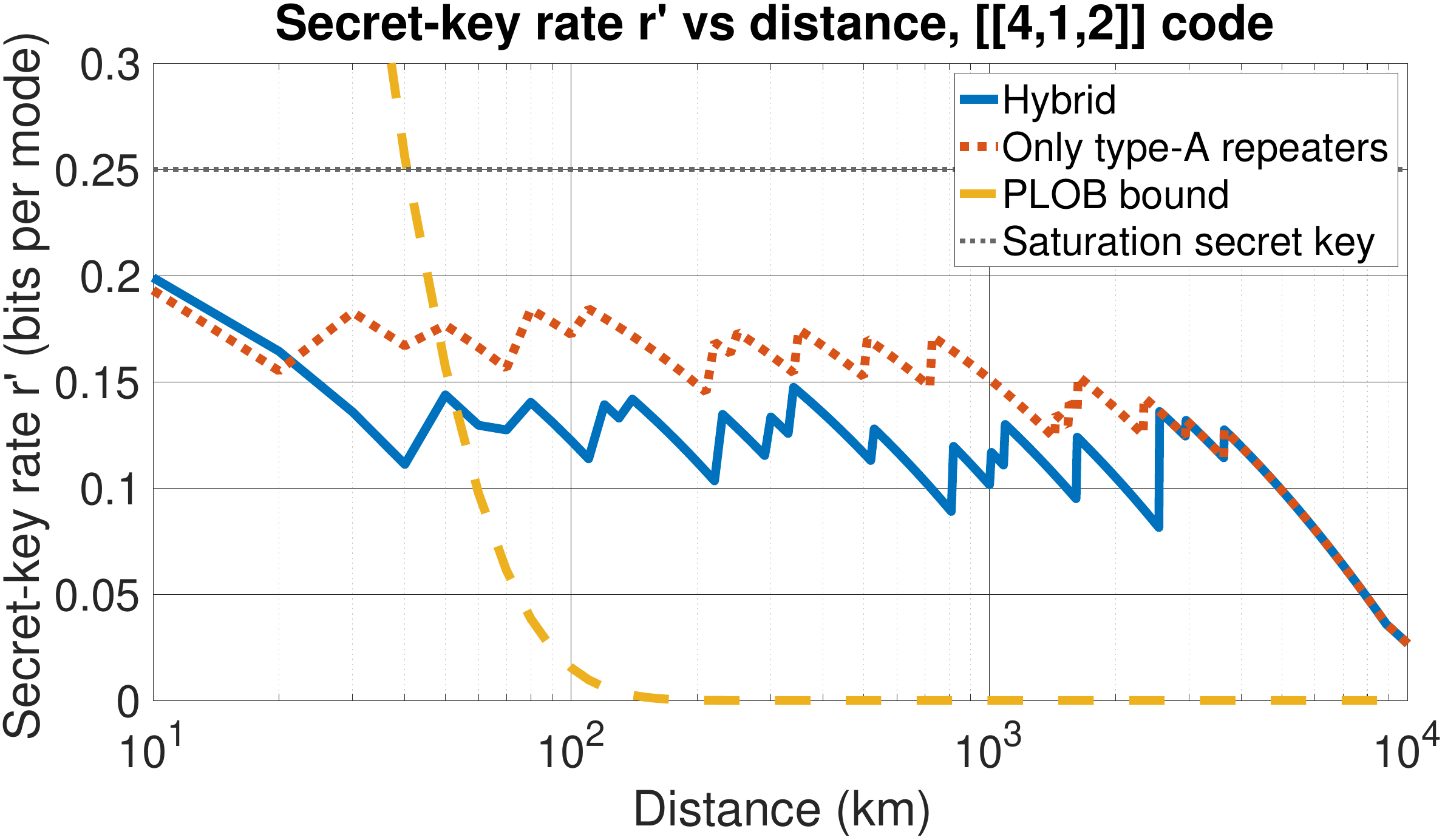}\\
\vspace{0.25cm}%
\includegraphics[width = \columnwidth]{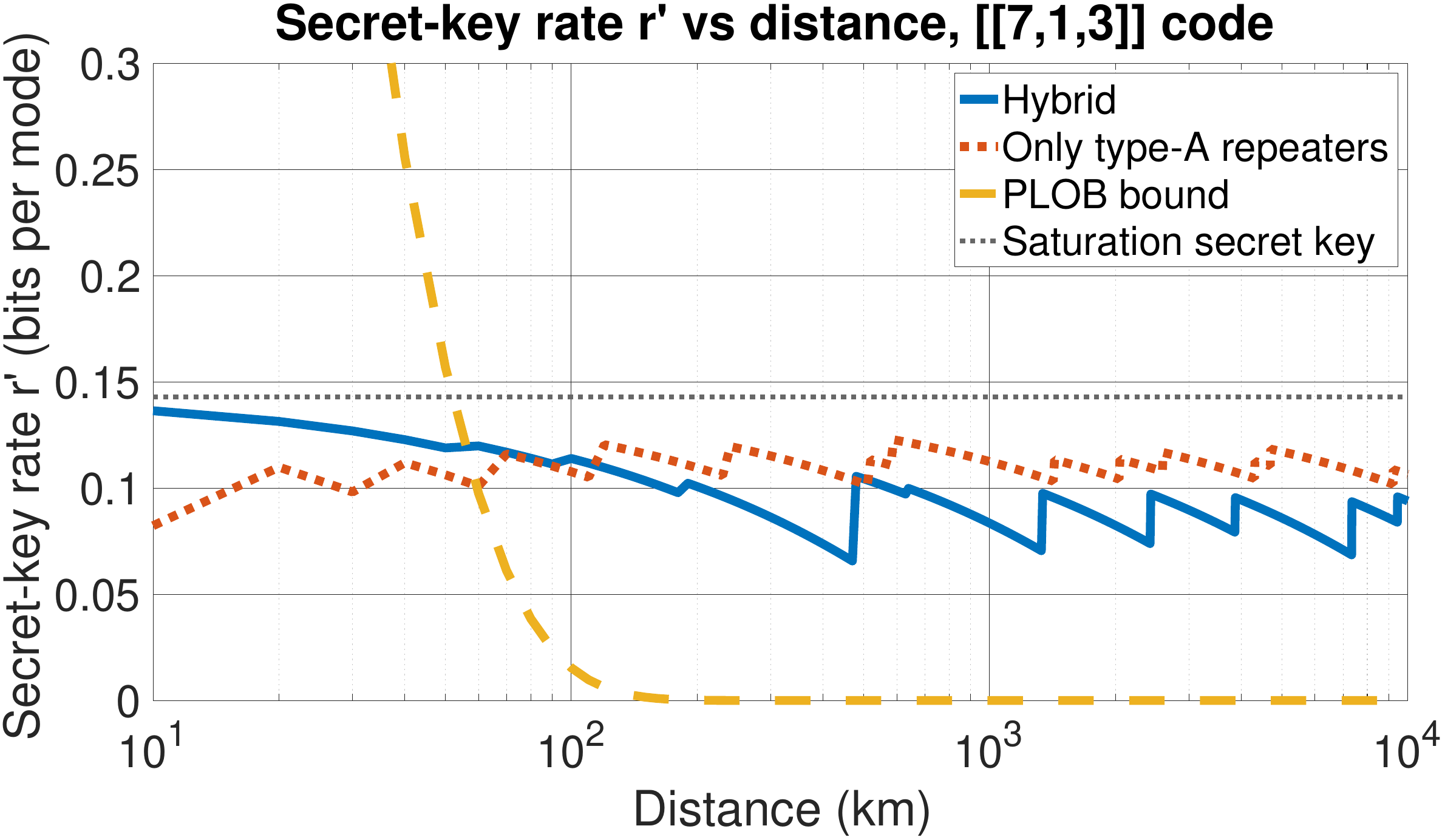}
\caption{\textbf{Secret-key rate in bits per optical mode $r'$ versus distance (km) for the concatenated-coded schemes.} We consider four schemes: two schemes based on the [[4,1,2]] code and two schemes based on the [[7,1,3]] code for $\eta_0 = 0.97$ and $\sigma_{\text{GKP}} = 0.09$. The blue solid lines correspond to the strategy with two types of repeaters, type-A and type-B stations, while the red dotted lines correspond to the schemes for which only multi-qubit repeaters are used. We see that the hybrid schemes exhibit similar performance as the schemes with only type-A stations. We also plot the PLOB bound, corresponding to the two-way assisted secret-key capacity of the pure-loss channel and therefore the ultimate limit of repeater-less quantum communication with perfect devices. Finally, we mark the saturation secret key, which is the maximum value $r'_{\text{max}} = 1/n$ attainable with zero QBER. For the [[4,1,2]] code it is 1/4 and for the [[7,1,3]] code it is 1/7. We observe that the overall trend for the [[7,1,3]] code schemes is that they maintain a steady secret-fraction $r'$ for all the considered distances while for the [[4,1,2]] code architectures the key starts decreasing for larger distances. All the schemes easily overcome the PLOB bound already for distances much smaller than 100 km. The effect of the simulation error is described in Section~\ref{sec:MonteCarlo}.}
\label{fig:secfrac}
\end{figure}

We also describe the behaviour of the secret key under the cost function minimisation. Again, for each of the two second-level codes, we consider two architectures, one using only type-A stations and the second one using both types of repeaters. Let us then consider the amount of secret key in bits per mode $r'$ that can be generated by each of these schemes. We plot $r'$ for all the four architectures in FIG.~\ref{fig:secfrac}. We see that the architectures based on the [[4,1,2]] code ([[7,1,3]] code) achieve $r'> 0.02$ ($r'> 0.06$) for all the distances up to 10000 km. While under the cost function minimisation the hybrid schemes generate for most distances slightly less key than the corresponding schemes based only on type-A stations, there is no significant difference in performance trend with distance between these two schemes. The overall ``zig-zag'' shape of the curves is caused by discrete changes in the optimal repeater placement configurations with changing distance. We also see that for most distances the [[4,1,2]] code architectures can generate more key per optical mode than the corresponding [[7,1,3]] code architectures since the former ones need less modes to transmit a logical qubit. However, we see that after around 4000 km the key starts decaying for the architectures based on the [[4,1,2]] code. This reveals that the [[4,1,2]] code schemes will not be able to sustain secret-key generation for distances much larger than 10000 km. On the other hand the [[7,1,3]] code architectures maintain a steady $r'$ for all the distances.

This conclusion can be also drawn from the consideration of the simulation error. Specifically, since the simulation data has at most 10\% relative error, we have also investigated the corresponding behaviour for the upper-bound on the simulated logical $X$ and $Z$ flip probabilities. In particular we have minimised the cost-function and investigated the resulting secret-key rate for the scenario when the obtained $X$ and $Z$ flip probabilities are increased by 10\% for all the repeater placement configurations. We find that this does not have any significant effect on the architectures based on the [[7,1,3]] code, that is the secret-key rate still stays such that $r'> 0.06$ for all the distances. However, for the [[4,1,2]] code schemes $r'$ drops below 0.02 for 10000 km now. This supports the observation that for the considered parameters the [[7,1,3]] code architectures remain robust even at such large distances. On the other hand, the [[4,1,2]] code schemes, which for distances close to 10000 km require placement of type-A repeaters almost every 250 m, become sensitive to noise at these distances.

We note that for comparison in FIG.~\ref{fig:secfrac} we also plot the PLOB bound~\cite{pirandola2017fundamental}. We see that for the considered set of parameters all our architectures overcome the PLOB bound already for distances much smaller than 100 km. 

\begin{figure}
\includegraphics[width = \columnwidth]{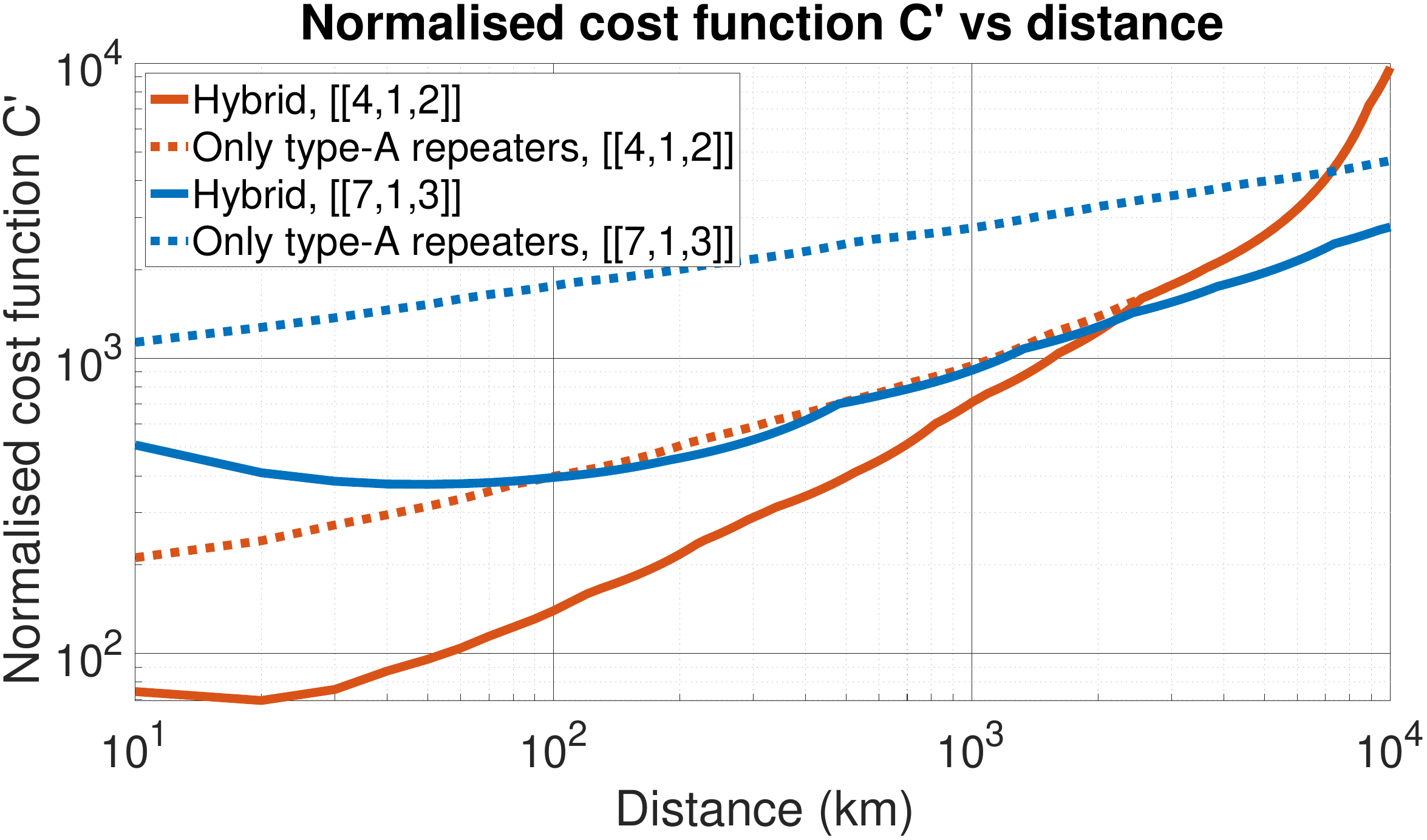}
\caption{\textbf{Normalised cost function $C'$ versus distance (km) for the concatenated-coded architectures.} The considered parameters are $\eta_0 = 0.97$ and $\sigma_{\text{GKP}} = 0.09$. We see that the hybrid scheme enables us to reduce the cost function with respect to the corresponding scheme that uses only type-A repeaters. We also see that for shorter distances it is more efficient to use the architecture based on the [[4,1,2]] code (solid red) since the type-A repeaters of the [[7,1,3]] code are expensive while their large error correcting capabilities are not needed for these distances. For larger distances the cost function is smaller for the [[7,1,3]] code architecture (solid blue), because these high error correcting capabilities allow for achieving better performance-cost trade-off than the use of the cheaper but less efficient [[4,1,2]] repeaters. The visible initial decrease of the cost function with distance for the solid blue line is caused by the initial cost of Alice's encoding station. The effect of the simulation error is described in Section~\ref{sec:MonteCarlo}.}
\label{fig:costfunction}
\end{figure}

The final question is how the costs of these different schemes compare. We plot the normalised cost function for all these schemes in FIG.~\ref{fig:costfunction}. Since we have already verified that the performances of the hybrid scheme and the scheme utilising only type-A repeaters are similar, we conclude from this plot that the hybrid architecture enables us to save a lot of resources in comparison to the architecture based only on the more expensive type-A repeaters. This is because the performance benefits of using the more expensive type-A repeaters can be maintained by replacing some of them with the cheaper type-B repeaters. Secondly we see that for shorter distances it is more resource-efficient to use the architecture based on the [[4,1,2]] code while for larger distances the [[7,1,3]] architecture is preferable. This is linked to the fact that the type-A repeaters in the [[7,1,3]] code architecture are more expensive but also more powerful than the type-A repeaters in the [[4,1,2]] code architecture. For smaller overall losses at shorter distances these larger powerful repeaters are not necessary while for larger distances they are more efficient at overcoming the overall high losses.

\section{Discussion}
\label{sec:discussion}

Let us now discuss the experimental challenges related to our scheme. These will naturally depend on its physical implementation. In our cost-function analysis we have assumed an architecture where all the GKP modes need to be placed in effective quantum memories for the duration of the error-correction operations. A possible implementation of such a CV-quantum memory that allows for preparation of the highly non-classical GKP state as well as GKP error correction is a superconducting microwave cavity as experimentally demonstrated in~\cite{campagne2019stabilized}. Storing GKP data and ancilla modes in coupled microwave cavities at the repeater nodes would clearly require an efficient transduction between the telecom optical channel and the microwave regime~\cite{andrews2014bidirectional, han2020cavity, zhong2020proposal, rueda2019electro, lambert2020coherent}.

On the other hand one could also consider an all-optical implementation where all the repeaters perform error correction online on the flying GKP qubits stored and coupled to ancilla GKP modes directly in the optical fibre. Such an implementation would be similar in spirit to the all-photonic repeaters~\cite{Azuma_15, pant2017rate}. Since storage of additional GKP modes in the same spool of optical fibre does not require additional quantum memories, the resource cost analysis for such an all-optical implementation will clearly differ from the one using the microwave-cavity-based repeaters. Since the all-optical preparation of GKP states remains a significant experimental challenge~\cite{su2019conversion, pirandola2004constructing, motes2017encoding, vasconcelos2010all, weigand2018generating}, the cost analysis presented here and the corresponding scheduling of operations discussed in Appendix~\ref{sec:ScheduleAndCost} are performed with respect to a model which is more suitable for the microwave cavity implementation.

Let us now compare the hardware requirements of our scheme with respect to other error-correction-based repeater proposals. The requirement on the photon coupling efficiency for our scheme is similar as for the proposed error-correction-based repeater architectures utilising discrete-variable encoding using tree codes and parity codes~\cite{muralidharan2014ultrafast, ewert2017ultrafast, lee2019fundamental, borregaard2019one, Azuma_15, pant2017rate}. On the other hand the need for operating on large number of modes/qubits which is required for these schemes is removed in our scheme at the expense of the requirement for being able to prepare highly non-classical and highly squeezed GKP states in each of these few modes. We note that while for the discrete-variable encodings large number of entangled qubits are needed, the number of required entangled photons can be significantly reduced by multiplexing which allows for encoding multiple qubits in a single photon~\cite{piparo2020resource}. In fact, if the average number of photons needed for the encoding is considered to be the main resource, then the relative cost of utilising such multiplexed schemes versus single-mode GKP encoding for combating photon loss depends on the channel transmissivity, see~\cite{piparo2020resource}.

The experimentally demonstrated amount of GKP squeezing is in the regime 7.5 - 9.5 dB~\cite{fluhmann2019encoding, campagne2019stabilized, de2020error}. Therefore more experimental progress is needed in order to achieve the required levels of squeezing predicted by our analysis. It would also be beneficial to study the effect of finite gate fidelity and finite storage time in our architecture~\cite{noh2020fault}. Furthermore, Steane error correction for GKP qubits has been demonstrated using an ancilla transmon qubit~\cite{campagne2019stabilized}. Hence experimental procedures for using ancilla GKP modes need to be developed as well as the procedures for encoding and decoding the proposed two-level coded qubits.

We note that our motivation for using the metric of secret-key rate in bits per optical mode as a way of assessing the repeater performance comes from the fact that this figure of merit has a clear operational meaning and quantifies the feasibility of a specific quantum communication task. Moreover, closely related figure of merits are also throughput and latency which quantify how much secret key can be generated per unit time and how long it takes to generate the first raw key bit respectively. We discuss the performance of our schemes with respect to these metrics in Appendix~\ref{sec:ThroughputLatency}. However, the considered repeater schemes can also enable and facilitate implementation of various tasks other than QKD. Specifically, the deterministic nature of these schemes could enable deterministic quantum state transfer as well as deterministic remote entanglement generation.

It is important to mention that the GKP encoding/decoding strategy based on effectively converting the pure-loss channel into the Gaussian random displacement channel using phase-insensitive amplification is an achievable strategy but not an optimal one. There exists a numerical proof based on semi-definite optimisation, that a more efficient strategy of using GKP codes against the pure-loss channel exists~\cite{albert2018performance}. It is plausible that under the optimal decoding strategy, the single-mode GKP architecture with coupling efficiency and squeezing levels similar to the ones considered in our concatenated-coded schemes will be sufficient for long-distance quantum communication. However, the numerical nature of this proof makes it difficult to extract from the solution the corresponding decoding procedure. In particular such an optimal decoding procedure might require much more complex operations than phase-insensitive amplification as well as the need to use large number of ancilla systems. Therefore, further study is needed to establish the optimal decoding procedure for correcting loss errors using GKP code and to evaluate its complexity and experimental feasibility.

Let us now summarise the future outlook of this work. Firstly, utilisation of microwave cavities in long-distance quantum communication will require experimental realisation of highly efficient transduction between the microwave and optical regimes~\cite{lambert2020coherent}. An alternative solution would be an all-optical realisation which would require implementation of GKP state preparation directly in the optical regime~\cite{su2019conversion, pirandola2004constructing, motes2017encoding, vasconcelos2010all, weigand2018generating}. Secondly, it could also be beneficial to incorporate autonomous GKP error correction into our procedure which does not require active measurements and feedback~\cite{royer2020stabilization,de2020error}. This technique could potentially be more efficient than the considered GKP Steane error correction, though it would not provide us with the additional analog information which we have seen plays a crucial role in the performance of the concatenated-coded schemes. Thirdly, additional improvements could come from investigation of the optimal decoding strategy for GKP code used against the action of the pure-loss channel~\cite{albert2018performance}. Fourthly, our analysis shows that more experimental progress on GKP squeezing is needed as well as implementation of high photon coupling efficiency in order for the considered schemes to become practical. Finally, given the nature of the concatenated-coded repeater architecture in which multiple GKP qubits from a single outer-code encoding block are transmitted in sequence, it could be valuable to investigate the use of a quantum convolutional code~\cite{ollivier2003description} as the outer code, as this would allow for easier online error correction, e.g. in the all-photonic implementation.

\section{Methods}

\subsection{Realistic GKP states}
\label{sec:realGKP}

Since $\ket{q}$ and $\ket{p}$ eigenstates are unphysical and require infinite amount of squeezing and energy, the ideal GKP states defined in Eq.~\eqref{eq:GKPStandardBasis} and Eq.~\eqref{eq:GKPXBasis} are also unphysical. Therefore we will consider imperfect GKP states corresponding to a finite amount of squeezing. Let $\ket{\psi_{\text{GKP}}}$ denote an ideal GKP state. Then an approximate GKP state can be obtained by applying a Gaussian envelope operator $\exp(-\Delta^2 \hat{n})$ to the perfect GKP state. Here $\hat{n}$ is the photon number operator and $\Delta$ describes the width of each peak in the grid-structure of the GKP Wigner function. We can use displacement operators $\hat{D}(\alpha) = \exp[\alpha \hat{a}^\dag - \alpha^* \hat{a}]$ to rewrite the approximate GKP state $\ket{\psi^{\Delta}_{\text{GKP}}}$ as~\cite{noh2020fault}:
\begin{equation}
\begin{aligned}
\ket{\psi^{\Delta}_{\text{GKP}}}  &\propto \int \frac{d^2\alpha}{\pi} \Tr[\exp(-\Delta^2 \hat{n})\hat{D}^\dag(\alpha)]\hat{D}(\alpha)\ket{\psi_{\text{GKP}}} \\
						&\propto \int d^2\alpha \exp\left[-\frac{\abs{\alpha}^2}{2\sigma^2_{\text{GKP}}}\right] \hat{D}(\alpha)\ket{\psi_{\text{GKP}}}.
\end{aligned}
\label{eq:impGKPsup}
\end{equation}
Here $\sigma_{\text{GKP}}^2 = (1- e^{-\Delta^2})/(1 + e^{-\Delta^2})$. We see that an imperfect GKP state can be described as a coherent superposition of randomly displaced ideal GKP states with a Gaussian envelope centred at zero displacement. Similarly as in~\cite{noh2020fault}, in our simulation we consider a more conservative error model for imperfect GKP states. Specifically, let us define the Gaussian random displacement channel as:
\begin{equation}
\mathcal{N}_{\text{disp}}[\sigma](\rho) = \frac{1}{\pi \sigma^2}\int d^2\alpha \exp\left[-\frac{\abs{\alpha}^2}{\sigma^2}\right] \hat{D}(\alpha)\rho \hat{D}^\dag(\alpha).
\label{eq:randomgaussiandisplChannel}
\end{equation}
Then by adding further twirling noise to the state in Eq.~\eqref{eq:impGKPsup} we can remove the coherences between the superposition terms with different values of the displacement amplitude $\alpha$. This can be done by applying random displacements by an integer multiple of $2\sqrt{\pi}$ in each quadrature, such that for the relevant amount of GKP squeezing considered here, we obtain a state that can be described as~\cite{noh2020fault}:
\begin{equation}
\rho_{\text{GKP}}[\sigma_{\text{GKP}}] = \mathcal{N}_{\text{disp}}[\sigma_{\text{GKP}}](\dyad{\psi_{\text{GKP}}}).
\end{equation}
Hence we can simulate the imperfect GKP state by sampling displacement values $\xi_q^{\text{GKP}}$ and $\xi_p^{\text{GKP}}$ along the $\hat{q}$ and $\hat{p}$ quadratures respectively from the normal distribution centred at zero and with standard deviation $\sigma_{\text{GKP}}$: $\xi_q^{\text{GKP}}, \xi_p^{\text{GKP}} \sim \mathcal{N}(0,\sigma_{\text{GKP}})$. We then consider an ideal GKP state that has been displaced according to these values.

\subsection{Cost function}
\label{sec:Cost}

In this section we make the notion of the repeater cost mathematically precise by defining a cost function whose minimisation aims at finding the best trade-off between the repeater performance and the resource cost. We also propose a specific scheduling procedure for the operations in all the repeaters and aim to minimise the cost function under this scheduling model.

The resource cost as well as duration and time scheduling of all the operations performed within the proposed repeaters will naturally depend on the physical implementation of our scheme. Two possible implementations for which the scheduling of the operations and the natural way of counting the resources would be very different are repeaters that store GKP modes inside microwave cavities and all-optical stations in which all the operations are performed on the fly while the GKP data and ancilla qubits are stored in the spools of optical fibre. The main difference between the two implementations from the perspective of the scheduling of operations as well as estimating resource cost is the fact that the first implementation entails the use of effective quantum memories that are required for storing the GKP modes. Hence if the number of such memories (e.g. microwave cavities) is limited, then not all the GKP data modes can be operated on simultaneously, while increasing the number of such available memories will clearly increase the resource cost of the stations. On the other hand, the all-optical implementation does not involve the concept of such quantum memories, as all the modes are operated on in the optical fibre. Hence the main limiting factor with respect to the delay between the consecutive GKP qubits will be in this case the repetition rate of the GKP source. As discussed in Section~\ref{sec:discussion} here we perform the analysis under the model of the first implementation involving the CV-quantum memories, motivated by the experimental demonstration of GKP error correction in a superconducting microwave cavity~\cite{campagne2019stabilized}.

For the considered model, the cost of the resources can be measured by the amount of GKP storage modes times the storage time in all the repeaters needed for communication over the distance $L_{\text{tot}}$. Let $t_{\text{GKP}}$ denote the cost of the single GKP repeater and $t_{\text{multi-qubit}}$ the cost of the single multi-qubit repeater. Then for each of these repeater types
\begin{equation}
t = \sum_{i = 1}^m k_i.
\end{equation}
Here $m$ denotes the number of storage modes (both for data and ancilla GKP qubits) that are required in a given repeater. Then the mode $i$ in that repeater needs to be able to store a GKP qubit for $k_i$ time steps defined below.

These repeater costs depend on the specific scheduling scheme of the operations performed inside the repeaters. Here we consider a specific scheduling scheme based on the following assumptions:
\begin{enumerate}
\item We assume full connectivity, that is a two-qubit gate can be performed between any two GKP qubits inside every repeater.
\item We measure time of performing all the operations inside repeaters in time steps. We assume that one time step is the time of performing each of the following procedures:
\begin{itemize}
\item preparing an ancilla GKP qubit,
\item applying a two-qubit gate between a data and an ancilla GKP qubit followed by a homodyne measurement of the ancilla and a subsequent feedback displacement of the data qubit; for the first or last operation on a given data mode inside a given repeater, the process of receiving or sending out a GKP qubit is also incorporated in this time step.
\end{itemize}
Clearly the second procedure involves more steps, but consists solely of Gaussian operations which are experimentally much easier to realise than the first procedure of GKP state preparation. Preparing GKP states requires a source of optical non-linearity as the GKP state is highly non-classical.
\end{enumerate} 

The detailed scheduling of all the operations performed inside the repeaters is described in Appendix~\ref{sec:ScheduleAndCost}, where it is shown that within our model the cost of the type-B repeater is $t_{\text{GKP}} = 4$, the cost of the type-A repeater based on the [[4,1,2]] code is $t_{\text{4-qubit}} = 68$, while the cost of the type-A repeater based on the [[7,1,3]] code is $t_{\text{7-qubit}} = 311$.

Now we can define the cost function that measures both the performance and cost for the concatenated-coded schemes as:
\begin{widetext}
\begin{equation}
C(L_{\text{tot}}, N_{\text{multi-qubit}}, N_{\text{all}}) = \frac{\frac{L_{\text{tot}}}{10} * \left( t_{\text{GKP}} (N_{\text{all}} - N_{\text{multi-qubit}}) + t_{\text{multi-qubit}} N_{\text{multi-qubit}}\right) + t_{\text{multi-qubit}}}{r'(L_{\text{tot}},N_{\text{multi-qubit}}, N_{\text{all}})} \, .
\label{eq:CostFunc}
\end{equation}
\end{widetext}
Here $r'$ is the secret-key rate per optical mode defined in Appendix~\ref{sec:seckeyRateAD} and $t_{\text{multi-qubit}}$ can correspond to $t_{\text{4-qubit}}$ or $t_{\text{7-qubit}}$ depending on the considered architecture. Moreover, the total distance $L_{\text{tot}}$ is expressed in km, $N_{\text{all}}$ is the number of all repeaters per 10 km and $N_{\text{multi-qubit}}$ is the number of multi-qubit repeaters per 10 km. That is, e.g. if after a type-A repeater there is a type-B repeater 5 km away and then another type-A repeater after another 5 km, such that the repeater types oscillate every 5 km, then $N_{\text{all}} = 2$ and $N_{\text{multi-qubit}} = 1$. Such convention has the nice feature that it is then reasonable to only consider cases when $N_{\text{all}}$ is a multiple of $N_{\text{multi-qubit}}$. This is because we can think of our architecture as firstly placing multi-qubit repeaters equidistant to each other and then adding GKP repeaters in between such that all the neighbouring type-B stations are also equidistant to each other. Moreover, the separation between the consecutive multi-qubit repeaters is then $10/N_{\text{multi-qubit}}$ km while the spacing between any two neighbouring repeaters (independently of their types) is $10/N_{\text{all}}$ km. The choice of 10 km as a reference distance is motivated by the fact that for all parameter regimes that we consider and for $L_{\text{tot}}$ of at least 500 km, it turns out that the optimal repeater configuration requires more than one type-A repeater per 10 km, see FIG.~\ref{fig:stations}. Hence we then only consider configurations in which $N_{\text{multi-qubit}}$ and $N_{\text{all}}$ are positive integers. We see that in Eq.~\eqref{eq:CostFunc} we include a residual $t_{\text{multi-qubit}}$ term to account for Alice's encoding station which we expect to have a similar cost as the type-A repeater. Bob's decoding station performs multi-qubit error correction and therefore counts as a type-A repeater and is implicitly included in $N_{\text{multi-qubit}}$ for the last 10 km segment.

Now, we aim to optimise our repeater configuration by minimising this cost function over $N_{\text{multi-qubit}}$ and  $N_{\text{all}}$ for each distance $L_{\text{tot}}$. We note that for practical terms it can be more informative to minimise a normalised cost function which for a given distance $L_{\text{tot}}$ counts resources per km rather than for the total distance:
\begin{widetext}
\begin{equation}
\begin{aligned}
C'(L_{\text{tot}}, N_{\text{multi-qubit}}, N_{\text{all}}) &= \frac{C(L_{\text{tot}}, N_{\text{multi-qubit}}, N_{\text{all}})}{L_{\text{tot}}} \\
										&= \frac{L_{\text{tot}}(t_{\text{GKP}} (N_{\text{all}} - N_{\text{multi-qubit}}) + t_{\text{multi-qubit}} N_{\text{multi-qubit}}) + 10 t_{\text{multi-qubit}}}{10 L_{\text{tot}}* r'(L_{\text{tot}},N_{\text{multi-qubit}}, N_{\text{all}})} \, \, .
\end{aligned}
\label{eq:normalisedcostfunct}
\end{equation}
\end{widetext}
This normalised cost function is plotted in FIG.~\ref{fig:costfunction}.

\subsection{Monte-Carlo simulation}
\label{sec:MonteCarlo}
Analytical modelling of the performance of the concatenated-coded repeater architectures is challenging. This is due to multiple effects. Firstly, even assuming the use of ideal GKP states, the analog information does not allow us to correct all the single-qubit errors for the [[4,1,2]] code and all the single- and two-qubit errors for the [[7,1,3]] code on the higher level. Specifically, from the simulation we see that the performance of the repeater in correctly identifying those errors depends on the channel parameters, that is, depending on the standard deviation $\sigma$ of the effective Gaussian random displacement channel between the repeaters we observe a different fraction of the single- and two-qubit logical GKP errors which the higher-level code fails to identify correctly and therefore fails to correct. In general, the larger $\sigma$, the higher the probability of misidentifying the erroneous qubit(s) on the second level. Moreover, the use of imperfect GKP ancillas together with the rescaling coefficients for the feedback displacement make it impractical to model errors after GKP correction as discrete Pauli errors on GKP qubits. Therefore we evaluate the performance of our scheme using a Monte-Carlo simulation for specific parameters.

We perform numerical Monte-Carlo simulation by tracking the evolution of the errors in the $\hat{q}$ and $\hat{p}$ quadratures.  While at the end of the simulation we would like to identify logical errors on the second level, the actual quadrature shifts on all the data qubits are continuous. Specifically, the imperfect squeezing in the GKP ancillas means that the final state of the qubits will in general be neither in the GKP code space nor in the second-level code space. Therefore we finish by applying first a round of virtual perfect GKP correction on all the GKP qubits and then a round of virtual perfect second-level correction, to bring the state to the code space of both codes. As here we just want to identify the closest logical state, we do not consider any analog information for this virtual corrections. These perfect corrections can be thought of as being performed using perfect infinitely squeezed ancilla GKP modes. The perfect GKP correction brings all the GKP data qubits to the nearest state in the code space such that we can now assign to them discrete values quantifying whether a logical $X$ and/or $Z$ error has taken place. Then the perfect multi-qubit correction (now without using the analog information) brings the state to the nearest logical state on the second level such that we can now count whether a logical error on the second level has taken place.

We find that it is not enough to simulate a single link between two consecutive type-A repeaters. When simulating only a single such link, there is a high probability that due to finite ancilla squeezing, before the perfect virtual multi-qubit correction we will be in a state that is outside of the code space on the second level and has e.g. a single GKP data qubit flipped. In the real-life scenario such a residual error on the second level would carry over to the next elementary link. Therefore the perfect virtual multi-qubit correction after a single link could significantly underestimate the error rate by effectively removing all such residual errors. However, we can simulate multiple consecutive links with the virtual correction only at the very end. In that case such a residual error on the second level before the final perfect virtual correction will only occur if there is a failure in correctly identifying the second-level stabilisers in the last link before the simulation end. This probability is always the same, that is, it is independent of the number of links we are simulating. Yet if we simulate multiple links then the total error accumulates so the probability of logical error after large number of links is much larger than after a single link. Therefore we simulate a chain of 100 such links before the virtual correction and in this way we make the probability of such a residual error negligible relative to the probability of the actual logical error.

For each quadrature we start the simulation directly after the multi-qubit correction in that quadrature in the first repeater assuming that each GKP data qubit carries a residual Gaussian random displacement error from a channel with variance $c_{\text{opt}}\sigma_{\text{GKP}}^2$ coming from the last GKP correction from the preceding link. Here $c_{\text{opt}}$ is the optimal coefficient used to rescale the GKP syndrome during error correction, see Appendix~\ref{sec:GKPrescale} for more details. We then evolve this quadrature following all the error channels and correction operations as described in Appendix~\ref{sec:GKrep}, Appendix~\ref{sec:2ndLevel}, Appendix~\ref{sec:multiqubitrep} and Appendix~\ref{sec:ScheduleAndCost}. We finish the simulation directly after the multi-qubit correction in that quadrature after the 100th link. We note that this means that the simulation of the $\hat{q}$ quadrature evolution has its beginning and end shifted with respect to the simulation of the $\hat{p}$ quadrature given that the multi-qubit syndrome measurements in those two quadratures happen at different times.  Finally, after the virtual corrections, we read off whether a logical error on the second level has occurred in any of the two quadratures. Hence from the simulation we extract the probabilities of $X$ and $Z$ logical errors $p_{\text{err,X/Z}}(\eta_0, \sigma_{\text{GKP}}, N_{\text{multi-qubit}}, N_{\text{all}})$ over such 100 elementary links. We can then discretise these errors, assigning a well-defined probability of a logical error for a single elementary link given by $P_{\text{err,X/Z}} = \frac{1-(1-2p_{\text{err,X/Z}})^{1/100}}{2}$. Hence we consider that the logical error over 100 links is given by an odd number of these effective discrete logical errors over single links. We can then use the values of $p_{\text{err,X}}$ and $p_{\text{err,Z}}$ obtained from the simulation to calculate the total probabilities of $X$ and $Z$ errors over the total distance $L_{\text{tot}}$ by considering the probabilities of odd number of such errors over the entire channel. These can be obtained by substituting $P_{\text{err,X/Z}}$ into the equation:
\begin{equation}
Q_{\text{err,X/Z}} = \frac{1-(1-2 P_{\text{err,X/Z}})^{L_{\text{tot}}/L}}{2}.
\end{equation}
where $L$ is the length of the single link given by $L = 10/N_{\text{multi-qubit}}$. As a result we have that:
\begin{widetext}
\begin{equation}
Q_{\text{err,X/Z}}(\eta_0, \sigma_{\text{GKP}}, N_{\text{multi-qubit}}, N_{\text{all}}, L_{\text{tot}}) = \frac{1-(1-2p_{\text{err,X/Z}})^{N_{\text{multi-qubit}} L_{\text{tot}}/1000}}{2} \, .
\end{equation}
\end{widetext}
Then this leads to an effective channel over $L_{\text{tot}}$ given by:
\begin{equation}
\mathcal{D}(\rho) = (1-q_X-q_Z-q_Y)\rho + q_X X \rho X + q_Z Z \rho Z + q_Y Y \rho Y.
\label{eq:channelpXpYpZMain}
\end{equation}
with $q_X = Q_{\text{err,X}}(1-Q_{\text{err,Z}})$, $q_Z = Q_{\text{err,Z}}(1-Q_{\text{err,X}})$, and $q_Y = Q_{\text{err,X}}Q_{\text{err,Z}}$. This enables us to calculate the secret-key rate per mode as described in Appendix~\ref{sec:seckeyRateAD} and then the normalised cost function given in Eq.~\eqref{eq:normalisedcostfunct}.

For each considered setting of the experimental parameters $\eta_0$ and $\sigma_{\text{GKP}}$ we run the simulation for multiple configurations of $\{N_{\text{multi-qubit}}, N_{\text{all}}\}$. That is, we start with $N_{\text{multi-qubit}} = N_{\text{all}} = 1$ and then rerun the simulation for the configurations for which $N_{\text{all}}$ is a multiple of $N_{\text{multi-qubit}}$, where we place a limit of 250 m on the minimum repeater spacing ($N_{\text{all}} \leq 40$ and $N_{\text{multi-qubit}} \leq 40$). In order to find the achievable distances presented in FIG.~\ref{fig:table} and in FIG.~\ref{fig:RepChainAnalVsSim} we maximise this secret-key rate for each distance by choosing the setting of $\{N_{\text{multi-qubit}}, N_{\text{all}}\}$ and the corresponding $p_{\text{err,X/Z}}$ which gives the highest secret-key rate for that distance $L_{\text{tot}}$. Then we look for the largest distance for which such secret-key rate per mode stays above 0.01.
We proceed similarly when calculating the optimal resource-cost trade-off. Then for each distance we minimise the cost function by choosing this setting of $\{N_{\text{multi-qubit}}, N_{\text{all}}\}$ and the corresponding $p_{\text{err,X/Z}}$ which gives the smallest cost function for that distance $L_{\text{tot}}$. We also evaluate the cost function for the architecture based solely on multi-qubit stations by imposing the additional constraint $\{N_{\text{multi-qubit}} = N_{\text{all}}\}$.

In a similar spirit we also run a Monte-Carlo simulation for a GKP repeater chain to verify the analytical model described in Appendix~\ref{sec:GKrep}. To include the effect of the residual displacements after GKP correction in a given repeater on the probability of successful correction in the next repeater, also in this case we simulate a chain of 100 elementary links, where this time an elementary link is a single link between the neighbouring type-B repeaters. We again simulate the errors in the two quadratures independently, where the simulation in each quadrature starts directly after the corresponding GKP correction, so that the initial error displacement comes from a distribution with variance $c_{\text{opt}}\sigma_{\text{GKP}}^2$. We then simulate 100 elementary links and apply the virtual perfect GKP correction at the end. We read-off whether there was a logical $X$ or $Z$ error, so that from the statistics we can extract the value of $p_{\text{err,X/Z}}$. Here $p_{\text{err,X/Z}}$ is the $X$ or $Z$ logical error probability over a chain of 100 GKP repeaters. Analogously to the concatenated-coded architecture, we calculate the logical $X$ and $Z$ error probability for the total distance $L_{\text{tot}}$:
\begin{widetext}
\begin{equation}
Q_{\text{err,X/Z}}(\eta_0, \sigma_{\text{GKP}}, N_{\text{GKP}}, L_{\text{tot}}) = \frac{1-(1-2p_{\text{err,X/Z}})^{N_{\text{GKP}} L_{\text{tot}}/1000}}{2} \, ,
\end{equation}
\end{widetext}
where $N_{\text{GKP}}$ is the number of GKP stations per 10 km. We can then extract the corresponding secret-key rate per optical mode $r'$ in the same way as described above for the concatenated-coded scheme. The results of our Monte-Carlo simulation for the GKP repeater chain verify the analytical model from Appendix~\ref{sec:GKrep} as shown in FIG.~\ref{fig:RepChainAnalVsSim}.

Now we also specify how we determine the accuracy of the simulations. When simulating a chain of 100 elementary links, we start with the sample of size $k=10$ and calculate the estimates of the standard error for the probability of logical $X$ and $Z$ flips $p_{\text{err,X/Z}}$ as:
\begin{equation}
\Delta p_{\text{err,X/Z}} = \sqrt{\frac{p_{\text{err,X/Z}}(1 - p_{\text{err,X/Z}})}{k}} \, ,
\label{eq:standardError}
\end{equation}
where the numerator is the standard deviation of the Bernoulli distribution. Then we calculate the relative error $\Delta p_{\text{err,X/Z}}/p_{\text{err,X/Z}}$ and check whether it is smaller than a threshold $b$ which we set. If not we increase $k$ by a factor of 10 and repeat the procedure. We iterate until the relative error becomes smaller than the threshold $b$. We then estimate the upper- and lower-bound on the achievable distance presented in FIG.~\ref{fig:table} and in FIG.~\ref{fig:RepChainAnalVsSim} by performing the above described optimisation of the secret-key rate per mode for each distance $L_{\text{tot}}$ not only for the logical error probabilities $\{p_{\text{err,X}},p_{\text{err,Z}}\}$ but also for the values $\{(1-b)p_{\text{err,X}},(1-b)p_{\text{err,Z}}\}$ leading to the upper-bound and $\{(1+b)p_{\text{err,X}},(1+b)p_{\text{err,Z}}\}$ leading to the lower-bound. For the simulations of the concatenated-coded schemes in FIG.~\ref{fig:table} we set $b=0.1$ and for the simulations of the GKP repeater chain in FIG.~\ref{fig:RepChainAnalVsSim} we set $b=0.02$. The optimal repeater placement configuration presented in FIG.~\ref{fig:stations}, the minimised cost function presented in FIG.~\ref{fig:costfunction} and the behaviour of secret-key rate per mode under cost function minimisation shown in FIG.~\ref{fig:secfrac} have all been obtained from the simulated data with the accuracy given by $b=0.1$.

Let us now briefly discuss the effect of minimising the cost function for the concatenated-coded schemes when taking the simulation error into account, that is when we increase or decrease $\{p_{\text{err,X}},p_{\text{err,Z}}\}$ by 10\% for the parameters $\eta_0 = 0.97$ and $\sigma_{\text{GKP}} = 0.09$. We have already mentioned in Section~\ref{sec:concatenatedrep} that increasing the logical error probabilities by 10\% leads to a visible decrease of the secret-key rate for the [[4,1,2]] scheme for 10000 km. We also find that decreasing the logical error probabilities by 10\% for this scheme leads to a visible increase of the secret-key rate above $r' = 0.03$ for that distance. Moreover, when varying the logical error probabilities within this confidence interval we find a visible change in behaviour for the first 200 km for both concatenated-coded schemes based on only type-A repeaters. Specifically, for distances up to around 100 km when the logical error probabilities are decreased by 10\%, the optimal repeater configuration for these schemes requires only a single multi-qubit repeater per 10 km, which also significantly lowers the achievable secret-key rate in that regime, yet allows to significantly decrease the cost function relative to the values shown in FIG.~\ref{fig:costfunction}. Finally, there is also a visible change in behaviour for the hybrid schemes when varying the logical error probabilities within the confidence interval, yet again only for the first 200 km. This change corresponds to the change of the optimal number of type-B repeaters within that distance regime which for the scheme based on the [[4,1,2]] code also affects the amount of generated secret key and the cost function in that distance regime.

Finally, we note that we also run a simple simulation of a single elementary link with channel loss $\gamma$ and ideal GKP and higher-level correction at the end in order to obtain the data presented in FIG.~\ref{fig:analog}. As in this case there are no residual errors after correction, it is then sufficient to simulate only a single such elementary link rather than 100 consecutive links. For this single link we extract from the simulation the probabilities of logical $X$ and $Z$ flips, $p_{\text{err,X/Z}}$. These probabilities are then used to calculate the maximum infidelity given by:
\begin{equation}
\epsilon_{\text{max}} = q_X + q_Z \, ,
\end{equation}
where: $q_X = p_{\text{err,X}}(1-p_{\text{err,Z}})$ and $q_Z = p_{\text{err,Z}}(1-p_{\text{err,X}})$, as shown in the Appendix~\ref{sec:infidelity}. We similarly use Eq.\eqref{eq:standardError} to calculate the standard error on $p_{\text{err,Z}}$ and $p_{\text{err,X}}$ and again require the corresponding relative errors to be smaller than $b$. We can then calculate the relative error on $\epsilon_{\text{max}}$ as follows. Firstly, the relative error on $q_X$ is bounded by:
\begin{equation}
\begin{aligned}
\frac{\Delta q_X}{q_X}	&= \sqrt{\left(\frac{\Delta  p_{\text{err,X}}}{ p_{\text{err,X}}}\right)^2 + \left(\frac{\Delta  p_{\text{err,Z}}}{ 1-p_{\text{err,Z}}}\right)^2} \\
					&\leq b \sqrt{1 + \left(\frac{p_{\text{err,Z}}}{1-p_{\text{err,Z}}}\right)^2} \, .
\end{aligned}
\end{equation}
Similarly:
\begin{equation}
\frac{\Delta q_Z}{q_Z} \leq b \sqrt{1 + \left(\frac{p_{\text{err,X}}}{1-p_{\text{err,X}}}\right)^2} \, .
\end{equation}
Let us define now $u = 1 + \left(\frac{p_{\text{err,Z}}}{1-p_{\text{err,Z}}}\right)^2$ and $v = 1 + \left(\frac{p_{\text{err,X}}}{1-p_{\text{err,X}}}\right)^2$. From here we can bound the relative error on the maximum infidelity as:
\begin{equation}
\frac{\Delta \epsilon_{\text{max}}}{\epsilon_{\text{max}}} \leq b \frac{\sqrt{u q_X^2 + v q_Z^2}}{q_X + q_Z} \, .
\end{equation}
Since $u$ and $v$ are close to one, we see that the relative error on $\epsilon_{\text{max}}$ is smaller than $b$. In the simulation we set $b = 0.1$. We then run the simulation for 101 values of $\gamma$ in the interval $[0.08,0.2]$. We find that the relative error on $\epsilon_{\text{max}}$ is around 7\% for all the data points.

\section*{Data availability}
No data sets were generated or analysed during the current study.

\section*{Code availability}
The code used for obtaining the presented numerical results as well as for generating the plots is available at \url{https://github.com/filiproz/ConcatenatedCodedRepeaters}.

\section*{Acknowledgements}

The authors would like to thank Benjamin D'Anjou, Prajit Dhara, Kaushik Seshadreesan, Shraddha Singh and Changchun Zhong for useful discussions and Narayanan Rengaswamy for feedback on the manuscript. FR, KN, QX and LJ acknowledge funding support from ARO (W911NF-18-1-0020, W911NF-18-1-0212), ARO MURI (W911NF-16-1-0349), AFOSR MURI (FA9550-19-1-0399), DOE (DE-SC0019406), NSF (EFMA-1640959, OMA-1936118, EEC-1941583), NTT Research and the Packard Foundation (2013-39273). SG would like to acknowledge funding support from U.S. Department of Energy UT-Battelle/Oak Ridge National Laboratory (4000178321), and the National Science Foundation (NSF) RAISE-EQuIP program, grant number 1842559. The authors are also grateful for the support of the University of Chicago Research Computing Center for assistance with the numerical simulations carried out in this work. This work was done before KN joined AWS Center for Quantum Computing.

We note that during the preparation of this paper we have become aware of the work of Fukui et al.~\cite{fukui2020all} which also considers the use of GKP encoding in quantum repeater architectures. That work complements our results. It investigates various decoding strategies for the GKP code for quantum communication and proposes multiple GKP-based quantum repeater schemes, including two-way schemes and the cluster-state-based repeater architecture utilising GKP encoding on the lower level.

\section*{Competing Interests}
The authors declare no competing interests.

\section*{Author Contribution}
LJ conceived the project. FR and LJ designed the proposed repeater architectures. KN and LJ provided expertise on quantum error correction. SG and LJ provided expertise on physical realisation of the error-correction based quantum communication. FR designed and implemented the analytical model of the GKP repeater chain and the Monte-Carlo simulations. KN provided high-level advice on the simulation and QX assisted with numerical implementation. FR, SG and LJ wrote the manuscript.
\bibliographystyle{naturemag}
\bibliography{library}{}
\onecolumngrid
\appendix

\section{GKP error correction} 
\label{sec:GKPErrCorr}

In this appendix we will describe the GKP error correction procedure. We will first describe how measuring the two stabilisers $\hat{S}_q$ and $\hat{S}_p$ allows us to correct errors caused by small displacements in phase space. Then we will describe how to perform these stabiliser measurements using ancilla GKP states. We will first focus on the scenario of error correction with perfect ancillas and then describe the imperfections arising from the use of ancillas with finite squeezing.

\subsection*{Correcting random displacement errors}

GKP code allows us to correct small random displacement errors from the logical subspace. The ideal error correction procedure for square lattice GKP code can be performed by measuring the two stabilisers $\hat{S}_q$ and $\hat{S}_p$. Such measurements are equivalent to measuring $\hat{q}$ and $\hat{p}$ modulo $\sqrt{\pi}$ producing the outcomes $\{q_0,p_0\}$, each of which belongs to the interval $[-\sqrt{\pi}/2, \sqrt{\pi}/2)$. The errors are then corrected by implementing the displacement $(\hat{q}, \hat{p}) \rightarrow (\hat{q} - q_0, \hat{p} - p_0)$. Hence, if the error displacement had the two components smaller in magnitude than $\sqrt{\pi}/2$, that is if the measured $\{q_0,p_0\}$ are exactly the error displacements, then the errors are successfully corrected. However, if e.g. the error displacement $\xi_q$ ($\xi_p$) was in the interval $[\sqrt{\pi}/2, 3 \sqrt{\pi}/2)$, then $q_0 = \xi_q - \sqrt{\pi}$ ($p_0 = \xi_p - \sqrt{\pi}$) and so the error channel followed by the correction operation would evolve the quadrature as: $\hat{q} \rightarrow \hat{q} + \xi_q - q_0 = \hat{q} + \sqrt{\pi}$ ($\hat{p} \rightarrow \hat{p} + \xi_p - p_0 = \hat{p} + \sqrt{\pi}$). This net shift by $\sqrt{\pi}$ would then lead to a logical $X$ ($Z$) error. In general, we see that if for the error displacement $\xi_q$ ($\xi_p$) we have that $\abs{\xi_q - n\sqrt{\pi}} < \sqrt{\pi}/2$ ($\abs{\xi_p - n\sqrt{\pi}} < \sqrt{\pi}/2$) for an odd integer $n$ then the correction operation $\hat{q} \rightarrow \hat{q} - q_0$ ($\hat{p} \rightarrow \hat{p} - p_0$) will result in the logical $X$ ($Z$) error.

Let us consider the scenario where the errors come from the Gaussian random displacement channel, such that $\rho  = \mathcal{N}_{\text{disp}}[\sigma](\ket{\psi_{\text{GKP}}} \bra{\psi_{\text{GKP}}})$ and let us focus on the $\hat{q}$ quadrature in the following example. Then after measuring the syndrome $q_0$ we can calculate the probability that the correction operation will result in the logical error. Specifically, we know that although by construction of the measurement, the measured $q_0$ belongs to the interval $[-\sqrt{\pi}/2, \sqrt{\pi}/2)$, it actually corresponds to a sample from the Gaussian distribution centred at $n\sqrt{\pi}$ for some integer $n$. If $n$ is even, then the correction operation successfully corrects the error otherwise it results in a logical $X$ error. Hence the probability of the logical $X$ for a given measured $q_0$ and for the strength of the Gaussian displacement channel given by the standard deviation $\sigma$ is given by~\cite{noh2020fault}:
\begin{equation}
p[\sigma](q_0) = \frac{\sum_{n \in \mathbb{Z}} \exp[- (q_0 - (2n+1)\sqrt{\pi})^2/(2\sigma^2)]}{\sum_{n \in \mathbb{Z}} \exp[- (q_0 - n\sqrt{\pi})^2/(2\sigma^2)]}.
\label{eq:analogInfoprob}
\end{equation}
For a given $\sigma$ the error probability increases with the $\abs{q_0}$ until it reaches maximum at $\abs{q_0} = \sqrt{\pi}/2$. Intuitively the error correction is less reliable if the measured syndrome is close to the decision boundary, that is the distance to $\ket{\psi_{\text{GKP}}}$ and to $X \ket{\psi_{\text{GKP}}}$ is similar. We note that in this work we only consider deterministic repeater schemes without post-selection. In such case this information about the probability of error during the correction operation cannot help us during the single mode GKP correction. However, it can facilitate error correction at the second level. Specifically, if after the GKP correction we measure the syndrome of the second-level code and we find that some GKP qubits have undergone a logical error, this analog information from the GKP correction can help us identify which GKP qubits need to be corrected on the second level~\cite{fukui2017analog,fukui2018high,vuillot2019quantum,noh2020fault}.

\subsection*{Stabiliser measurement using ancilla GKP}

\begin{figure}
\includegraphics[trim={0 0 0 0}, width=\columnwidth]{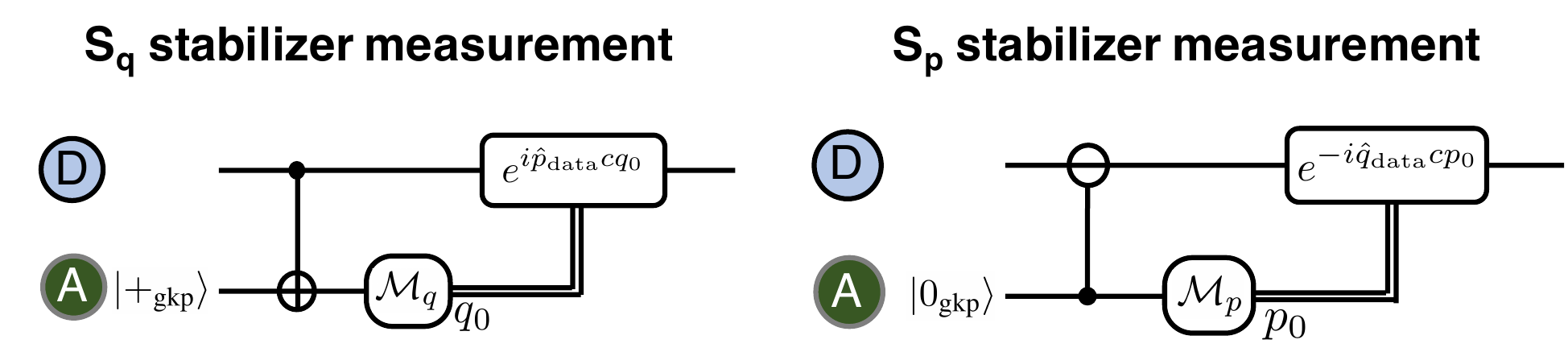}
\caption{\textbf{GKP Steane error correction on the data qubit ``D'' using GKP ancilla ``A''.} To measure the $\hat{S}_{\text{q}}$ stabiliser (left) we prepare the ancilla in state $\ket{+_{\text{GKP}}}$. Then we apply the $\text{SUM}_{D \rightarrow A}$ gate defined in Eq.~\eqref{eq:sumgate} and marked as controlled-$\oplus$ from the data qubit onto the ancilla. After that we measure the $\hat{q}$ quadrature of the ancilla using a homodyne measurement. The outcome modulo $\sqrt{\pi}$ belonging to the interval $[-\sqrt{\pi}/2, \sqrt{\pi}/2)$ is denoted by $q_0$. This value is then rescaled by the factor $c$ which reduces the error when noisy ancillas are used, as discussed in Appendix~\ref{sec:GKPrescale}. Finally the correction operation is performed by applying the displacement operator $e^{i\hat{p}_{\text{data}} c q_0}$ which implements the transformation $\hat{q}_{\text{data}} \rightarrow \hat{q}_{\text{data}} - c q_0$. The procedure to measure the $\hat{S}_{\text{p}}$ stabiliser (right) is similar but requires an ancilla in state $\ket{0_{\text{GKP}}}$ and the inverse SUM gate applied from the ancilla onto the data qubit, $\text{SUM}_{A \rightarrow D}^\dag$, denoted by controlled-$\ominus$. Finally the obtained $p_0$ outcome is used in the correction operation $e^{-i\hat{q}_{\text{data}} c p_0}$ which implements the transformation $\hat{p}_{\text{data}} \rightarrow \hat{p}_{\text{data}} - c p_0$.}
\label{fig:GKPSyndromeMeas}
\end{figure}

To measure the stabilisers $\hat{S}_q$ and $\hat{S}_p$ we require ancilla GKP qubits as shown in FIG.~\ref{fig:GKPSyndromeMeas}. Specifically to measure $\hat{S}_q$ we prepare the ancilla in the state $\ket{+_{\text{GKP}}}$ and apply a sum gate:
 \begin{equation}
 \text{SUM}_{j \rightarrow k} = \exp[-i \hat{q}_j \hat{p}_k] \, ,
 \label{eq:sumgate}
 \end{equation} 
 from the data qubit onto the ancilla. This gate shifts the $\hat{q}_{\text{anc}}$ quadrature of the ancilla qubit by the amount $\hat{q}_{\text{data}}$ of the data qubit. At the same time it shifts the $\hat{p}_{\text{data}}$ quadrature of the data qubit by the amount $-\hat{p}_{\text{anc}}$. Since $\ket{+_{\text{GKP}}} = \sum_{n \in \mathbb{Z}} \ket{p = 2n\sqrt{\pi}}$, the effect on the data mode is effectively the application of $\hat{S}_q$, under which the data mode is invariant. At the same time, since $\ket{+_{\text{GKP}}} = \sum_{n \in \mathbb{Z}} \ket{q = n\sqrt{\pi}}$, measuring the ancilla mode after the action of the sum gate using homodyne detection, results in precisely measuring $\hat{q}_{\text{data}}$ modulo $\sqrt{\pi}$. This can be seen as follows: since after preparation the ancilla was periodic $\hat{q}_{\text{ancilla}} = 0$ modulo $\sqrt{\pi}$, the only information we can infer from the outcome is the deviation from this periodicity, that is $q_0$, which is the measured value modulo $\sqrt{\pi}$ and which has been imprinted on the ancilla by the data qubit. Therefore from now we will refer to $q_0$ as the outcome of this measurement.  As described before, the data mode is subsequently displaced back by the measured amount $q_0$. The procedure to measure $\hat{S}_p$ is similar. It requires a GKP ancilla in the state $\ket{0_{\text{GKP}}}$ and the action of the inverse SUM gate $\text{SUM}^\dag_{j \rightarrow k} = \exp[i \hat{q}_j \hat{p}_k]$, this time from the ancilla onto the data mode. Subsequent measurement of the ancilla allows us now to measure $\hat{p}_{\text{data}}$ modulo $\sqrt{\pi}$.

Let us now consider the scenario where the ancilla GKP states are imperfect. As explained in Section~\ref{sec:realGKP} we can model such states as ideal GKP states that have been displaced by some small amount $(\xi_q^{\text{GKP}},\xi_p^{\text{GKP}})$, where both of these quantities follow a Gaussian distribution with mean zero and standard deviation $\sigma_{\text{GKP}}$. These imperfections cause additional errors. Let us consider the measurement of $\hat{S}_q$. Firstly, measuring the ancilla results now in measuring $(\hat{q}_{\text{data}} + \xi_q^{\text{GKP}})$ modulo $\sqrt{\pi}$. Thus now, even if the error displacement on the data mode and the measured $q_0$ were smaller in magnitude than $\sqrt{\pi}/2$, the correction displacement will correct the original error but will also imprint a new residual displacement of  $-\xi_q^{\text{GKP}}$ on the data mode. Secondly, there will be a back-action onto the $\hat{p}_{\text{data}}$ quadrature of the data mode, which after the SUM gate will undergo a transformation $\hat{p}_{\text{data}} \rightarrow \hat{p}_{\text{data}} - \xi_p^{\text{GKP}}$. Similarly measuring $\hat{S}_p$ with imperfect ancilla will result in measuring $\hat{p}_{\text{data}} + \xi_p^{\text{GKP}}$ modulo $\sqrt{\pi}$ and in an back-action error of $\hat{q}_{\text{data}} \rightarrow \hat{q}_{\text{data}} - \xi_q^{\text{GKP}}$.

\section{Overcoming finite GKP squeezing by rescaling the GKP syndrome}
\label{sec:GKPrescale}

In this appendix we describe a procedure of reducing the noise contributed by finite GKP squeezing during the correction feedback displacement. The error resulting from measuring $\hat{q}_{\text{data}} + \xi_q^{\text{GKP}}$ modulo $\sqrt{\pi}$ rather than $\hat{q}_{\text{data}}$ modulo $\sqrt{\pi}$ (and similarly for the $\hat{p}$ quadrature) can be significantly reduced by rescaling the subsequent displacement operation, as shown in FIG.~\ref{fig:GKPSyndromeMeas}. Specifically, let
\begin{equation}
R_{s}(x) = x - s \floor*{\frac{x}{s} + \frac{1}{2}}.
\end{equation}
That is $R_{\sqrt{\pi}}(x)$ denotes the value of $x$ modulo $\sqrt{\pi}$ shifted to the interval $[-\sqrt{\pi}/2,\sqrt{\pi}/2)$.
Then the measured value of the syndrome in the $\hat{q}$ quadrature is:
\begin{equation}
q_0 = R_{\sqrt{\pi}}(\hat{q}_{\text{data}} + \xi_q^{\text{GKP}}).
\label{eq:q0}
\end{equation}
We can now improve the performance of our error correction procedure by applying the displacement in the $\hat{q}$ quadrature not by $-q_0$ but rather by $-c q_0$~\cite{noh2019encoding}. Here $c$ is the constant that depends on the distribution governing the data and ancilla errors. The idea of rescaling the syndrome to compensate for the ancilla noise was also adapted in~\cite{fukui2019high, yamasaki2020polylog} where the authors combine it with post-selection based on the syndrome outcome to reduce the error probability. In this work, by contrast, we always perform the recovery operation and output the corrected state. We know that the ancilla error $ \xi_q^{\text{GKP}}$ comes from a normal distribution with mean zero and standard deviation $\sigma_{\text{GKP}}$. Similarly let us assume that the error on the data qubit comes also from an action of a Gaussian random displacement channel with standard deviation $\sigma_{\text{data}}$ and let us denote it as $\xi_q^{\text{data}}$. Then we want to find $c$ that minimises the variance of the error after correction, that is we want to minimise $\text{Var}(\xi_q^{\text{data}} - c q_0)$. Since for the error distributions that we consider in our simulations, in most cases the sum of the samples of the data and ancilla errors will be in the interval $[-\sqrt{\pi}/2,\sqrt{\pi}/2)$, and since we are only looking for a heuristic method for optimizing $c$, for the purpose of this calculation we shall approximate $q_0 \approx \xi_q^{\text{data}} + \xi_q^{\text{GKP}}$. Since the error distributions for the data and ancilla errors are independent we can easily establish that $\text{Var}(\xi_q^{\text{data}} - c (\xi_q^{\text{data}} + \xi_q^{\text{GKP}}))$ is minimised for
\begin{equation}
c_{\text{opt}} = \frac{\sigma_{\text{data}}^2}{\sigma_{\text{data}}^2 + \sigma_{\text{GKP}}^2},
\label{eq:optC}
\end{equation}
and such minimum variance is:
\begin{equation}
\text{Var}(\xi_q^{\text{data}} - c_{\text{opt}} (\xi_q^{\text{data}} + \xi_q^{\text{GKP}})) = c_{\text{opt}} \sigma_{\text{GKP}}^2.
\label{eq:varianceOptC}
\end{equation}
We note that since $\xi_q^{\text{data}}$ and $\xi_q^{\text{GKP}}$ follow a normal distribution with mean zero, $\xi_q^{\text{data}} - c_{\text{opt}} (\xi_q^{\text{data}} + \xi_q^{\text{GKP}})$ also follows a normal distribution with mean zero. Of course the actual distribution of the residual displacements after correction is given by $\xi_q^{\text{data}} - c_{\text{opt}} R_{\sqrt{\pi}}(\xi_q^{\text{data}} + \xi_q^{\text{GKP}})$ which is not Gaussian. From the formulas given in Eq.~\eqref{eq:optC} and in Eq.~\eqref{eq:varianceOptC}, we see that the reduction of errors by using $c_{\text{opt}}$ rather than $c= 1$ is greater, the smaller $\sigma_{\text{data}}$ relative to $\sigma_{\text{GKP}}$. We have also numerically verified in our simulation that the above heuristic is a good estimate for the optimal $c$ and enables us to significantly improve our GKP error correction procedure.

We have seen that the value of the optimal rescaling coefficient depends on $\sigma_{\text{data}}$, the standard deviation of the effective Gaussian random displacement channel describing all the processes that could have shifted the state away from the code space. If we have a sequence of error correction rounds, then $\sigma_{\text{data}}$ will include the residual displacement error from the previous correction round. In other words the residual displacement with variance $c_{\text{opt}} \sigma_{\text{GKP}}^2$ after a given error correction round can be seen as coming from a noise process contributing to $\sigma_{\text{data}}$ used to calculate $c_{\text{opt}}$ for the next correction round. Therefore the optimal rescaling coefficients at consecutive rounds depend on each other. Below we provide a recipe how to track these coefficients for a sequence of error correction rounds.

\subsection*{Translationally invariant sequence}

Firstly let us consider a simple scenario where we have a translationally invariant symmetry in our sequence of error correction operations. That is, the error channel between two consecutive GKP correction rounds is always the same and so the system is translationally invariant with respect to any number of correction rounds. In this case, the value of $c_{\text{opt}}$ will be the same at every correction round. This means that $\sigma_{\text{data}}(c_{\text{opt}})$ depends on exactly the same $c_{\text{opt}}$ that we want to calculate in Eq.~\eqref{eq:optC}. Hence the relation can be solved for $c_{\text{opt}}$ as follows. Firstly, we note that  $\sigma_{\text{data}}^2$ will have a contribution from the residual displacement after imperfect error correction and from the noise added between the correction rounds whose standard deviation we will denote as $\sigma_{\text{noise}}$. Therefore:
\begin{equation}
\sigma_{\text{data}}^2 = c_{\text{opt}} \sigma_{\text{GKP}}^2 + \sigma_{\text{noise}}^2.
\end{equation}

Substituting this formula into Eq.~\eqref{eq:optC} and solving for $c_{\text{opt}}$ we obtain:
\begin{equation}
c_{\text{opt}} = \frac{\sigma_{\text{noise}}}{2\sigma_{\text{GKP}}^2} \left(-\sigma_{\text{noise}} + \sqrt{\sigma_{\text{noise}}^2 + 4\sigma_{\text{GKP}}^2}\right)
\label{eq:optCTransInvar}
\end{equation}

This is the value of $c_{\text{opt}}$ that should be then used during each correction round.

\subsection*{General case}

The difficult situation arises if the system is not translationally invariant with respect to each GKP correction round as in that case $c_{\text{opt}}$ will assume a different value after each correction round. In this scenario the optimisation problem is more complex due to all the dependencies of $c_{\text{opt}}$ between consecutive correction rounds.

\begin{figure}
\includegraphics[trim={0 0 0 0}, width=\columnwidth]{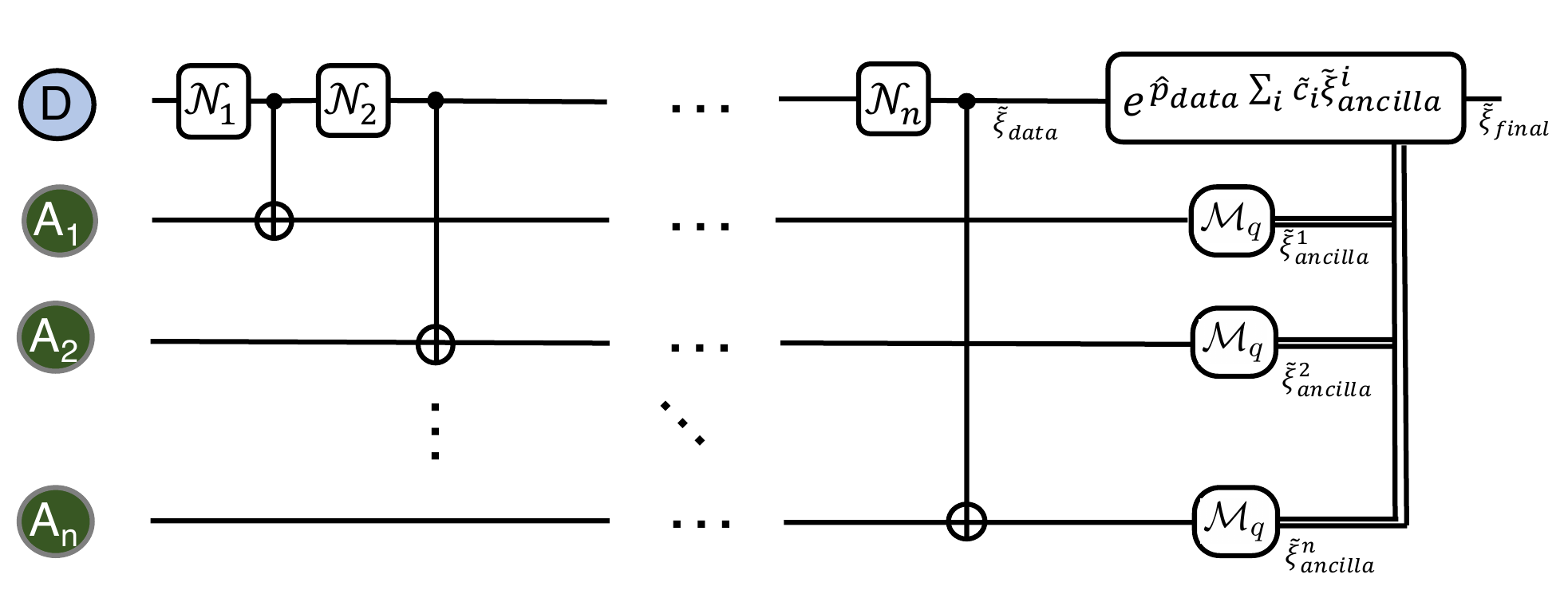}
\includegraphics[trim={0 0 0 0}, width=\columnwidth]{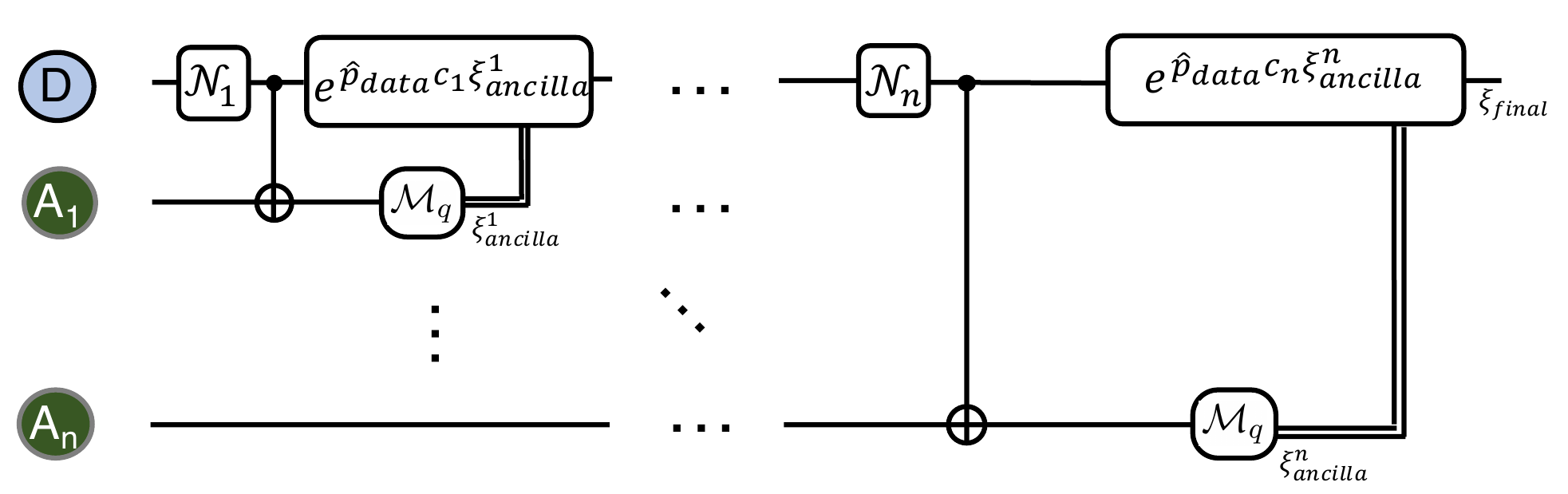}
\caption{\textbf{Sequence of GKP syndrome measurements with feedback displacement using rescaling coefficients.} We depict the postponed correction scenario (top) and the real-time correction scenario (bottom). For the purpose of calculating the optimal rescaling coefficients we consider an approximation $R_{\sqrt{\pi}}(\xi) \approx \xi$ and the circuits are shown for the evolution of the $\hat{q}$ quadrature. In both scenarios a single GKP data qubit (D) is subjected to a sequence of Gaussian random displacement channels $\{\mathcal{N}_i\}_i$. After the action of each channel $\mathcal{N}_i$, the GKP syndrome measurement is performed using an ancilla GKP mode ($\text{A}_i$). In the postponed scenario all the syndromes $\tilde{\xi}_i$ are collected at the end and a single correction feedback displacement is performed to remove the $\tilde{\xi}_{\text{data}}$ error accumulated on the data mode. In the real-time scenario the measured syndromes are immediately fed into the feedback displacements on the data mode. The two scenarios can be mapped onto each other such that the residual error after correction is the same in both cases, that is $\text{Var}(\xi_{\text{final}})  = \text{Var}(\tilde{\xi}_{\text{final}})$.    }
\label{fig:RescalingCoefffs}
\end{figure}

We will start by solving a simpler problem, in which we have a chain of GKP syndrome measurements but we postpone the correcting displacement operation until the end of the chain and we aim to minimise the quadrature noise variance only at the end of the chain. For each quantity $w$ for a scenario in which the feedback displacement is applied immediately after syndrome measurement, we will use $\tilde{w}$ to denote the corresponding quantity for such a postponed scenario. Then we will show how to minimise the variance $\text{Var}(\tilde{\xi}_{\text{final}})$, where
\begin{equation}
\tilde{\xi}_{\text{final}} = \tilde{\xi}_{\text{data}} - \sum_{i=1}^n \tilde{c}_i \tilde{\xi}_{\text{ancilla}}^{(i)}.
\end{equation}
Here $\tilde{\xi}_{\text{data}}$ denotes the error displacement accumulated until the end of the chain of $n$ error channels and syndrome extractions and $\tilde{\xi}_{\text{ancilla}}^{(i)}$ denotes the total noise on the $i$'th ancilla GKP mode which consists of the consecutive GKP error shifts accumulated along the chain from the beginning of the chain until the $i$'th syndrome extraction together with the imperfect ancilla GKP noise. We depict the corresponding quantum circuit on the top in FIG.~\ref{fig:RescalingCoefffs}. We see that at the end we apply a displacement operation where the displacement value is given by the sum of $\tilde{\xi}_{\text{ancilla}}^{(i)}$ weighted by the rescaling coefficients which we want to optimise over. We again note that we linearised the problem by assuming that we know $\tilde{\xi}_{\text{ancilla}}^{(i)}$ for each $i$, while in fact we only measure the noise displacements modulo $\sqrt{\pi}$.  In the postponed scenario this is not a valid approximation, while in the scenario where the displacement operation is applied directly after each round of syndrome extraction this approximation can be justified by the fact that by giving the immediate feedback we keep each $\xi_{\text{ancilla}}^{(i)}$ small. However, we will show that the simpler problem with a postponed displacement scenario can be in fact mapped onto a real-time-feedback scenario. Therefore we will start by solving the postponed scenario.

\subsubsection*{Quadratic problem in the postponed scenario}

We note that the variance $\text{Var}(\tilde{\xi}_{\text{final}})$ can be written as:

\begin{equation}
\text{Var}(\tilde{\xi}_{\text{final}}) = \text{Var}(\tilde{\xi}_{\text{data}})  - \tilde{c}_i\sum_{i=1}^n \text{Cov}(\tilde{\xi}_{\text{data}} , \tilde{\xi}_{\text{ancilla}}^{(i)} ) - \tilde{c}_i\sum_{i=1}^n \text{Cov}(\tilde{\xi}_{\text{ancilla}}^{(i)}, \tilde{\xi}_{\text{data}}) + \sum_{i,j=1}^n \tilde{c}_i \tilde{c}_j \text{Cov}(\tilde{\xi}_{\text{ancilla}}^{(i)} , \tilde{\xi}_{\text{ancilla}}^{(j)} ).
\end{equation}

By noting that $\text{Cov}(\tilde{\xi}_{\text{data}} , \tilde{\xi}_{\text{ancilla}}^{(i)} ) = \text{Cov}(\tilde{\xi}_{\text{ancilla}}^{(i)}, \tilde{\xi}_{\text{data}})$ we can write this function as:
\begin{equation}
\text{Var}(\tilde{\xi}_{\text{final}}) = a + \vec{b}^T \vec{x} + \vec{x}^T A \vec{x} \, ,
\end{equation}
where
\begin{equation}
\begin{aligned}
a &= \text{Var}(\tilde{\xi}_{\text{data}}), \\
\vec{b} &= -2
\begin{pmatrix}\text{Cov}(\tilde{\xi}_{\text{data}} , \tilde{\xi}_{\text{ancilla}}^{(1)} ) \\ \text{Cov}(\tilde{\xi}_{\text{data}} , \tilde{\xi}_{\text{ancilla}}^{(2)}) \\ ...\end{pmatrix},
\end{aligned}
 \end{equation}
and $A$ is the covariance matrix between all the $\tilde{\xi}_{\text{ancilla}}^{(i)}$. Finally $\vec{x}$ is the vector of the rescaling coefficients $\{\tilde{c}_i\}_i$ over which we want to minimise the total variance. Now we note that

\begin{equation}
\begin{aligned}
\text{Cov}(\tilde{\xi}_{\text{data}} , \tilde{\xi}_{\text{ancilla}}^{(i)} ) &= \text{Var}( \tilde{\xi}_{\text{ancilla}}^{(i)}) - \sigma_{\text{GKP}}^2, \\
a &=  \text{Var}( \tilde{\xi}_{\text{ancilla}}^{(n)}) - \sigma_{\text{GKP}}^2.
\end{aligned}
\label{eq:quadraticproblemrelations}
\end{equation}
The first line in Eq.~\eqref{eq:quadraticproblemrelations} follows from the fact that the noise $\tilde{\xi}_{\text{ancilla}}^{(i)}$ accumulated on the ancilla includes a part of $\tilde{\xi}_{\text{data}}$, namely the noise transferred using a SUM gate during the $i$'th syndrome measurement, and the noise from the preparation of the ancilla GKP state with standard deviation $\sigma_{\text{GKP}}$. The transferred noise appears both on the data and the ancilla qubits, while the preparation noise contributes only to the ancilla. Hence the covariance is exactly the noise transferred using the SUM gate, whose variance is given by $\text{Var}(\tilde{\xi}_{\text{ancilla}}^{(i)}) - \sigma_{\text{GKP}}^2$. The second line in Eq.~\eqref{eq:quadraticproblemrelations} follows the same argument with respect to the last syndrome measurement denoted with index $n$.

Now we need to establish some basic properties of $A$. Since $A$ is a covariance matrix, it is symmetric and positive-semidefinite. Moreover, it is the covariance matrix between the GKP ancillas and each GKP ancilla has a contributing noise factor from an imperfect GKP state preparation. These preparation noise contributions are independent for each ancilla. Therefore the total noise distributions on all the ancillas are independent which means that $A$ is positive definite. This implies that the minimised variance is give by:
\begin{equation}
\min \text{Var}(\tilde{\xi}_{\text{final}}) = a - \frac{1}{4} \vec{b}^T A^{-1} \vec{b},
\label{eq:minVar}
\end{equation}
and the values of optimal $\{\tilde{c}_{i, \text{opt}}\}_i$ coefficients are given by:
\begin{equation}
\vec{x}_0 = \frac{-1}{2}A^{-1} \vec{b}.
\end{equation}

\subsubsection*{Real-time rescaling factors}
Now we will show that the above problem of finding postponed coefficients enables us to also find the real-time coefficients for the case in which the displacement operation is applied directly after each GKP syndrome measurement as shown on the bottom in FIG.~\ref{fig:RescalingCoefffs}.
Hence we will show how to find the optimal real-time coefficients $c_{\text{opt}}$ as a function of the optimal postponed coefficients $\tilde{c}_{\text{opt}}$.

We will start by finding $\{\tilde{c}_i\}_i$ as a function of $\{c_i\}_i$ and then invert the relation. We will also denote $\xi_{\text{ancilla}}^{(i)}$ the total noise on the GKP ancilla $i$ in the real-time scenario. To be able to map the real-time problem onto the postponed problem we need to be able to express $\xi_{\text{ancilla}}^{(i)}$ in terms of $\tilde{\xi}_{\text{ancilla}}^{(i)}$. This can be done using the following recurrence relation:
\begin{equation}
\xi_{\text{ancilla}}^{(k)}= \tilde{\xi}_{\text{ancilla}}^{(k)} - \sum_{i=1}^{k-1} c_i \xi_{\text{ancilla}}^{(i)}.
\label{eq:recrel}
\end{equation}

That is the noise accumulated on the ancilla in the real-time scenario is given by the corresponding noise in the postponed scenario plus all the corrections that have been done real time up to the $k$'th syndrome measurement (corrections have a negative sign). Those corrections again depend on the real-time syndromes measured in the previous GKP syndrome measurement rounds. Since the noise channels $\{ \mathcal{N}_i \}_i$ acting on the data GKP qubit are the same in both scenarios, we can think that $\xi_{\text{data}}=\tilde{\xi}_{\text{data}}$ stays the same, but the effective sum of all the correction displacements has value $-\sum_{i=1}^n c_i \xi_{\text{ancilla}}^{(i)}$ for the real-time scenario now. To find the relation $\tilde{c}_{k} = f_k(\{ c_i \}_i)$, we need to find the functions $f_k$ such that:
\begin{equation}
\sum_{i=1}^n c_i \xi_{\text{ancilla}}^{(i)} = \sum_{i=1}^n f_i(\{ c_j \}_j) \tilde{\xi}_{\text{ancilla}}^{(i)}.
\end{equation}

Substituting $\xi_{\text{ancilla}}^{(i)}$ from Eq.~\eqref{eq:recrel} into $\sum_{i=1}^n c_i \xi_{\text{ancilla}}^{(i)}$, we now investigate the coefficients in front of the $\tilde{\xi}_{\text{ancilla}}^{(k)}$ term. Firstly there will be the term $c_k$ coming from the term with $i=k$ in the sum $\sum_{i=1}^n c_i \xi_{\text{ancilla}}^{(i)}$. Then there will be contributions from the terms in which $i>k$ in that sum. In this case we will have the terms that look like $-c_i c_k$ times the product of $-c$'s with subscripts in the range from $k+1$ to $i-1$ and without repetition. The sum includes the terms with all the distinct choices of the subscripts as well as the term with the choice of no subscripts in which case $-c_i c_k$ is just multiplied by 1. If $i=k+1$ we only have a single term with just $-c_i c_k$.

Hence, the functions $f_k$ can be written as:
\begin{equation}
\begin{aligned}
\tilde{c}_k &= f_k(\{ c_i \}_i) = c_k \left(1- \sum_{i = k+1}^{n} c_i \sum_{j=1}^{2^{i-1-k}} \prod_{l \in W^j_{k,i}} l \right), \, \, \, \text{for} \, \, \,  k<n \, , \\
\tilde{c}_n &= c_n \, .
\end{aligned}
\label{eq:ckintermsoftildec}
\end{equation}
Here let us denote by $W_{k,i}$ the set such that:
\begin{equation}
W_{k,i} = 
\begin{cases}
\{1, -c_{k+1},  -c_{k+2},  ..., -c_{i-1}\} \, \, \, \text{if} \, \, \, i > k+1, \\
\{1\}  \, \, \, \text{if} \, \, \, i \leq k+1. \\
\end{cases}
\end{equation}
Then by $W^j_{k,i}$, where $j \in \{1, 2, ..., 2^{i-1-k}\}$ and $i \ge k+1$ we denote the $2^{i-1-k}$ distinct subsets of $W_{k,i}$ such that:
\begin{equation}
\begin{gathered}
W^j_{k,i} \subseteq W_{k,i} \, \, \, \text{and} \, \, \, 1 \in W^j_{k,i} \, \, \, \forall j \in \{1, 2, ..., 2^{i-1-k}\} \, , \\
W^j_{k,i} \neq W^l_{k,i} \, \, \, \text{for} \, \, \, j \neq l \, .
\end{gathered}
\end{equation}

The final step is to invert the relation in Eq.~\eqref{eq:ckintermsoftildec}, to find the function $c_k = g_k(\{\tilde{c}_i\}_i)$. For $k<n$ this can be done as follows:
\begin{equation}
\begin{aligned}
\tilde{c}_k 	&= c_k \left(1 - c_{k+1} - \sum_{i = k+2}^{n} c_i \sum_{j=1}^{2^{i-1-k}} \prod_{l \in W^j_{k,i}} l \right) \\
	&= c_k \left(1 - c_{k+1} - \sum_{i = k+2}^{n} c_i \left( - c_{k+1}\sum_{j=1}^{2^{i-2-k}} \mathop{\prod \; \; l}_{l \in W^j_{k+1,i}}  + \sum_{j=1}^{2^{i-2-k}} \mathop{\prod \; \; l}_{l \in W^j_{k+1,i}}  \right)\right) \\
	&= c_k \left(1 - c_{k+1}\left(1 - \sum_{i = k+2}^{n} c_i \sum_{j=1}^{2^{i-2-k}} \mathop{\prod \; \; l}_{l \in W^j_{k+1,i}} \right) -  \sum_{i = k+2}^{n} c_i \sum_{j=1}^{2^{i-2-k}} \mathop{\prod \; \; l}_{l \in W^j_{k+1,i}}  \right) \\
	&= c_k \left( 1 - \tilde{c}_{k+1} - \sum_{i = k+2}^{n} c_i \sum_{j=1}^{2^{i-2-k}} \mathop{\prod \; \; l}_{l \in W^j_{k+1,i}}  \right)  \\
	&= ... \\
	& =  c_k \left( 1 - \tilde{c}_{k+1} - \tilde{c}_{k+2} - \sum_{i = k+3}^{n} c_i \sum_{j=1}^{2^{i-3-k}} \mathop{\prod \; \; l}_{l \in W^j_{k+2,i}}  \right) \\
	&= c_k \left( 1 - \sum_{i=k+1}^n \tilde{c}_i  \right).
\end{aligned}
\end{equation}
Here in the second line we split the $\sum_{j=1}^{2^{i-1-k}} \prod_{l \in W^j_{k,i}} l$ expression into two terms: the term that includes $- c_{k+1}$ and the one which does not. For the term which includes $- c_{k+1}$, we take it out and for both terms the sum over $j$ now involves only the sets $W^j_{k+1,i}$. In the fourth line, we notice that the second term inside the outer brackets in the third line is just $\tilde{c}_{k+1}$.

Hence:
\begin{equation}
c_k = \frac{\tilde{c}_k}{ 1 - \sum_{i=k+1}^n \tilde{c}_i}.
\label{eq:realintermsofpostponed}
\end{equation}

Since we have established one-to-one relations:
\begin{equation}
\begin{aligned}
c_k =  g_k(\{\tilde{c}_i\}_i), \\
\tilde{c}_k =  f_k(\{c_i\}_i),
\end{aligned}
\end{equation}
it follows that the optimal real-time rescaling factors are given by $c_{k,\text{opt}} =  g_k(\{\tilde{c}_{\text{i,opt}}\}_i)$. This is because if we assume that there exists a set of real-time coefficients $\{c_i\}_i$ which leads to smaller variance than $g_k(\{\tilde{c}_{\text{i,opt}}\}_i)$, then for this $\{c_i\}_i$ we could calculate $\tilde{c}_k =  f_k(\{c_i\}_i)$ for all $k$ which would give the same variance in the postponed scenario and that variance would be smaller than given in Eq.~\eqref{eq:minVar} which is a contradiction. Therefore the optimal real-time rescaling factors are given by $c_{k,\text{opt}} =  g_k(\{\tilde{c}_{\text{i,opt}}\}_i)$.

Moreover, although we know that the obtained rescaling coefficients minimise the final variance at the end of the chain of GKP syndrome measurements and GKP correction operations, we know from Eq.~\eqref{eq:optC} and Eq.~\eqref{eq:varianceOptC}, that the minimum variance at the end is obtained for the scenario in which the final $c_{n, \text{opt}}$ is the smallest. This is achieved when $\sigma_{\text{data}}^2$ before the last correction round was minimised for the fixed $\sigma_{\text{GKP}}^2$. Since this $\sigma_{\text{data}}^2$ has a contribution from $c_{n-1} \sigma_{\text{GKP}}^2$, minimising $\sigma_{\text{data}}^2$ corresponds to minimising the residual variance $c_{n-1} \sigma_{\text{GKP}}^2$ after the penultimate correction. This recursive dependency shows that minimising the variance at the end in the real-time scenario corresponds to actually minimising it after each GKP correction round. This guarantees than the variance stays as low as possible all the time and justifies the application of our linearised model to the actual scenario in which we do not measure the actual noise displacement, but only its value modulo $\sqrt{\pi}$. When the variance is kept low at all times, we make sure that the probability that the actual noise displacement differs from the noise displacement modulo $\sqrt{\pi}$ is as low as possible, hence making the optimal real-time rescaling coefficients obtained in the linearised model applicable to the actual scenario.

\subsubsection*{Comment about numerical evaluation}
An important observation is that the optimal rescaling coefficients in the postponed scenario can differ by many orders of magnitude. In general, the last coefficient is of the order of magnitude of $\tilde{c}_{n, \text{opt}} \sim 10^{-1}$, while for around $n=40$ syndrome measurements in the chain, the first optimal coefficient will be of the order $\tilde{c}_{1, \text{opt}} \sim 10^{-31}$. This behaviour is intuitive, since for a long chain of noisy channels and syndrome measurements, in the postponed scenario more weight will be given to the last syndromes which carry more information about the total noise than the initial syndromes. When converting the postponed coefficients to the real-time ones, all the real-time coefficients are again of the order $10^{-1}$. This is because for the first few real-time coefficients, in the Eq.~\eqref{eq:realintermsofpostponed} we divide a small $\tilde{c}_k$ by an also small $1 - \sum_{i=k+1}^n \tilde{c}_i$. This means that the rescaling factors $\{\tilde{c}_i\}_i$ need to be calculated with very high level of accuracy in order to obtain reliable values of $\{c_i\}_i$. Mathematically, the difficulty arises at the step of taking the inverse of the matrix $A$ whose eigenvalues differ by four orders of magnitude. Hence the crucial step is to calculate $A$ to very high accuracy before calculating its inverse. In our analysis we find one scenario in which this issue becomes a limitation, namely for the parameters considered in FIG.~\ref{fig:stations} we find that we can reliably calculate a chain of maximum 48 rescaling coefficients. This means that for the hybrid architecture based on the [[7,1,3]] code, for the scenario with one type-A repeater per 10 km, we can directly calculate the chain of only up to 34 type-B repeaters per 10 km. Therefore for the cases with 35 to 39 type-B repeaters per 10 km we obtain the near-optimal rescaling coefficients by extrapolating from shorter chains.

\section{Application of the GKP code for quantum communication}
\label{sec:GKPagainstLoss}

We have already shown the usefulness of the GKP code for correcting errors arising from the action of the Gaussian random displacement channel. Yet the absorption or scattering of photons in the optical fibre does not correspond to such a channel. In this appendix we therefore describe how we can effectively transform the action of a lossy optical fibre into such a Gaussian random displacement channel. 

Transmission through the optical fibre can be modelled by a pure-loss channel whose action is given by:
\begin{equation}
\mathcal{N}_{\text{loss}}[\eta](\rho) = \Tr_B[\hat{B}(\eta)(\rho_A \otimes \dyad{0}_B)\hat{B}^\dag(\eta)].
\end{equation}
Here $\hat{B}(\eta)$ is the unitary corresponding to the action of a beam-splitter with transmissivity $\eta$ acting on the two input modes corresponding to states $\rho$ and the vacuum state $\ket{0}$.
For optical fibre the transmissivity decays exponentially with communication distance $L$, such that we can write the transmissivity as:
\begin{equation}
\eta = \eta_0 e^{-L/L_0} \, .
\end{equation}
Here $\eta_0$ is the efficiency of coupling the photons in and out of the fibre and $L_0$ is the attenuation length of the fibre. For commercially available fibre at telecom frequency we have $L_0 = 22$ km.
Let us now also define a quantum-limited amplification channel whose action can be described as:
\begin{equation}
\mathcal{A}_{\text{amp}}[G](\rho) = \Tr_B[\hat{S}_2(G)(\rho_A \otimes \dyad{0}_B)\hat{S}^\dag_2(G)].
\end{equation}
Here $\hat{S}_2(G)$ is the two-mode squeezing unitary with gain $G$.
Now, if the sender knows the transmissivity of the pure-loss channel and before sending the state $\rho$, they first apply the quantum-limited amplification with $G = 1/\eta$, then the effective action of both channels is precisely the Gaussian random displacement channel~\cite{kim1996quantum,sabapathy2011robustness,ivan2011operator,noh2018quantum}:
\begin{equation}
\mathcal{N}_{\text{loss}}[\eta] \cdot \mathcal{A}_{\text{amp}}[1/\eta] =\mathcal{N}_{\text{disp}}[\sigma = \sqrt{1 - \eta}].
\end{equation}
This shows the usefulness of the GKP code in our quantum communication scenario.

\section{Approximations for performance estimates of GKP repeater chain}
\label{sec:approxAncIntoChannel}

Analytical modeling of the quantum repeater chain based on concatenated GKP and discrete-variable codes would be a very challenging task. Therefore we have developed a detailed numerical simulation to investigate this complex behaviour, see Section~\ref{sec:MonteCarlo}. On the other hand the scenario of a GKP repeater chain in which we only consider one level of encoding based on the GKP code is much simpler to analyse. Therefore for that case we develop an analytical model that enables us to efficiently evaluate its performance for different parameter sets. In this appendix we describe the details of this analytical model and describe the approximations that allow us to easily integrate finite GKP squeezing into the model.

\subsection*{Logical error probability for ideal GKP states}

Here we provide simple estimates of the logical error probability of the square lattice GKP code for error correction. Let us first focus on ideal correction using infinitely squeezed ancilla GKP states. For the errors resulting from the Gaussian random displacement channel with standard deviation $\sigma$ defined in Section~\ref{sec:realGKP}, the logical error probability can be approximated by finding the probability that in one of the $(\hat{q}, \hat{p})$  quadratures the displacement is larger in magnitude than the critical value $\frac{\sqrt{\pi}}{2}$:
\begin{equation}
P_{\text{err}}(\sigma) =  \frac{1}{2\pi \sigma^2}\int_{\abs{q}>\sqrt{\pi}/2 \lor \abs{p}>\sqrt{\pi}/2}   dq dp \exp\left[-\frac{q^2 + p^2}{2\sigma^2}\right].
\label{eq:pfail}
\end{equation}
In fact this formula is an upper bound on the logical error probability, since displacements in $\hat{q}$ or $\hat{p}$  whose magnitude lies in the interval $[\frac{3\pi}{2}, \frac{5\pi}{2})$ or any other subsequent interval of length $\pi$ with an even multiple of $\pi$ at its centre can also be corrected. However, the probability of so large error displacements is generally negligible for the values of $\sigma$ that we will consider here.

The function $P_{\text{err}}(\sigma)$ can be approximated by integrating it over a circular rather than square boundary. In particular, we can overestimate the error by performing the integration over the region outside of a circle centred at the origin with radius $\sqrt{\pi}/2$. This leads to an upper bound on $P_{\text{err}}(\sigma)$ which has been calculated in~\cite{albert2018performance}. On the other hand by considering a larger circle with radius equal to half of the diagonal of the square, $\sqrt{\pi/2}$ we can establish a lower bound on this logical error probability. Unfortunately, we find that the gap between these two bounds is too large to be useful for our practical estimates. Therefore we will consider here a numerical evaluation of the integral. Specifically, similarly to~\cite{gottesman2001encoding}, we will consider separately the probability of $X$ and $Z$ error. Since by symmetry this probability is the same for both errors we will denote it as:
\begin{equation}
P_{\text{err,XZ}}(\sigma) =  \frac{1}{\sqrt{2\pi} \sigma}\int_{\abs{x}>\sqrt{\pi}/2} dx \exp\left[-\frac{x^2}{2\sigma^2}\right] = \text{erfc}\left(\sqrt{\frac{\pi}{8\sigma^2}}\right),
\label{eq:pfailbasis}
\end{equation}
where $\text{erfc}(z) = \frac{2}{\sqrt{\pi}}\int_z^\infty e^{-t^2} dt $ is the complementary error function.

\subsection*{Incorporating finite GKP squeezing into the channel noise}

Now let us consider error correction using imperfect GKP ancillas. We will show that this case can be approximated by a scenario in which perfect GKP correction is performed using an infinitely squeezed ancilla, however with additional Gaussian random displacement channels preceding and following the error correction procedure.
Let us start by considering error correction using a finitely squeezed ancilla with its squeezing parameterised by the standard deviation $\sigma_{\text{GKP}}$. Let us consider error correction in the $\hat{q}$ quadrature and let $\xi _q^{\text{GKP}}$ denote the displacement error on the GKP ancilla due to finite squeezing. Then after the error correction procedure, the $\hat{q}$ quadrature has undergone the transformation:

\begin{equation}
EC_{\text{exact}}: \, \hat{q}_{\text{data}} \rightarrow \hat{q}_{\text{data}} - c R_{\sqrt{\pi}}(\hat{q}_{\text{data}} + \xi _q^{\text{GKP}}) \, .
\label{eq:errorIntoChannel1}
\end{equation}

Here we will show that in the relevant case, we can approximate this evolution by a different transformation, namely:
\begin{equation}
EC_{\text{approx}}: \,\hat{q}_{\text{data}} \rightarrow (\hat{q}_{\text{data}} + \xi_q^{\text{GKP}}) - R_{\sqrt{\pi}}(\hat{q}_{\text{data}} + \xi _q^{\text{GKP}}) - \alpha \xi_q^{'\text{GKP}} \, .
\label{eq:errorIntoChannel2}
\end{equation}

Here $\xi_q^{\text{GKP}}$ and $\xi_q^{'\text{GKP}}$ are independent Gaussian random variables with zero mean and standard deviation $\sigma_{\text{GKP}}$ and $\alpha$ is a coefficient we will optimise. The importance of this approximation step is that the transformation $EC_{\text{approx}}$ can be interpreted as follows. Before the correction operation, the quantum state has been subjected to a Gaussian random displacement channel with standard deviation $\sigma_{\text{GKP}}$. Then ideal GKP correction has been performed, such that the quantum state has been brought back into the code space. Finally, the quantum state has been subjected to another Gaussian random displacement channel with standard deviation $\alpha \sigma_{\text{GKP}}$. Hence, under this model we transfer the imperfection of the GKP ancillas into the channel acting on the data mode.

In order to make the approximation reliable we want to find $\alpha$ such that the approximate distribution is as close to the exact distribution  as possible. Clearly the mean of both distributions is the same and is equal to zero. Yet both the exact and the approximate distributions are non-Gaussian because of the $R_{\sqrt{\pi}}$ component, which however differs from identity function only in the less likely scenarios in which its argument has an absolute value larger than $\sqrt{\pi}/2$. Therefore we will look for the optimal $\alpha$ again under the assumption that we can approximate $q_0$ given in Eq.~\eqref{eq:q0} as $q_0 \approx \xi_q^{\text{data}} + \xi_q^{\text{GKP}}$. In this case both the exact and approximate distributions become Gaussian and so to make them the same we need to find $\alpha$ such that they both have the same variance. The variance of the approximate distribution is then simply: $\text{Var}_{\text{approx}} = \alpha^2 \sigma_{\text{GKP}}^2$. In general the variance of the exact distribution is $\text{Var}_{\text{exact}} =(1-c)^2 \sigma_{\text{data}}^2 + c^2 \sigma_{\text{GKP}}^2$. However, we have already seen that for the optimal rescaling coefficient $c = c_{\text{opt}}$ given in Eq~\eqref{eq:optC}, the variance is given by Eq.~\eqref{eq:varianceOptC}. Since here we are only interested in the case where the optimal rescaling coefficient is used, we can obtain the same variance for the approximate distribution by setting $\alpha = \sqrt{c_{\text{opt}}}$. Hence, the considered approximation leads to a probability distribution with the same mean as the exact transformation, it has the same condition for the logical $\sqrt{\pi}$ shift as the exact transformation and under the linearisation approximation it also has the same variance as the exact distribution.

Thus we have constructed an analytical model of the GKP error correction with imperfect ancilla modes by approximating it with perfect correction that brings the state back to the code space. In this approximation however, before and after the correction the data mode is additionally subjected to Gaussian random displacement channels with standard deviations $\sigma_{\text{GKP}}$ and $\sqrt{c_{\text{opt}}} \sigma_{\text{GKP}}$ respectively, where $c_{\text{opt}}$ is the optimal rescaling coefficient used during the correction operation. We numerically verify the reliability of this approximation in the relevant parameter regime. Specifically, the analytical model of a GKP repeater chain which is based on this approximation is validated against a numerical Monte-Carlo simulation. We refer the reader to Appendix~\ref{sec:GKrep} for the details of the analytical model and to Section~\ref{sec:MonteCarlo} for the details on the corresponding Monte-Carlo simulation.

\section{GKP repeater chain architecture}
\label{sec:GKrep}

\begin{figure}
\includegraphics[trim={0 0 0 0}, width = \columnwidth]{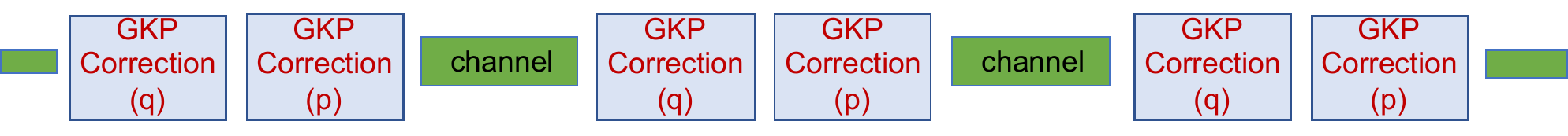}
\caption{\textbf{Schematic diagram of the GKP repeater chain architecture.} Each repeater station receives a single GKP data mode from the channel, performs firstly correction in the $\hat{q}$ and then in $\hat{p}$ quadrature followed by amplification and then forwards the GKP state to the next repeater.}
\label{fig:GKPrepchain}
\end{figure}

Now we will consider a specific repeater architecture in which each GKP data qubit is transmitted between subsequent GKP correction stations. These stations in sequence perform first correction in $\hat{q}$ quadrature and then in $\hat{p}$ quadrature as shown in FIG.~\ref{fig:GKPrepchain}. We can now establish the total noise channel that combines the effects of both the pure-loss channel during transmission and error correction using finitely squeezed ancillas. We note that we use the approximation of transferring the noise from the imperfect ancilla modes into the channel described in Appendix~\ref{sec:approxAncIntoChannel} only for the purpose of obtaining estimates of the performance of the GKP repeater chain. In the Monte-Carlo simulations of our repeater schemes we consider the finite squeezing of the ancilla modes inside repeaters and the lossy communication channels separately.

Let us now consider the $\hat{q}$ quadrature and all the errors arising between two consecutive corrections in that quadrature. In our GKP repeater station we firstly correct errors in $\hat{q}$ quadrature and then in $\hat{p}$ quadrature. Hence, the first error process that takes the state outside of the code space is the noisy channel directly after the correction as described in Appendix~\ref{sec:approxAncIntoChannel}. The random displacement corresponding to this noise process has magnitude $\sqrt{c_{\text{opt}}}\xi_q^{\text{GKP}}$, where $\xi_q^{\text{GKP}}$ comes from a Gaussian distribution with standard deviation $\sigma_{\text{GKP}}$. Then we have another $\xi_q^{'\text{GKP}}$ from the back-action during the $\hat{p}$ quadrature correction followed by the error from the transmission channel $\xi_q^{\text{trans}}$ and finally the last $\xi_q^{''\text{GKP}}$ again due to the imperfect ancilla in the $\hat{q}$ quadrature correction as described in Appendix~\ref{sec:approxAncIntoChannel}. Since all these four errors come from four independent Gaussian random displacement channels, they could be combined into a single channel by noting that a sum of independent normally distributed random variables is itself a normal random variable with the mean and the variance given by the sum of the means and variances respectively of the individual distributions. Hence we can model our repeater architecture as consisting of ideal GKP (type-B) repeaters that correct errors arising due to the action of a Gaussian random displacement channel with variance $\sigma_{\text{eff}}^2$ given by:
\begin{equation}
\sigma_{\text{eff}}^2 = \sigma_{\text{trans}}^2 + (2 +  c_{\text{opt}})\sigma_{\text{GKP}}^2 =  (1 - \eta_0 e^{-L/L_0}) + (2 +  c_{\text{opt}}) \sigma_{\text{GKP}}^2,
\label{eq:effectiveChannel}
\end{equation}
where
\begin{equation}
\sigma_{\text{trans}}^2 = 1 - \eta_0 e^{-L/L_0}.
\end{equation}
Here $\eta_0$ is the photon coupling efficiency at the repeaters, $L$ is the repeater spacing and $L_0$ is the attenuation length of the optical fibre. Now let us evaluate the rescaling coefficient $c_{\text{opt}}$. The considered GKP repeater chain is an example of a translationally invariant scenario described in Appendix~\ref{sec:GKPrescale}. Hence, to establish $c_{\text{opt}}$ we need to establish $\sigma_{\text{noise}}^2$ which is the variance of the effective Gaussian noise contributed by the environment between two consecutive GKP correction rounds. We note that $\sigma_{\text{noise}}^2$ is not evaluated under the model of transferring noise from the GKP ancilla into the channel, as $c_{\text{opt}}$ is calculated by comparing the error accumulated on the data qubit up to the correction operation relative to the error on the imperfect ancilla. Hence we have that:
\begin{equation}
\sigma_{\text{noise}}^2 = \sigma_{\text{GKP}}^2 + \sigma_{\text{trans}}^2.
\label{eq:SIgmaDataBetweenRep}
\end{equation}
The two terms include the error due to back-action of the GKP correction in the other quadrature and the error due to the communication channel. 

Plugging this into the formula for $c_{\text{opt}}$ in Eq.~\eqref{eq:optCTransInvar} gives:
\begin{equation}
c_{\text{opt}} = \frac{-\sigma_{\text{GKP}}^2 - \sigma_{\text{trans}}^2 + \sqrt{(\sigma_{\text{GKP}}^2 + \sigma_{\text{trans}}^2)(5\sigma_{\text{GKP}}^2 + \sigma_{\text{trans}}^2 )}}{2\sigma_{\text{GKP}}^2}.
\label{eq:optimalCchannel}
\end{equation}
Thus when using $c_{\text{opt}}$ in our model of transferring the ancilla noise into the channel, the effective variance becomes:
\begin{equation}
\sigma_{\text{eff}}^2 = \sigma_{\text{trans}}^2 + (2 +  c_{\text{opt}})\sigma_{\text{GKP}}^2 =  \frac{1}{2} \left( 3\sigma_{\text{GKP}}^2 + \sigma_{\text{trans}}^2 + \sqrt{(\sigma_{\text{GKP}}^2 + \sigma_{\text{trans}}^2)(5\sigma_{\text{GKP}}^2 + \sigma_{\text{trans}}^2 )}\right).
\label{eq:effectiveChannelCopt}
\end{equation}
Clearly this effective variance is the same for correction both in $\hat{q}$ and $\hat{p}$ quadratures. This is because the channel between two consecutive $\hat{q}$ corrections involves the same noise processes as the one between two consecutive $\hat{p}$ corrections. The order of these noise processes is different in the two cases, as for the correction in $\hat{q}$ quadrature we firstly have the back action from the syndrome measurement in the opposite quadrature and then the communication channel while in $\hat{p}$ quadrature the order of these two processes is reversed. However, the overall effective channel depends only on the variances of these processes and not on their order.

The effective logical error rate in $\hat{q}$ or $\hat{p}$ can be obtained by substituting the above $\sigma_{\text{eff}}^2$ for the variance in Eq.~\eqref{eq:pfailbasis}. In this way we obtain a relation for the estimate of the logical $X$ or $Z$ flip for a single link as a function of $\eta_0$, $\sigma_{\text{GKP}}$ and the repeater spacing $L$:
\begin{equation}
P_{\text{err,XZ}}(\eta_0, \sigma_{\text{GKP}}, L) = \text{erfc}\left(\sqrt{\frac{\pi}{8\sigma_{\text{eff}}(\eta_0, \sigma_{\text{GKP}}, L)^2}}\right).
\end{equation}

Since an even number of $X/Z$ errors can cancel the error, the probability of $X/Z$ error for the total distance $L_{\text{tot}}$ is the probability of odd number of errors given by:
\begin{equation}
Q_{\text{err,XZ}}(\eta_0, \sigma_{\text{GKP}}, L, L_{\text{tot}}) = \frac{1-(1-2 P_{\text{err,XZ}}(\eta_0, \sigma_{\text{GKP}}, L))^{L_{\text{tot}}/L}}{2}.
\end{equation}
Then the total channel acting on our quantum state during transmission over distance $L_{\text{tot}}$ is:
\begin{equation}
\mathcal{D}(\rho) = (1- Q_{\text{err,XZ}})^2 \rho + Q_{\text{err,XZ}}(1- Q_{\text{err,XZ}})(X\rho X + Z \rho Z) +  Q_{\text{err,XZ}}^2 Y \rho Y.
\label{eq:effchannel}
\end{equation}

We note that in quantum key distribution we estimate the quantum bit error rate (QBER) which is the probability that Alice's and Bob's bits do not agree in the given basis. We consider three bases $\{X,Y,Z\}$ and the logical Pauli error $P_i$, where $i \in \{1,2,3\}$ and $P_1 = X, P_2 = Y, P_3 = Z$, will flip a bit encoded in one of the two $P_j$ bases, where $j \neq i$ and will leave the bit encoded in the $P_i$ basis unaffected. Hence the QBER in the three bases is given by:
\begin{equation}
\begin{aligned}
e_X &= e_Z = Q_{\text{err,XZ}}(1- Q_{\text{err,XZ}}) +  Q_{\text{err,XZ}}^2 = Q_{\text{err,XZ}} \, , \\
e_Y &= 2 Q_{\text{err,XZ}}(1- Q_{\text{err,XZ}}) \, .
\end{aligned}
\label{eq:QBER}
\end{equation}
We provide details how to calculate secret-key rate in bits per mode $r'$ in Appendix~\ref{sec:seckeyRateAD}. After establishing the secret-key rate as a function of the repeater spacing $L$ for a given total distance $L_{\text{tot}}$ and for fixed parameters $\eta_0, \sigma_{\text{GKP}}$, we numerically maximise the secret-key rate over $L$ restricted to the interval between 250 meters and 1.5 kilometre. Then we look for the largest total distance $L_{\text{tot}}$ for which with the optimal $L$ the secret-key rate in bits per mode is still larger than $r'=0.01$. We search for this achievable distance using the binary search method.

\begin{figure}
\includegraphics[trim={0 0 0 0}, width = \columnwidth]{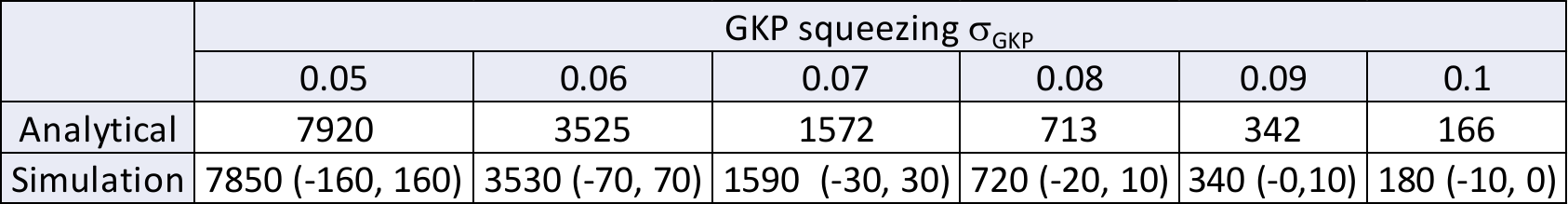}
\caption{\textbf{Achievable distance in kilometres for the GKP repeater chain, over which secret-key rate in bits per optical mode stays above $r'=0.01$ as predicted by the analytical model and the numerical Monte-Carlo simulation.} We fix photon coupling efficiency to $\eta_0 = 0.98$ and vary the amount of GKP squeezing $\sigma_{\text{GKP}}$. The repeater spacing is fixed to 250 m for both the simulation and the analytical model, as the latter predicts this minimum allowed repeater separation to be the optimal one from the range of allowed values for all the parameters considered in this table. The errors on the values obtained from the simulation are quoted in brackets. The values obtained from the analytical model on the other hand have an accuracy of around 10 km, as the achievable distance is obtained through 10 iterations of the binary search method over the distance range $[0, 10000]$ km. We observe a good agreement between the predictions of the analytical model and the Monte-Carlo simulation.}
\label{fig:RepChainAnalVsSim}
\end{figure}

Finally, to verify the validity of the above described analytical method, we compare its estimate to the Monte-Carlo simulation of the GKP repeater chain (the details of the simulation are described in Section~\ref{sec:MonteCarlo}). In this verification we consider scenarios with $\eta_0 = 0.98$ and multiple different values of $\sigma_{\text{GKP}}$. For each case we consider the minimum allowed repeater spacing of 250 m which based on the analytical model is the optimal spacing for all these parameters. In FIG.~\ref{fig:RepChainAnalVsSim} we show the maximum achievable distance over which secret-key rate in bits per optical mode stays above $r'=0.01$ as predicted by the analytical model and the Monte-Carlo simulation. We see that the analytical model closely agrees with the simulation for the considered parameters.

\section{Second level coding}
\label{sec:2ndLevel}

In this appendix we explain in more detail how to concatenate the GKP code with the higher-level qubit codes. Specifically, we consider two small qubit codes, namely the [[4,1,2]] code and the [[7,1,3]] Steane code. That is the logical qubit is now encoded in a quantum state consisting of respectively four and seven GKP qubits.

\subsection*{[[4,1,2]] code}

The [[4,1,2]] code enables us to detect a single-qubit Pauli error by measuring its stabilisers \{$Z_1 Z_2$, $Z_3 Z_4$, $X_1X_2X_3X_4$\}. The logical operations on this code-space are defined such that: $X_L = X_1X_2 = X_3X_4$ and $Z_L = Z_1Z_3 = Z_1Z_4 = Z_2Z_3 = Z_2Z_4$. However, on its own this code does not allow us to correct any errors. That is, if measuring one of the stabilisers triggers an error, we do not know which qubit needs to be corrected, e.g. if measuring $Z_1 Z_2$ results in the measurement outcome -1, we do not know whether the $X$ error happened on the first or the second qubit. If we apply the $X$ correction on the wrong qubit, we effectively apply $X_1X_2 = X_L$ hence causing a logical error. However, as we have discussed in Appendix~\ref{sec:GKPErrCorr} we can use analog information from lower-level GKP corrections to help us identify errors on this higher level. Specifically let us consider a repeater architecture in which four GKP qubits encoding one logical qubit through the [[4,1,2]] code are sent through a Gaussian random displacement channel to a repeater station which firstly performs GKP correction on all the four GKP qubits and then measures the joint stabilisers of the [[4,1,2]] code. Let us first assume the scenario with perfect GKP correction. Then, if the [[4,1,2]] code detects an error, this must have been caused by a failure to correct an error on the GKP level, e.g. if $Z_1 Z_2$ measurement gives -1 outcome, then this $X$ error on one of the first two qubits must have been caused by the fact that the Gaussian displacement channel resulted in the $\hat{q}$ quadrature shift on the first or second qubit which was larger in magnitude than $\frac{\sqrt{\pi}}{2}$. This was an uncorrectable error which resulted in a logical $X$ flip after the GKP correction.
However, as discussed in Appendix~\ref{sec:GKPErrCorr} the probability that the GKP correction failed and caused the logical error given the GKP syndrome $q_0$ is known and is given by the function given in Eq~\eqref{eq:analogInfoprob}. This error probability $p[\sigma](q_0)$ grows with the magnitude of $q_0$ reaching its maximum at the decision boundaries $\pm \sqrt{\pi}/2$. Hence it is clear that in this case on the second level there is a higher probability of error on this one of the first two qubits for which $\abs{q_0}$ during the GKP correction was larger. This is the qubit which we decide to correct in this case. We proceed similarly during the measurement and error correction of the other two stabilisers.

\subsection*{[[7,1,3]] Steane code}

The [[7,1,3]] Steane code can detect two errors and correct one by measuring the stabilisers \{$Z_4Z_5Z_6Z_7$, $Z_2 Z_3 Z_6 Z_7 $, $Z_1 Z_3 Z_5 Z_7$, $X_4X_5X_6X_7$, $X_2 X_3 X_6 X_7 $, $X_1 X_3 X_5 X_7$\}. The logical Pauli operators take a simple form $X_L = X^{\otimes 7}$ and $Z_L = Z^{\otimes 7}$. For this code every syndrome for $Z$ and $X$ stabilisers can uniquely identify a single-qubit error if such occurred. If a two-qubit $Z$ or $X$ error occurs, this cannot be distinguished from a single-qubit error happening on another qubit. Since single-qubit errors are more likely, one will in this case proceed to correct the single-qubit error thus causing effectively a three-qubit error which will implement a logical $X/Z$ operation on the encoded qubit. Although single-qubit errors are in general more likely, one can again use the analog information from the preceding GKP correction. Then again considering the scenario in which some of the $Z$ stabilisers produced outcome -1, we can now use the GKP correction syndrome $q_0$ from all the seven qubits to identify whether the single or two-qubit error is more likely in this scenario. Specifically for every syndrome for $Z$ ($X$) stabilisers there is 1 single and 3 two-qubit $X$ ($Z$) errors consistent that could have produced this syndrome. We can then calculate the probability of each of these four events using the analog information $q_0$ $(p_0)$ and correct the most likely one. Though the information provided by the analog syndrome is probabilistic, we find that in most cases it allows for successful correction of both single- and two-qubit errors.

\subsection*{Higher-level stabiliser measurement using ancilla GKP}

\begin{figure}
\includegraphics[trim={0 0 0 0}, width=\columnwidth]{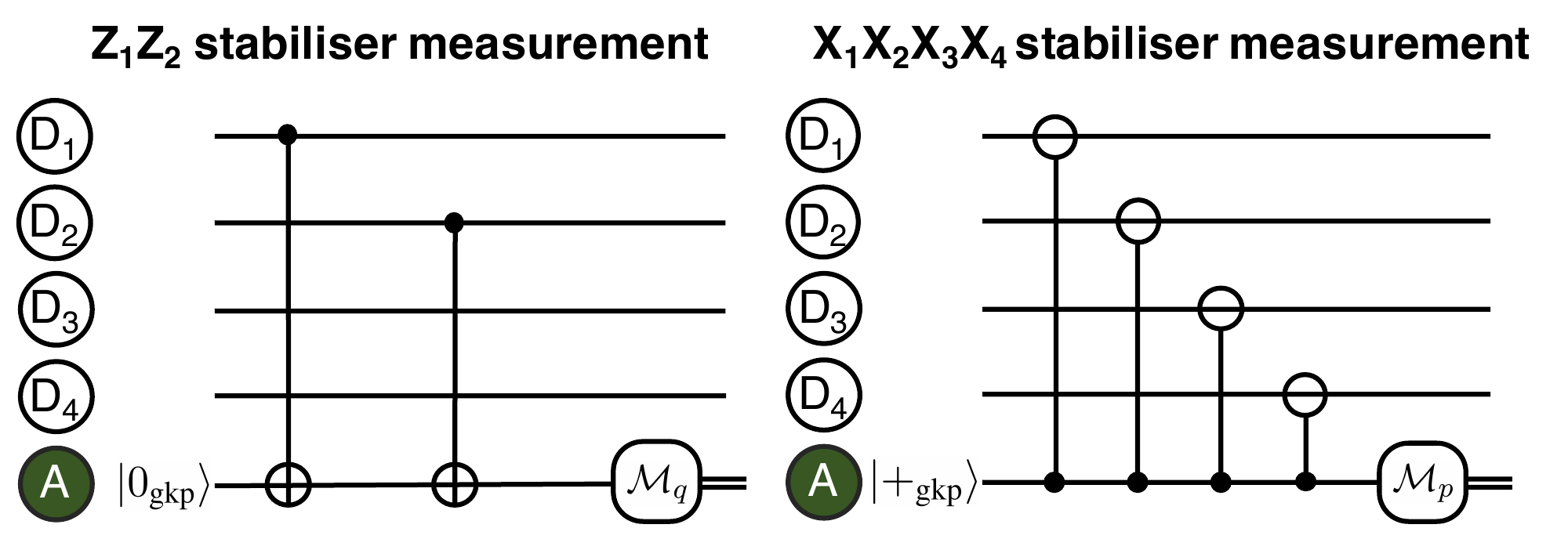}
\caption{\textbf{Quantum circuits for the measurements of the outer-code stabilisers.} We depict the circuit for the measurement of the $Z$-type $Z_1Z_2$ stabiliser (left) and an $X$-type $X_1X_2X_3X_4$ stabiliser (right) for the second-level [[4,1,2]] code with GKP encoding on the first level. The four data qubits are denoted by ``$D_i$'' and the ancilla qubit by ``A''.}
\label{fig:2ndLevStabMeas}
\end{figure}

Now we will briefly describe how to measure the stabilisers of the higher-level code. In fact this can be done in a similar way as in the GKP correction, using a GKP ancilla and the SUM gate for measuring $Z$ stabilisers for identifying $X$ errors due to $\hat{q}$ quadrature shifts and the inverse SUM gate for measuring $X$ stabilisers for identifying $Z$ errors due to $\hat{p}$ quadrature shifts as shown in the examples in FIG.~\ref{fig:2ndLevStabMeas}.

For measuring the $Z$ stabilisers we in sequence apply the SUM gate from the corresponding data qubits onto the GKP ancilla and then measure the ancilla. The main difference is that in the GKP correction we aimed for correcting only small shifts without revealing the GKP encoded information. For that purpose we needed to measure the data qubit modulo $\sqrt{\pi}$. Now measuring the $Z$ stabilisers amounts to measuring whether an even or odd number of $X$ errors occurred on the corresponding qubits. Since such individual errors are caused by a $\sqrt{\pi}$ shift in the $\hat{q}$ quadrature, to determine the parity, we need to be able to measure the ancilla modulo $2\sqrt{\pi}$ in the interval $[-\sqrt{\pi}, \sqrt{\pi})$. Such measurement can be performed by preparing the ancilla in the state $\ket{0_{\text{GKP}}} = \sum_{n \in \mathbb{Z}} \ket{q = 2n\sqrt{\pi}}$ which is periodic modulo $2\sqrt{\pi}$ in the $\hat{q}$ quadrature. Moreover, since $\ket{0_{\text{GKP}}} = \sum_{n \in \mathbb{Z}} \ket{p = n\sqrt{\pi}}$, the back-action of the sum gate is that it also shifts the $\hat{p}$ quadrature of the data mode by a superposition of integer multiples of $\sqrt{\pi}$. However, this back-action is the same on all the data qubits involved in the stabiliser measurement and shifting the $\hat{p}$ quadrature of all these data qubits by the same integer multiple of $\sqrt{\pi}$ effectively implements the action of exactly the $Z$ stabiliser which we are measuring. Since our encoded state is by construction invariant under this transformation, the data modes remain untouched (we note that even though the individual GKP qubits in the block could be displaced from the eigenspace of the stabilisers of the second-level code, the displacement operators commute up to a phase which in this case would be global and hence irrelevant, thus we can see the back-action as acting on the eigenstates of the second-level stabilisers followed by the residual displacements). Finally, the ancilla is measured using homodyne detection and we will consider the measured value modulo $2\sqrt{\pi}$ which we will denote here as $q_0^{\text{SL}} \in [-\sqrt{\pi}, \sqrt{\pi})$, where SL refers to the second level of coding in our scheme. The stabiliser value is inferred from $q_0^{\text{SL}}$  according to the rule:
\begin{equation}
S(q_0^{\text{SL}}) = \left.
\begin{cases}
        1, & \text{for } \abs{q_0^{\text{SL}}}\leq \sqrt{\pi}/2 \, ,\\ 
        -1, & \text{for } \abs{q_0^{\text{SL}}}\geq \sqrt{\pi}/2 \, .
\end{cases}
\right.
\label{eq:multiSyndromeRule}
\end{equation}
We note that for perfect GKP ancillas, if before the multi-qubit syndrome measurement the GKP correction was performed, then $q_0^{\text{SL}}$ can only take one of two discrete values $\{-\sqrt{\pi}, 0\}$. However if GKP ancillas are noisy then $q_0^{\text{SL}}$ can take any value from the interval $[-\sqrt{\pi}, \sqrt{\pi})$.
Finally, after all the $Z$ stabilisers are measured, and using the syndrome of the second-level code as well as the analog information it is established which GKP qubits need to be corrected, the $\sqrt{\pi}$ shift in the $\hat{q}$ quadrature is applied to these GKP qubits to correct the errors. The procedure to measure $X$-stabilisers involves inverse SUM gates, as in the case of individual GKP correction, and is analogous to the $Z$-stabiliser measurement.

We note that the use of finitely squeezed ancillas leads two types of errors. Firstly, the back-action from the ancilla onto the data qubits means that the small residual displacement on the ancilla due to its finite squeezing will propagate onto all the data qubits involved in the measurement of that stabiliser. However, as we will see these small displacements can later be corrected by subsequent GKP corrections. Secondly the measurement outcome of the ancilla used to measure the higher-level stabilisers is given by:
\begin{equation}
q_0^{\text{SL}} = R_{2\sqrt{\pi}}\left(\sum_i\hat{q}_{\text{data}, i} + \xi_{q,\text{anc}}^{\text{GKP}}\right) \, .\label{eq:measError}
\end{equation}
The summation index $i$ runs over all the data qubits measured by the specific stabiliser. We have already established in Eq.~\eqref{eq:varianceOptC} the variance on each GKP data qubit after GKP correction for the most likely scenario in which there was no logical error on the GKP level, only a small residual displacement. For this most likely case of no logical error on any of the GKP qubits during the GKP correction directly preceding the measurement of the second-level stabiliser, we can approximate the variance of $q_0^{\text{SL}}$ as:
\begin{equation}
\text{Var}(q_{0,\text{no error}}^{\text{SL}}) \approx \left(\sum_i c_{\text{opt,i}} + 1\right) \sigma_{\text{GKP}}^2 \, .
\end{equation} 

Clearly for the no-error case we ideally would like to have that $S(q_{0,\text{no error}}^{\text{SL}}) = 1$. However, we see that the higher the weight of the stabiliser measured, the larger the summation range of the index $i$ and hence the larger the variance. This large variance means that the residual errors accumulated on the ancilla after all the SUM gates can result in a flip of the higher level stabiliser so that we erroneously observe $S(q_{0,\text{no error}}^{\text{SL}}) = -1$. Here we find that already for weight-four stabilisers this measurement error probability can become a hindrance and therefore we will later discuss how we can overcome this problem by repeating the measurement of the same syndrome twice.

\section{Multi-qubit repeaters}
\label{sec:multiqubitrep}

\begin{figure}
\includegraphics[trim={0 0 0 0}, width = \columnwidth]{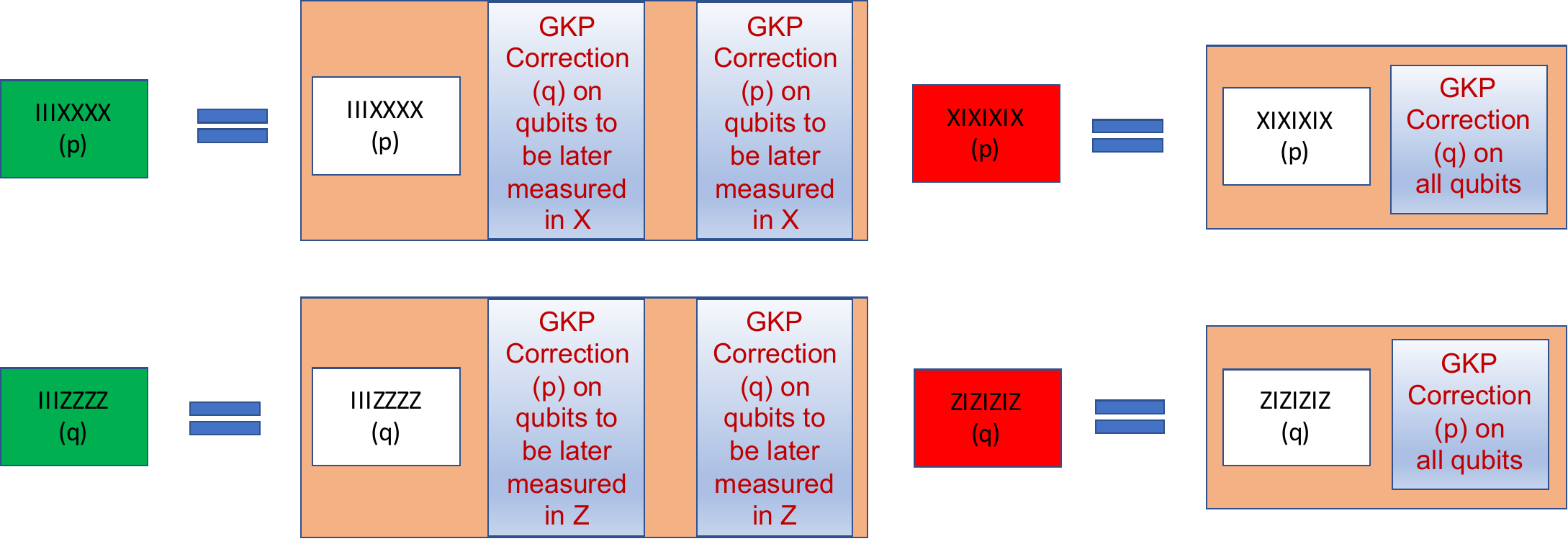}
\caption{\textbf{A schematic diagram of the building block operations performed inside type-A (multi-qubit) repeaters.} Green blocks denote multi-qubit stabiliser measurements which will be later followed by other stabiliser measurements in the same quadrature. They describe multi-qubit stabiliser measurements followed by the GKP correction firstly in the opposite and then in the same quadrature as the one measured during the multi-qubit correction. The GKP correction is applied to these data qubits out of the qubits participating in the given multi-qubit stabiliser measurement, which will later participate in other multi-qubit stabiliser measurements in the same quadrature in the same repeater. Red blocks denote the last multi-qubit stabiliser measurements of a given quadrature. They describe the given multi-qubit stabiliser measurement followed by the GKP correction in the opposite quadrature on all the data qubits.}
\label{fig:buldingBlockOperations}
\end{figure}

\begin{figure}
\includegraphics[width = \columnwidth]{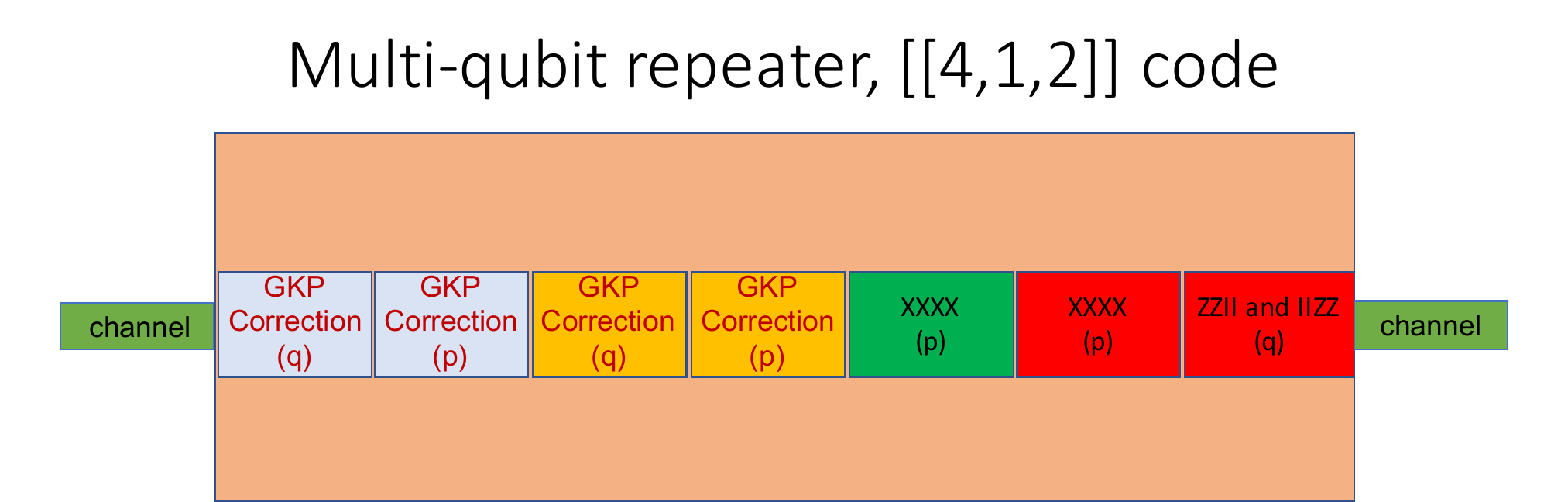}
\caption{\textbf{A schematic diagram of the operations performed inside a multi-qubit repeater based on the [[4,1,2]] code.} Firstly we perform GKP correction to overcome the errors from the communication channel (blue). After performing this correction both in $\hat{q}$ and in $\hat{p}$ quadrature in that order, we repeat it, as this allows us to reduce the residual displacements (yellow). Now the $\hat{p}$ quadrature has less residual displacement than the $\hat{q}$ quadrature so we proceed to perform second-level correction by measuring the $X$ stabiliser to correct errors in $\hat{p}$ quadrature. We perform the measurement of this single $X$ stabiliser twice (green and red). Then we proceed to do second-level correction using the measurement of $Z$ stabilisers to correct logical GKP errors in $\hat{q}$ quadrature (red).}
\label{fig:4QbitRepeater}
\end{figure}

Now we will describe the repeater architecture which uses multi-qubit (type-A) repeater nodes that aim to correct both lower and higher-level errors. Specifically such stations will firstly perform single-mode GKP correction on all the physical GKP qubits followed by the measurement of the second-level stabilisers and the corresponding higher-level correction. However, as already discussed, the use of imperfect ancillas results in errors that can accumulate and therefore it is vital to perform these correction operations in such a way as to minimise the overall error probability. Specifically, residual displacements accumulated on the data qubits can effectively flip the ancilla used for the measurement of the higher-level stabiliser, thus leading to the measurement error. During such a measurement, these residual errors from all the measured data qubits are transferred onto that ancilla. Therefore measurements of higher weight stabilisers are more prone to errors. This also justifies why it is vital to make the residual displacements on the data qubit as small as possible immediately before the measurement of the higher-level stabiliser. Therefore now we motivate and describe in detail all the operations performed in the multi-qubit repeaters that help us to increase the reliability of the higher-level stabiliser measurements. These operations follow a basic set of rules:
\begin{enumerate}
\item Whenever a data qubit participates in a higher-level stabiliser measurement $\text{M}_i(\hat{x}_1)$ for correcting errors in the quadrature $\hat{x}_1$, and in the same repeater that data qubit will later again participate in some higher-level stabiliser measurement $\text{M}_j(\hat{x}_1)$ again in the same quadrature $\hat{x}_1$, then after the measurement of $\text{M}_i(\hat{x}_1)$ on that data qubit we apply two GKP corrections to it. Firstly, we perform a GKP correction $\text{GKP}(\hat{x}_2)$ in the quadrature $\hat{x}_2$, that is the opposite quadrature to $\hat{x}_1$, in order to eliminate the back-action error from that higher-level measurement. Then we apply a GKP correction $\text{GKP}(\hat{x}_1)$ in the $\hat{x}_1$ quadrature, to eliminate the noise from the back-action during $\text{GKP}(\hat{x}_2)$ and hence to prepare the data qubit for the measurement of $\text{M}_j(\hat{x}_1)$. This is illustrated on the left in FIG.~\ref{fig:buldingBlockOperations}.
\item In a quadrature in which weight-four stabilisers need to be measured on the higher level, we measure all the stabilisers twice. That is we perform two consecutive measurement rounds where in each round we measure all the different stabilisers. We describe later how we use majority voting over these two rounds of outcomes to reliably estimate the value of the higher-level syndrome.
\item After measuring the last higher-level stabiliser of a given quadrature, say $\hat{x}_1$, all the data qubits are subjected to the GKP correction in the opposite quadrature $\hat{x}_2$ in order to remove all the back-action effects, e.g. after the last measurement of the last $X$ stabiliser for detecting $\hat{p}$ errors, we need to apply $\text{GKP}(\hat{q})$ to all the data qubits in order to remove the residual errors caused by the back-action during the $X$-stabiliser measurements. This is illustrated on the right in FIG.~\ref{fig:buldingBlockOperations}.
\item The first operations in the type-A repeater are the GKP corrections in both quadratures aiming at correcting the communication channel noise. If directly after these GKP corrections we proceed to measure weight-four stabilisers, then instead of performing these GKP corrections once, we repeat them again before measuring these higher-level stabilisers. That is, the GKP corrections are performed twice in the alternate fashion: $\text{GKP}(\hat{q})-\text{GKP}(\hat{p})-\text{GKP}(\hat{q})-\text{GKP}(\hat{p})$. Repeating such GKP corrections twice enables for better suppression of the residual displacements. This follows from Eqs.~\eqref{eq:varianceOptC} and~\eqref{eq:optC}. The residual displacement after GKP correction is smaller, if the contributed noise before the correction operation was smaller relative to $\sigma_{\text{GKP}}$. The communication channel adds significant amount of noise, hence e.g. the residual displacement after the second $\text{GKP}(\hat{p})$ will be smaller than after the first $\text{GKP}(\hat{p})$. This is because the amount of noise added between the two $\text{GKP}(\hat{p})$ is small comparing to the channel noise, namely it only comes from the residual displacement during $\text{GKP}(\hat{q})$.
\item When switching from measuring one quadrature to the other, if the quadrature to be measured includes measuring weight-four stabilisers, we repeat GKP corrections similarly to the previous case, e.g. when switching from measuring $X$ to $Z$, we will apply $\text{GKP}(\hat{q})-\text{GKP}(\hat{p})-\text{GKP}(\hat{q})$ instead of just $\text{GKP}(\hat{q})$. Similarly as in the previous case the residual displacements after second $\text{GKP}(\hat{q})$ will be smaller than after the first one.
\end{enumerate} 
The additional repeated GKP corrections described in points 4 and 5 are marked in yellow in FIG.~\ref{fig:4QbitRepeater} and in FIG.~\ref{fig:7QbitRepeater}.

We now focus on the multi-qubit repeater nodes based on the [[4,1,2]] code and then on those based on the [[7,1,3]] code. 

\subsection*{Multi-qubit repeaters based on the [[4,1,2]] code}

We depict the operations performed inside the multi-qubit repeater in FIG.~\ref{fig:4QbitRepeater}. A type-A repeater based on the [[4,1,2]] code firstly performs GKP corrections on the four individual GKP modes. This operation is repeated to reduce the residual displacement before measuring the weight-four $X$-stabiliser twice. Then we proceed to measure the $Z_1Z_2$ and $Z_3Z_4$ stabilisers. This is followed by a decoding procedure which we describe in more detail below. Finally the four GKP qubits are subjected to phase-insensitive amplification and sent sequentially to the next repeater station.

\subsection*{Multi-qubit repeaters based on the [[7,1,3]] code}

\begin{figure}
\includegraphics[width = \columnwidth]{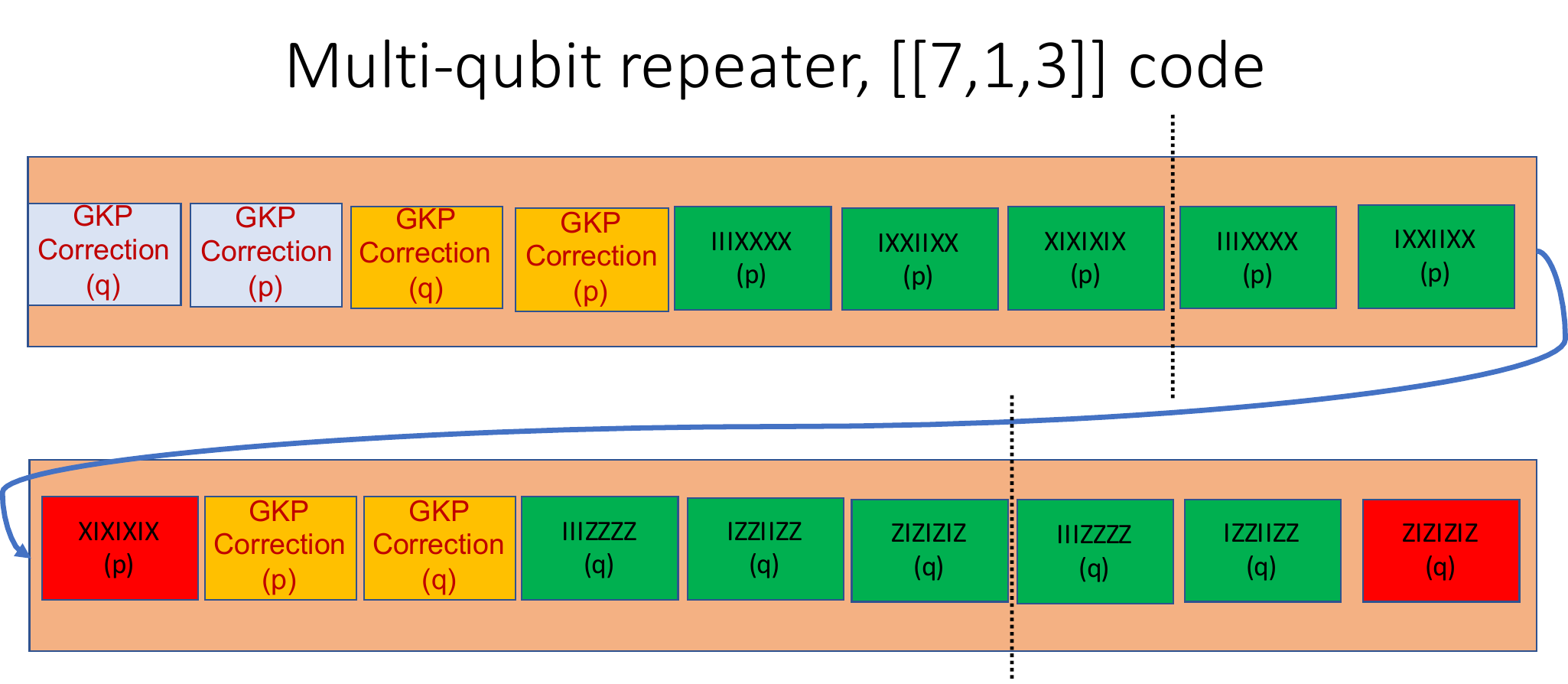}
\caption{\textbf{A schematic diagram of the operations performed inside a multi-qubit repeater based on the [[7,1,3]] code.} We start by performing the GKP correction (blue). We then do it again to reduce the residual displacement (yellow). Then we measure all the $X$-stabilisers in two rounds (green and red). Then we switch to measuring $Z$-stabilisers. To do that, we again add the additional GKP corrections first, in order to reduce the residual displacement in $\hat{q}$ (yellow). Finally we measure all the $Z$-stabilisers in two rounds (green and red). The black dashed lines separate the two rounds of measuring the stabilisers of a given quadrature.}
\label{fig:7QbitRepeater}
\end{figure}

The procedure inside the type-A repeater based on the [[7,1,3]] code is similar and is depicted in FIG.~\ref{fig:7QbitRepeater}. Here we measure weight-four stabilisers both in $X$ and in $Z$, therefore we add the additional GKP corrections before measuring both $X$ and $Z$ stabilisers (yellow). W also measure both the $X$ and the $Z$ stabilisers in two rounds.

\subsection*{Decoding procedure at multi-qubit repeaters using analog information}

Here we will describe the procedure of identifying the errors based on the information obtained from the GKP and multi-qubit syndrome measurements. Our decoding procedure consists of two steps, firstly we want to reliably estimate the second-level syndrome and secondly we want to identify the most likely errors consistent with that syndrome. Before we go into more detail into these two steps we expand on the ideas described in Appendix~\ref{sec:GKPErrCorr} and explain how we establish the error likelihood during a GKP and multi-qubit syndrome measurements.

We have already established in Eq.~\eqref{eq:analogInfoprob} how to calculate the error likelihood during the GKP correction for a given measured GKP syndrome $q_0$. For that we need to know standard deviation $\sigma$ of the distribution from which the measured GKP syndrome was effectively drawn. Hence $\sigma$ needs to include all the processes that contribute to $q_0$ which will include the residual displacement after the previous GKP correction (with variance $c_{\text{opt}}'\sigma^2_{\text{GKP}})$, the noise added by the environment before the GKP correction (with variance $\sigma^2_{\text{noise}})$ and the noise of the GKP ancilla used for measuring the GKP stabiliser (with variance $\sigma^2_{\text{GKP}}$). Hence:
\begin{equation}
\sigma^2 = (c_{\text{opt}}' + 1) \sigma^2_{\text{GKP}} + \sigma^2_{\text{noise}},
\end{equation}
where $c_{\text{opt}}'$ denotes the value of $c_{\text{opt}}$ used during the previous GKP correction round.

We can follow a similar strategy to establish an error likelihood during a multi-qubit syndrome measurement. In this case the syndrome is in the interval $q_0^{\text{SL}} \in [-\sqrt{\pi}, \sqrt{\pi})$ and the discrete stabiliser value is assigned according to the rule given in Eq.~\eqref{eq:multiSyndromeRule}. Hence we can calculate $R_{\sqrt{\pi}}(q_0^{\text{SL}})$ which does not reveal the information about the discrete stabiliser value, but similarly as in the case of GKP stabiliser measurement, can provide us with a likelihood of error during this assignment of the discrete stabiliser value. The error likelihood can then also be established using Eq.~\eqref{eq:analogInfoprob}, where we now replace $q_0$ with $R_{\sqrt{\pi}}(q_0^{\text{SL}})$, that is we want to calculate $p[\sigma](R_{\sqrt{\pi}}(q_0^{\text{SL}}))$. Since we always measure the multi-qubit stabilisers directly after a GKP correction in the same quadrature, the corresponding $\sigma$ will include the residual errors from all the data qubits involved and the errors from the ancilla. That is:
\begin{equation}
\sigma^2 = \left(\sum_i c_{\text{opt}, i} +1 \right) \sigma_{\text{GKP}}^2.
\label{eq:errorDistMulti}
\end{equation}
 Here $c_{\text{opt}, i}$ denotes the rescaling coefficient used at the last GKP correction on $i$'th data qubit involved in this multi-qubit stabiliser measurement and the sum is taken over all the data qubits involved in this measurement. The additional plus one term comes from the noise contributed by the ancilla.
 
 Having established how to calculate error likelihood $p[\sigma]$ during a GKP and multi-qubit syndrome measurement, we proceed to describe how this information can be used to identify the errors.
 
 \subsubsection*{Step 1: Estimating multi-qubit stabiliser values}
 The first step is to reliably assign the multi-qubit stabiliser values. For weight-two stabilisers measured for the case of the [[4,1,2]] code, in general the error distribution given in Eq.~\eqref{eq:errorDistMulti} will be narrow because the sum is taken only over two values of $i$. Hence the probability of error during a measurement of the weight-two stabiliser will be small and so we just keep the discrete value obtained by applying the rule from Eq.~\eqref{eq:multiSyndromeRule} to the measured $q_0^{\text{SL}}$.
 
 In the case of weight-four stabiliser, $\sigma$ will be much larger as the summation index $i$ will now run over four values corresponding to four data qubits rather than two. This means that the probability of measurement error is much higher, and therefore, as discussed, we make two rounds of measuring all such stabilisers. If in both rounds the discrete value of the stabiliser is the same, we assign that value. However, if the two values are different, we consider three error possibilities. Either there was a measurement error during the first or the second time this stabiliser was measured or there was a logical error during one of the GKP corrections on the relevant qubits in that quadrature somewhere between the two multi-qubit stabiliser measurements. We then use the analog error likelihood to establish the most likely possibility. This we can do through the following steps:
 \begin{enumerate}
 \item Firstly we compare $p[\sigma](R_{\sqrt{\pi}}(q_0^{\text{SL}}))$ from the two multi-qubit stabiliser measurements with the likelihood of an error on one of the intermediate GKP corrections given by $1-\prod_i(1-p[\sigma_i](q_{0,i}))$, where the product is taken over all the GKP corrections in the same quadrature and on the relevant qubits, performed between the two rounds of measuring a given multi-qubit stabiliser. If the most likely possibility is the error during one of the two rounds of the multi-qubit stabiliser measurements, we simply assign the discrete value corresponding to the outcome of the other measurement which we assume was not erroneous.
 \item If the most likely possibility is an error during one of the intermediate GKP corrections, then we aim to identify the most likely qubit on which the error happened. That is we compare the values of $1-\prod_i(1-p[\sigma_i](q_{0,i}))$, where now the product is taken over all the GKP corrections in the same quadrature but now only on a single qubit. We calculate this likelihood for all the qubits that were measured during the concerned multi-qubit stabiliser measurement and identify the qubit with highest error likelihood.
 \item This procedure is repeated for all the other stabilisers. If for some other stabiliser we also find that the two rounds of measurements resulted in different outcomes and that the most likely cause of this error is an error during an intermediate GKP correction, we assume that this is the same error which we have identified as the most likely for the case of the previous stabiliser. That is we assume that there was at most one logical error during all the intermediate GKP corrections performed after the first measurement of the first multi-qubit stabiliser and before the second measurement of the last multi-qubit stabiliser. Hence if for both stabilisers we find that there was an intermediate error, we assume that this is the same error which affected both of the stabilisers. In case we find that for both of the two discussed stabilisers the most likely is the error on an intermediate qubit but on a different qubit for each of these two stabilisers, we still assume that there was only a single error and that error was on the one of these two qubits for which the error likelihood is higher. However, we see in our simulation that the discrepancy between the most likely qubit with an error almost never arises, that is whenever the processing of the outcomes for two or more stabilisers points towards a high likelihood of an error during an intermediate GKP correction, almost always they all point to the error on the same data qubit.
 \item Finally, after identifying the most likely data qubit with an error during an intermediate GKP correction as well as the stabilisers it has affected, we flip the value of certain stabilisers so that we can effectively have that error reflected in the final syndrome. Specifically, we flip those stabilisers which involved measuring the qubit that we expect had an error and for which both of the measurement rounds occurred before this most likely intermediate GKP error. For the stabilisers which were affected by that intermediate error (that is that error happened on a relevant qubit between the first and the second round of measuring that stabiliser) we assign the value from the second measurement. In this way we effectively include that error in the final second-level syndrome.
 \end{enumerate}
 
\begin{figure}
\includegraphics[width = \columnwidth]{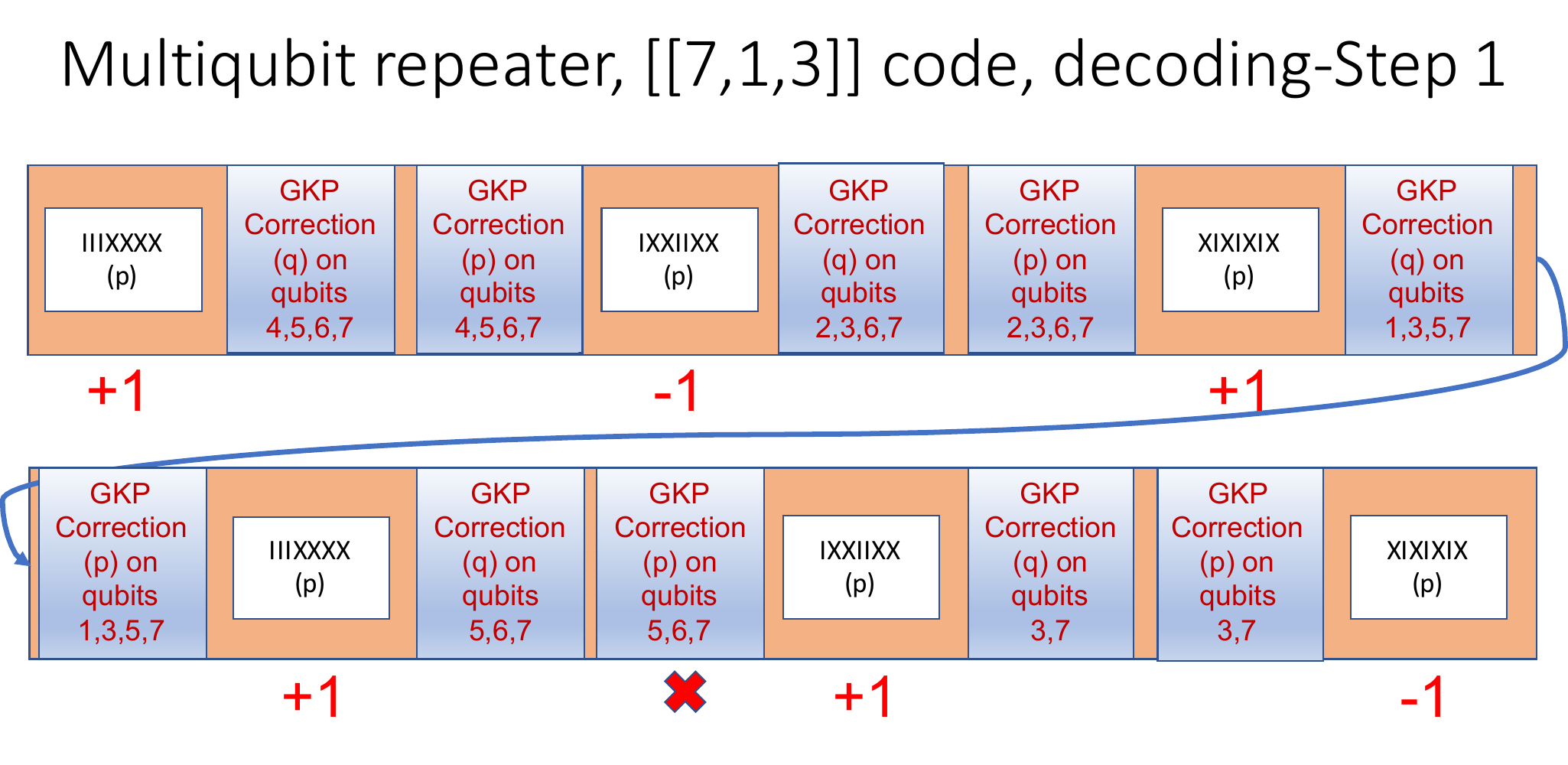}
\caption{\textbf{Example of multi-qubit stabiliser outcomes for the $X$-stabiliser measurements inside the multi-qubit repeater based on the [[7,1,3]] code.} The red numbers denote the discrete stabiliser values. The red cross denotes the most likely error location consistent with the multi-qubit stabiliser values in the scenario where the most likely error causing discrepancy between the two rounds of measuring the stabilisers $IXXIIXX$ and $XIXIXIX$ is a logical error during the intermediate GKP correction on qubit-7.}
\label{fig:DecodingStep1}
\end{figure}

 To better understand this procedure we provide an example based on the measurement of $X$-stabilisers for the [[7,1,3]] code. The specific scenario is depicted in FIG.~\ref{fig:DecodingStep1}. In the considered example we find that both measurements of the $IIIXXXX$ stabiliser produced the same value, while the two rounds of measurement of the $IXXIIXX$ and $XIXIXIX$ produced opposite outcomes as marked by the red $\pm 1$ numbers. Moreover, let us assume that the likelihood analysis shows that for both of these last two stabilisers the most likely source of this error is an error during an intermediate GKP correction on qubit 7. Therefore we conclude that there is a high likelihood of an error at the location marked with the red cross, as that error would affect the measurements of the stabilisers $IXXIIXX$ and $XIXIXIX$ but would not affect the measurements of the $IIIXXXX$ stabiliser which also measures qubit-7 but for which both measurements took place before the location marked with the red cross. Then we assign to the stabilisers $IXXIIXX$ and $XIXIXIX$ the values obtained from the second measurements that is $+1$ and $-1$ respectively, while we also flip the value of the $IIIXXXX$ stabiliser to $-1$. In this way we effectively include the error at location marked with the red cross into our syndrome.
 
 Here we want to comment why we simply do not apply a $Z$ flip on qubit-7 to correct the error if we know from our likelihood analysis its most likely location. In particular, we see from our simulation that the method described here that does not involve such a correction flip gives better performance. We believe that this is because in case our likelihood analysis has failed to correctly identified the error, the correcting $Z$ flip would result in a new error that we cannot correct. However, if we do not correct this error, there are specific configurations in which a misidentification of such an intermediate GKP error can still be corrected in the second decoding step described below. If the $Z$ flip for correction is wrongly applied, it cannot be corrected in this second step.

 \subsubsection*{Step 2: Finding the most likely errors}
 
 The second step in our decoding procedure is to find the most likely error consistent with our reliably established stabiliser values. For that we use the analog information from all the GKP corrections in the relevant quadrature from a single segment between two consecutive multi-qubit syndrome measurements in two consecutive type-A repeaters (including also the intermediate GKP corrections placed between multi-qubit stabiliser measurements which we used in Step 1 and including the GKP corrections in the type-B stations in the hybrid architecture). Specifically let $p[\sigma_{i,j}](q_{0,i,j})$ denote the error likelihood of $i$'th GKP correction on qubit $j$. We can then use these single-qubit error likelihoods to identify the most likely error qubit(s) consistent with the multi-qubit syndrome. For the case of the [[4,1,2]] code, we are looking to identify only a single-qubit error. Therefore for each of the relevant qubits we calculate the probability that there was no effective logical error after all the GKP corrections which corresponds to the probability of an even number of errors:
 \begin{equation}
 p_{\text{no error}, j} = \frac{1 + \prod_i (1-2p[\sigma_{i,j}](q_{0,i,j}))}{2} \, .
 \end{equation}
 Then we find $j$ for which $p_{\text{no error}, j}$ is the smallest. This is the qubit with the highest error likelihood and which we will correct.
 
 For the case of the [[7,1,3]] code, we also aim to correct two-qubit errors. Specifically for each multi-qubit syndrome there is one single-qubit and three two-qubit errors that are possible. To identify the most likely possibility, for each case we calculate the probability of odd number of errors on the erroneous qubits multiplied by the probability of the even number of errors on the non-erroneous qubits. This is our error likelihood for each of these cases. We then choose the case with largest error likelihood. We can then correct the relevant qubits corresponding to that case. 
 
\section{Repeater performance: secret key generation and advantage distillation}
\label{sec:seckeyRateAD}
 
The repeater performance is measured by its ability to generate shared secret key and here we provide the details of the specific quantum key distribution (QKD) protocol which we propose to implement on the considered repeater architectures. Specifically, we consider a prepare-and-measure version of the six-state QKD protocol~\cite{bruss1998optimal} supplemented with a two-way advantage distillation scheme~\cite{renner2008security,watanabe2007key}. In this protocol Alice chooses one of the three mutually unbiased qubit bases $\{X,Y,Z\}$ at random and then uniformly at random prepares one of the basis states of that chosen basis, encoding the bit value of 0 or 1. In our architectures this state becomes then encoded in the GKP or the two-level code and is sent to Bob through the repeater chain. After arriving to Bob, he decodes the state and measures it in one of the three $\{X,Y,Z\}$ bases at random. In this way Alice and Bob establish a shared raw bit of the secret key. Such raw bits need to be post-processed~\cite{scarani2009security}. The first step in such postprocessing is sifting in which Alice and Bob discard bits for the rounds in which they have chosen different bases. Yet it has been shown that in the asymptotic limit of $n \rightarrow \infty$, where $n$ is the number of protocol rounds, the key generation basis can be chosen with probability approaching one, without compromising security~\cite{lo2005efficient}. Thus effectively a negligible fraction of raw bits will be discarded during sifting. We call such a protocol a fully asymmetric protocol. In the next step Alice and Bob perform parameter estimation to evaluate the quantum bit error rate (QBER) $\{e_X, e_Y, e_Z\}$ which corresponds to the probability that in the given basis their bits have a discrepancy. Then Alice and Bob apply the advantage distillation procedure in which they use two-way communication in order to discard some sets of bits which contribute to the error. As a result the error in the key generation basis can be significantly decreased especially in the high-noise regime. Then Alice and Bob perform a standard one-way error correction and privacy amplification~\cite{scarani2009security}.

Both the GKP code and the multi-qubit codes which we use here have the nice feature that they aim to independently correct $X$ and $Z$ errors. This means that the $Y$ error is quadratically suppressed as $Y$ error will arise only if we failed to correct both the $X$ and the $Z$ error. The effective channel induced by our encoded transmission is the Pauli channel:
\begin{equation}
\mathcal{D}(\rho) = (1-q_X-q_Z-q_Y)\rho + q_X X \rho X + q_Z Z \rho Z + q_Y Y \rho Y.
\label{eq:channelpXpYpZ}
\end{equation}
As we have already discussed in Appendix~\ref{sec:GKrep} a flip in a given basis affects only the QBER in the other two six-state protocol bases. Hence:
\begin{equation}
\begin{aligned}
e_X &= q_Z +  q_Y \, , \\
e_Z &= q_X +  q_Y \, , \\
e_Y &= q_X +  q_Z \, .
\end{aligned}
\label{eq:QBERrelations}
\end{equation}
Since $q_Y$ is quadratically suppressed with respect to $q_X$ and $q_Z$, $e_Y$ is much larger than $e_X$ and $e_Z$. The fact that we have one basis with much higher QBER allows us to use the observation of~\cite{murta2020key} that when using advantage distillation, including the specific scheme of~\cite{watanabe2007key}, we can extract more key if the key is generated in the basis with the highest QBER.

We note here that in our analysis we assume that Bob also performs a final round of quantum error correction on both levels before measuring the encoded qubit. However, if he chooses to measure the logical qubit in the $X$ ($Z$) basis, then he does not need to correct $X$ ($Z$) Pauli errors. This means that for the considered GKP and multi-qubit codes, in these cases he in principle only needs to measure the $X$ ($Z$) stabilisers which can be done on the classical level. That is Bob could first measure all the GKP qubits in the relevant basis and then perform error correction (still using analog information) on the classical values which he obtained from these measurements. Such classical error correction would not require ancilla modes hence eliminating the noise coming from finite squeezing. Unfortunately that strategy does not work if he chooses to perform the measurement in the $Y$ basis as then he needs to correct both $X$ and $Z$ errors. Therefore in our analysis we treat Bob's station as a repeater that also performs full quantum error correction in both quadratures.

After error correction, Bob can measure the encoded qubit in a chosen basis. For the concatenated-coded schemes he needs to measure the logical $X$, $Z$ or $Y$ operator of the outer code. For the case of the [[7,1,3]] code this corresponds to measuring all the GKP qubits in the chosen basis and then applying XOR to all the 7 bits to obtain the final logical outcome. For the case of the [[4,1,2]] code the situation is more complex. Since, the logical $Z$ operator can be written as $Z_L = Z_1 Z_3$, measurement in the standard basis corresponds to measuring the first and third qubit in the standard basis followed by the XOR of the resulting two bits to obtain the final logical outcome. Similarly to perform the measurement in the $X$-basis, he needs to measure the first and second qubit in the $X$ basis since $X_L = X_1 X_2$. Logical $Y$ operator on the other hand is $Y_L = i X_L Z_L = Y_1 X_2 Z_3$. Hence measurement in the $Y$ basis corresponds to measuring the first qubit in $Y$ basis, second in $X$ basis and third in $Z$ basis followed by the XOR of the resulting three bits. Since these physical qubits are actually GKP qubits, measuring them in a given basis can be implemented by the homodyne detection along the $q$, $p$ or the diagonal axis in phase space corresponding to the measurement in the $Z$, $X$ and $Y$ basis respectively.

Now we describe how to calculate secret key using the advantage distillation protocol of~\cite{watanabe2007key}. The expressions below correspond to the entanglement-based scheme. However, they can also be used to describe a prepare-and-measure implementation as the entanglement-based scheme can be considered as the virtual purification of the prepare-and-measure scheme. That is, a possible way of implementing the prepare-and-measure scheme is for Alice to generate locally a maximally entangled state and measure one of the qubits in one of the protocol bases. This effectively prepares the second qubit in one of the prepare-and-measure protocol states, so that the second qubit can then be sent to Bob. Similarly to~\cite{watanabe2007key}, we denote the four Bell states as
\begin{eqnarray}
\ket{\psi(\san{x},\san{z})} =
\frac{1}{\sqrt{2}}(
\ket{0}\ket{0 + \san{x}} + (-1)^{\san{z}}
\ket{1} \ket{1 + \san{x}\,\,(\textrm{mod} \, 2)}
),
\end{eqnarray}
for $\san{x},\san{z} \in \{0,1\}$.
The Bell-diagonal state can then be written as
\begin{eqnarray}
\rho_{AB} = \sum_{\mathclap{\san{x},\san{z} \in \{0,1\} }}\; p_{\san{xz}}
\ket{\psi(\san{x},\san{z})}\bra{\psi(\san{x},\san{z})}\ .
\label{eq:belldiag}
\end{eqnarray}
Here the four Bell-diagonal coefficients $p_{\san{xz}}$ define the probability distribution $P_{\san{XZ}}$. If we implement the fully asymmetric six-state protocol for key extraction in the $Z$ basis and supplement this protocol with the two-way post-processing of~\cite{watanabe2007key}, then the resulting secret-key rate is given by:
\begin{eqnarray} 
r_{\text{six-state}} = \max \left\lbrace1 - H(P_{\san{XZ}}) + \frac{P_{\bar{\san{X}}}(1)}{2}
h\left( \frac{p_{00} p_{10} + p_{01} p_{11}}{
(p_{00} + p_{01})(p_{10} + p_{11})}\right), 
\frac{P_{\bar{\san{X}}}(0)}{2} [
1 - H(P_{\san{XZ}}^\prime) ] \right\rbrace ,
\label{eq-key-rate-vs-error-rate}
\end{eqnarray} 
where
\begin{equation}
\begin{aligned}
P_{\bar{\san{X}}}(0) &=& (p_{00} + p_{01})^2
+ (p_{10} + p_{11})^2\ , \\
P_{\bar{\san{X}}}(1) &=& 2 (p_{00} + p_{01})
(p_{10} + p_{11})\ , \\
p_{\san{00}}^\prime &=&
\frac{p_{00}^2 + p_{01}^2}{(p_{00} + p_{01})^2 + (p_{10}+p_{11})^2}, \\
p_{\san{10}}^\prime &=& 
\frac{2 p_{00} p_{01}}{(p_{00} + p_{01})^2 + (p_{10}+p_{11})^2}, \\
p_{\san{01}}^\prime &=&
\frac{p_{10}^2 + p_{11}^2}{(p_{00} + p_{01})^2 + (p_{10}+p_{11})^2}, \\
p_{\san{11}}^\prime &=&
\frac{2 p_{10}p_{11} }{(p_{00} + p_{01})^2 + (p_{10}+p_{11})^2} \ ,
\end{aligned}
\label{eq:ParamForSecKey}
\end{equation}
and $H(P_{\san{XZ}})$ denotes the Shannon entropy of the distribution $P_{\san{XZ}}$.

The Bell-diagonal coefficients $p_{\san{xz}}$ can be expressed in terms of the QBER $\{e_X, e_Y, e_Z\}$ established during the parameter estimation step of the protocol. For a prepare-and-measure scheme, the QBER in a given basis is the fraction of the bits encoded by Alice and measured by Bob in that basis for which there is a discrepancy between the bit value encoded by Alice and the one measured by Bob. For the purifying entanglement-based scheme we need to assume a specific target Bell-state. Let us assume here that the target state that Alice and Bob aim to generate is $\ket{\psi(\san{0},\san{0})}$. For key generation in the $Z$ basis, $e_i$ will be equal to the sum of the Bell-diagonal coefficients $p_{\san{xz}}$ which contribute to the error in the basis $i \in \{X,Y,Z\}$ relative to the correlations of $\ket{\psi(\san{0},\san{0})}$, e.g. $e_Z = p_{10} + p_{11}$, as the state $\ket{\psi(\san{0},\san{0})}$ is correlated in the $Z$ basis while the states $\ket{\psi(\san{1},\san{0})}$ and $\ket{\psi(\san{1},\san{1})}$ are anti-correlated. Note that $\ket{\psi(\san{0},\san{0})}$ is correlated both in the $Z$ and $X$ bases, but is anti-correlated in the $Y$ basis.

 However, as mentioned earlier, we would like to calculate the amount of key that can be extracted if the key is generated in the $Y$ basis. As discussed in~\cite{murta2020key}, the amount of generated key in a different protocol basis than the $Z$ basis can still be calculated using the same function of the Bell-diagonal coefficients $p_{\san{xz}}$. However, in the relation between the Bell-diagonal coefficients and the QBER obtained for the case when the key is extracted in the $Z$ basis we now need to permute the individual QBERs. As a result for the key generation in the $Y$ basis we obtain the following relations:
\begin{equation}
\begin{aligned}
p_{00} &= 1 - \frac{e_X + e_Z + e_Y}{2} \, ,\\
p_{01} &= \frac{e_X + e_Z - e_Y}{2}  \, ,\\
p_{10} &= \frac{-e_X + e_Z + e_Y}{2}  \, ,\\
p_{11} &= \frac{e_X - e_Z + e_Y}{2}  \, .
\end{aligned}
\end{equation}

Finally, we note that in information theory the crucial figure of merit with regard to generation of secret key is the amount of such key that can be generated per optical mode. Each of our GKP qubits occupies a single optical mode and in order to transmit and reconstruct one logical qubit, we need to transmit $n$ GKP qubits, where $n$ is the number of physical qubits encoding one logical qubit in the second-level code. Hence we can calculate the secret-key rate in bits per optical mode for our scheme as:
\begin{equation}
r' = \frac{r}{n},
\label{eq:secfracnormal}
\end{equation}
where $n=4$ for the architecture based on the [[4,1,2]] code and $n=7$ for the architecture based on the [[7,1,3]] code. For the GKP repeater chain $n=1$.

\section{Transmission inflidelity}
\label{sec:infidelity}

The performance metric we chose for FIG.~\ref{fig:analog} is the transmission infidelity for the worst case scenario. That is, it is the infidelity maximised over all the possible pure qubit states $\ket{\psi}$ which we may want to transmit. In this appendix we provide the expressions for this maximum transmission infidelity for all the encodings considered in FIG.~\ref{fig:analog}.

Let $\mathcal{D}$ denote the effective qubit channel comprising the encoding by the sender, transmission through the pure-loss channel and the decoding at the receiver. Then the transmission fidelity $F(\ket{\psi})$ between the input state $\ket{\psi}$ and the output state $\mathcal{D}(\dyad{\psi})$, is given by:
\begin{equation}
F(\ket{\psi}) = \bra{\psi}\mathcal{D}(\dyad{\psi})\ket{\psi} \, .
\end{equation}
Let us now define the minimum fidelity as:
\begin{equation}
F_{\text{min}} = \min_{\ket{\psi}} F(\ket{\psi}) \, .
\end{equation}
Then the maximum infidelity is simply $\epsilon_{\text{max}} = 1 - F_{\text{min}}$.

Let us now describe this maximum infidelity for all the strategies considered in FIG.~\ref{fig:analog}. Firstly, let us consider the discrete-variable encoding using the [[4,1,2]] code. This encoding can be used for approximate error correction against amplitude damping channel as proposed in~\cite{leung1997approximate}. It is shown there that for the specific considered decoding strategy, the maximum infidelity expressed in terms of the photon loss probability $\gamma$ is given by:
\begin{equation}
\epsilon_{\text{max}} = 5 \gamma^2 + \mathcal{O}(\gamma^3) \, .
\end{equation}

Now let us consider the maximum infidelity for the GKP-based schemes. Specifically, we shall calculate that maximum infidelity in terms of the logical $X$ and $Z$ flip probabilities $p_{\text{err,X/Z}}$ after a single elementary link, obtained from the simulation as discussed in Section~\ref{sec:MonteCarlo}. As we have already seen, for these encoding strategies the effective channel $\mathcal{D}$ is the Pauli channel given in Eq.~\eqref{eq:channelpXpYpZ}. In the simulation for FIG.~\ref{fig:analog}, we simulate only a single link, so $q_X = p_{\text{err,X}}(1-p_{\text{err,Z}}), \,\, q_Z = p_{\text{err,Z}}(1-p_{\text{err,X}}), \,\, q_Y = p_{\text{err,X}}p_{\text{err,Z}}$. The crucial feature of these coefficients is that the probability of a $Y$ flip is suppressed, that is $q_Y <q_X$ and $q_Y<q_Z$.

Now, let $\ket{\psi} = a \ket{0} + \sqrt{1-a^2}e^{i\theta} \ket{1}$ where $a \in [0,1]$ and $\theta \in [0, 2\pi)$. Then the infidelity $\epsilon = 1-F$ becomes:
\begin{equation}
\epsilon = q_X + q_Y + 4a^2 (1-a^2) \left(q_Z - q_X + (q_X - q_Y) \sin^2 \theta \right).
\end{equation}
Since $q_Y < q_X$ and $q_Y< q_Z$, the maximum infidelity can be obtained by setting $\theta = \pm \pi/2$ and $a = 1/\sqrt{2}$. The resulting maximum infidelity is then:
\begin{equation}
\epsilon_{\text{max}} = q_X + q_Z.
\end{equation}
We see that the maximum infidelity is obtained after transmission of the $Y$ basis states $\ket{\psi} = \frac{1}{\sqrt{2}}(\ket{0} \pm i \ket{1})$.

\section{Scheduling of operations}
\label{sec:ScheduleAndCost}

In this appendix we propose a specific scheduling procedure for the operations in all the repeaters and then establish the cost of each repeater type. These costs based on the number of GKP modes needed in each repeater as well as the number of time steps needed for storing the state are then used in the cost function minimisation as described in Section~\ref{sec:Cost}.

Let us first describe the scheduling of the operations inside a GKP repeater. Since we would like to use a minimal amount of resources for the operations performed inside this type of repeater, we consider an architecture that can store only a single data GKP mode at any given time. Such a repeater will then perform GKP correction in both quadratures on this data qubit. Since after measuring a GKP syndrome in one quadrature we will need a fresh GKP ancilla for measuring the syndrome in the other quadrature, we assume that the GKP repeater can simultaneously store two GKP ancilla modes such that when one is being used for syndrome measurement, the GKP ancilla is re-prepared in the other mode and vice versa. Hence such a repeater stores one data GKP mode for two time steps and uses two ancilla modes, each of which stores a GKP qubit for only one unit giving the total cost as $t_{\text{GKP}} = 4$. We depict this scheduling of the syndrome measurement inside a GKP repeater in FIG.~\ref{fig:RepSchedule}.

As we have seen in Appendix~\ref{sec:multiqubitrep} type-A repeaters require much more operations than type-B repeaters. In FIG.~\ref{fig:RepSchedule} we depict the operations performed inside the multi-qubit repeater based on the [[4,1,2]] code. The stairs-like shape of this diagram comes from the fact that the data qubits are received and sent in sequence, due to the fact that the type-B stations can operate only on individual GKP modes at a time (we maintain that scheduling also for the architecture which makes use of only type-A repeaters). We see that firstly we perform the GKP correction on each data qubit in each quadrature twice, then we measure the $XXXX$ stabiliser twice followed by the measurement of the two $Z$ stabilisers with additional GKP corrections in between. We see that each data mode needs to be stored for 11 time steps inside the repeater. Additionally we need 3 ancilla modes for measuring multi-qubit stabilisers. Two modes are used for measuring each round of $XXXX$ stabiliser, each requiring storage for 7 time steps and one mode for measuring the $IIZZ$ stabiliser requiring storage for 3 time steps. The ancilla mode used for the first measurement round of the $XXXX$ stabiliser can be reused to measure the $ZZII$ stabiliser. Additionally, 7 modes are needed to measure the GKP stabilisers. This can be seen by considering the largest number of GKP corrections in two neighbouring time steps. For GKP corrections in non-neighbouring time steps, we can reuse the previous storage mode as it takes one time step to re-prepare the ancilla GKP states. These ancilla modes for GKP correction need to be able to store the GKP state for only one time step. We also show that the repeater receives and starts operating on a data GKP qubit from a new block immediately after sending the GKP qubit from the old block out. This is important for the throughput consideration. This gives the total cost as $t_{\text{4-qubit}} = 68$.

The number of operations performed in the type-A repeater based on the [[7,1,3]] code is even larger as can be seen in FIG.~\ref{fig:RepSchedule2}. Firstly we measure the GKP stabilisers in both quadratures twice on each data qubit, then we perform two rounds of measuring $X$ stabilisers. After that we measure the $Z$-stabilisers twice. Again we place additional GKP corrections in between as discussed in Appendix~\ref{sec:multiqubitrep}. Since some of the data qubits are measured more often than the others, the less frequently measured qubits will need to stay idle for some time. We introduced these idle time slots at such locations so as to make the structure of the time ordering of the operations clear, i.e. the operations come in clear rounds marked with different colours. In this case we have 7 data modes stored for 40 time steps. We need 3 ancilla modes for measuring multi-qubit stabilisers. Two of them need to be stored for 7 time steps, one for 8 time steps. Finally we require 9 ancilla modes for measuring the GKP stabilisers. This gives the total cost as $t_{\text{7-qubit}} = 311$.

We note here that in FIG.~\ref{fig:RepSchedule} and in FIG.~\ref{fig:RepSchedule2} we have not marked when the feedback displacement for correcting second-level errors should be performed. This displacement is a displacement by $\sqrt{\pi}$ on the relevant qubits. After measuring $X$ stabilisers the corresponding displacement is along the $\hat{p}$ quadrature as it implements a $Z$ flip while after measuring the $Z$-stabilisers the displacement is along the $\hat{q}$ quadrature which implements an $X$ flip. Clearly it might not be possible to implement these displacements directly after the relevant multi-qubit stabiliser measurements, as we see in FIG.~\ref{fig:RepSchedule} and in FIG.~\ref{fig:RepSchedule2} that e.g. after finishing the measurement of the last stabiliser, some of the data qubits that should be corrected might have already been sent out. Fortunately, up to a global irrelevant phase, these $Z$ and $X$ correction flips commute with all the GKP corrections as well as Gaussian random displacement channels and multi-qubit stabiliser measurements in the opposite quadrature. Hence this displacement could be implemented later e.g. in one of the GKP stations or the following multi-qubit repeater. The only constraint is that it needs to be implemented before measuring the same stabilisers in the next type-A repeater. By counting the relevant time steps in FIG.~\ref{fig:RepSchedule} and in FIG.~\ref{fig:RepSchedule2}, one can verify that even if the two neighbouring repeaters are both type-A stations, there is enough time for the information which qubit(s) to correct to reach the second multi-qubit repeater before measuring the same stabilisers there, without the need to delay any of the data qubits. 

\begin{figure}
\includegraphics[trim={0 510 190 50}, clip, width = \columnwidth]{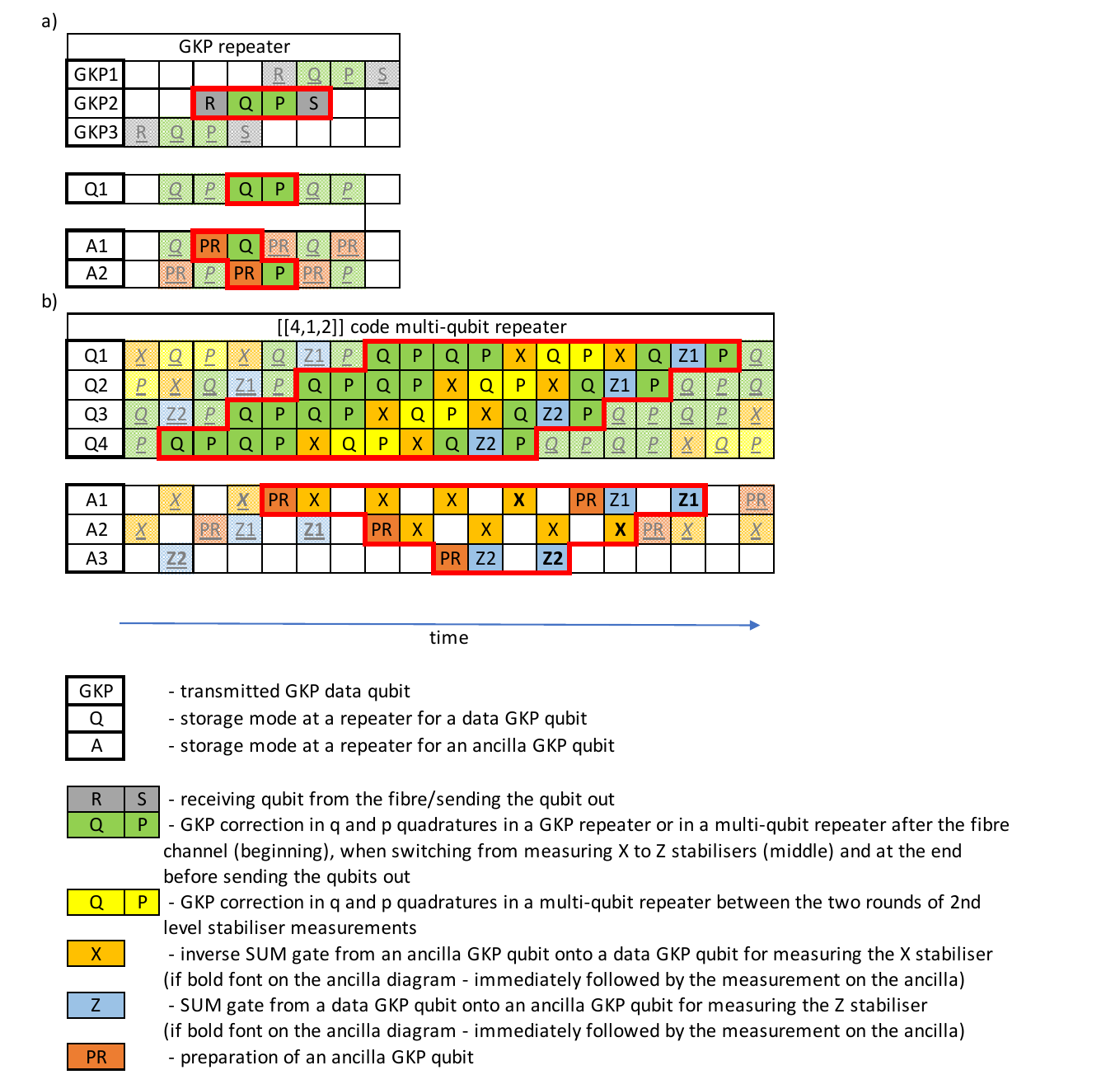}
\caption{\textbf{A schematic diagram of the time scheduling of the operations performed in the GKP and [[4,1,2]] code multi-qubit repeater.} For the GKP repeater (a) we first illustrate the GKP correction on three consecutively arriving GKP data qubits, marked with GKP. The grey boxes marking the time slots for receiving and sending out the modes are marked only for clarity, we assume that these processes are negligible in duration relative to all the other operations and can be incorporated into the beginning of the first operation time step (receiving) or into the end of the last operation time step (sending out).  We see that immediately after sending out the qubit GKP3, the qubit GKP2 is received and operated on. We then show the scheduling from the perspective of the storage modes at a repeater. A single storage mode (Q1) is used to store and correct errors on the GKP qubits using two ancilla modes (A1 and A2). For the [[4,1,2]] code type-A repeater (b) we illustrate the scheduling for the four needed storage modes and three ancilla modes used for measuring second-level stabilisers. Ancilla modes A1 and A2 need to be able to store the state for 7 time steps as they are used for the $X$ stabiliser measurements. Additionally they are also used for Z1 measurement. Ancilla mode A3 has a required storage of only 3 time steps as it is only used for measuring Z2. Additionally 7 ancilla modes are needed for the GKP corrections inside the [[4,1,2]] code multi-qubit repeater (not shown). For both (a) and (b) we mark with the red contour a single round of operations, which corresponds to a correction on a single GKP data qubit for the GKP repeater and all the operations on the four GKP data qubits from a single encoded block for the [[4,1,2]] code type-A repeater. The stair-like shape of the red contour is due to the sequential transmission of the consecutive GKP data modes.}
\label{fig:RepSchedule}
\end{figure}

\begin{figure}
\includegraphics[trim={0 650 190 50}, clip, width = \columnwidth]{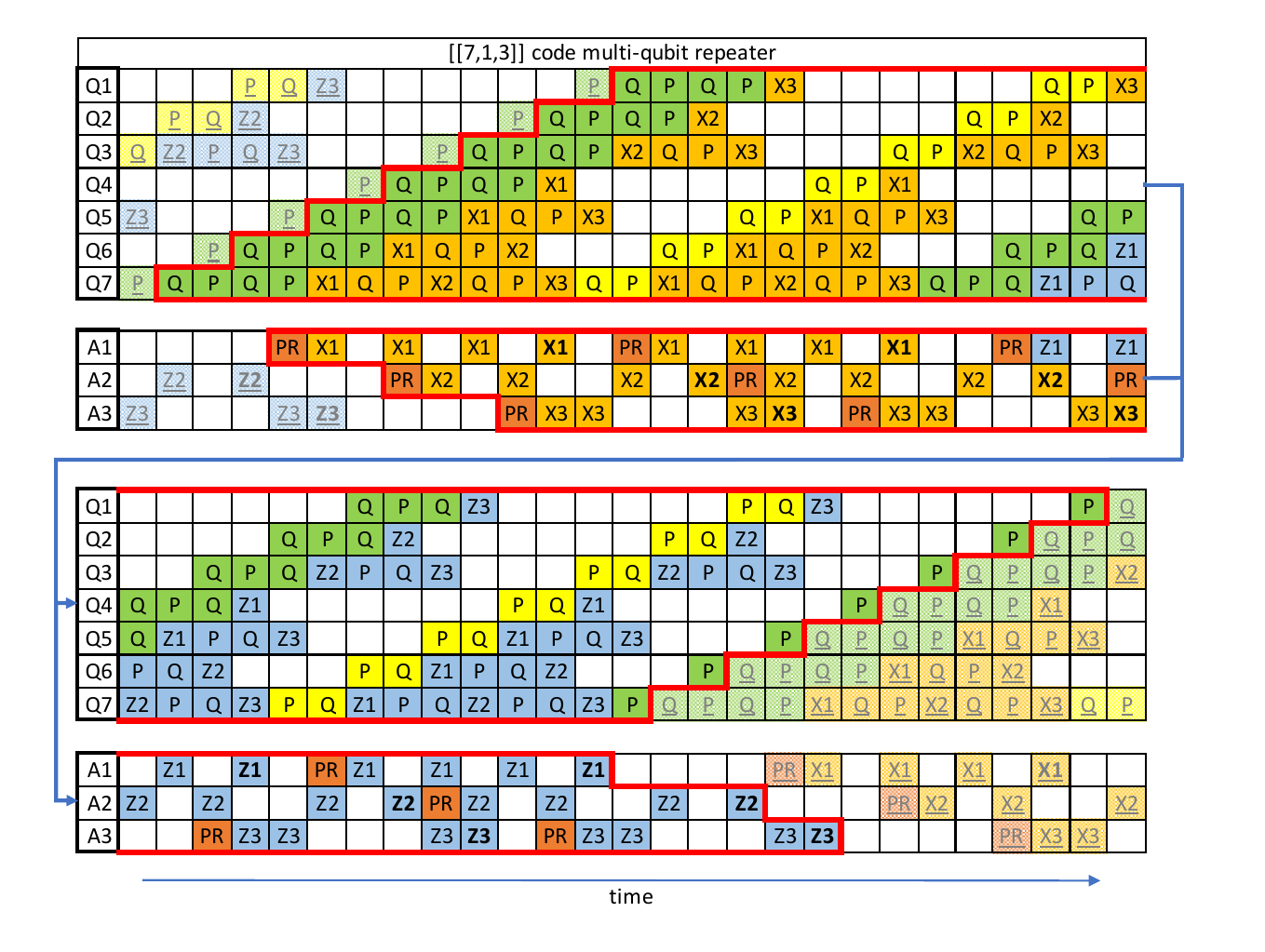}
\caption{\textbf{A schematic diagram of the time scheduling of the operations performed in the [[7,1,3]] code multi-qubit repeater.} The use of the symbols is the same as in FIG.~\ref{fig:RepSchedule}. We note that in this repeater we have additional GKP corrections within each round of measuring the higher-level syndrome. These corrections are marked with the same colour as the measurements of the corresponding higher-level stabilisers, that is the GKP corrections between the measurements of different $X$ stabilisers within a given round are marked with orange while the GKP corrections between the measurements of different $Z$ stabilisers within a given round are marked with blue. We now need 7 data storage modes and again 3 ancilla storage modes for the measurements of the 2nd level stabilisers. However two of the ancilla modes have a storage requirement of 7 time steps and one of 8 time steps. Additionally we require 9 ancilla modes for GKP corrections inside the repeater (not shown), which require storage of only one time step each. }
\label{fig:RepSchedule2}
\end{figure}

\section{Throughput and latency}
\label{sec:ThroughputLatency}

In our analysis we have focused on the performance of our scheme in its ability to generate secret key in secret bits per optical mode. From the practical perspective two other important figures of merit will be the throughput and latency. Throughput tells us how much secret key in secret bits can be generated per unit time and latency how long it takes to generate the first raw bit of key, which for our scheme amounts to the transmission time from Alice to Bob of a single encoded qubit. These figures of merit will depend on how fast the encoded states can be generated at Alice, decoded and measured at Bob and the duration of the operations in the type-B and type-A repeaters. Based on our considered time-scheduling model suitable for the microwave cavity implementation, we can make the estimate of these figures as a function of the duration of one time step introduced in Section~\ref{sec:Cost}. Clearly the duration of such a time step $\tau_0$ will depend on the experimental parameters.

To maximize throughput, we assume that each repeater receives a new GKP mode as soon as the corresponding old mode is sent out, e.g. as soon as the 7th GKP qubit of one [[7,1,3]]-code-encoded block is sent out, in its place a 7th GKP qubit of the next block is received, see Appendix~\ref{sec:ScheduleAndCost}. It is clear that the limiting factor will be the processing time $\tau_{\text{multi-qubit}}$ at type-A repeaters. This quantity is the duration over which we need to store each data mode in the type-A repeater and it is equal to $\tau_{\text{4-qubit}} = 11 \tau_0$ and $\tau_{\text{7-qubit}} = 40 \tau_0$. For the type-B repeaters the corresponding value is $\tau_{\text{GKP}} = 2 \tau_0$. Since the throughput of the concatenated-coded architecture could be approximated as $R \approx \frac{r}{\tau_{\text{multi-qubit}}}$, where $r$ is the secret-key rate, we see that using larger codes can be a significant limitation. If the high parameter requirements of the GKP repeater chain architecture could be achieved, then its throughput $R \approx \frac{r}{\tau_{\text{GKP}}}$ could be higher than for the concatenated-coded scheme. To overcome this limitation of the concatenated-coded scheme one can consider multiplexing such that every type-A repeater can start receiving and processing new GKP modes, before the corresponding old ones are sent out. In general this might significantly increase the repeater cost as in order to reduce the relevant time-scale from $\tau_{\text{multi-qubit}}$ by a factor of $k$, a type-A repeater will require $k$ times as many equivalent storage modes, which will increase the repeater cost $k$ times. We note that in this section we consider the secret-key rate $r$, not the secret-key rate per optical mode, as for this metric it is not important how many optical modes per logical qubit are required to generate the secret key.
   
Moreover, we expect then that latency of the schemes using two levels of coding will be limited by the number of the type-A repeaters. Specifically, the transmission time of the first encoded qubit can be written as:
\begin{equation}
t_l = \frac{L_{\text{tot}}}{c} + (N + 1) \tau_{\text{multi-qubit}}  + \left(mN +  (n-1)\right) \tau_{\text{GKP}}.
\label{eq:latency}
\end{equation} 
Here we have contribution from Alice's encoding, the communication time, storage time in all $N$ type-A and $mN$ type-B repeaters and the additional overhead due to sequential transmission of the GKP modes, so that after having processed the first GKP qubit from the block, Bob needs to wait additional  $(n-1)\tau_{\text{GKP}}$ to finish processing the remaining GKP qubits, where $n$ is the size of the outer code. We note that as described in the cost analysis in Section~\ref{sec:Cost}, we expect the storage time needed for encoding at Alice's station to be approximately $\tau_{\text{multi-qubit}}$ which is accounted for in the $+1$ pre-factor before the  $\tau_{\text{multi-qubit}}$ term in Eq.~\eqref{eq:latency}. Since Bob also performs quantum error correction, his decoding station is also counted as a type-A repeater (see Appendix~\ref{sec:seckeyRateAD}), and therefore is included in $N$. We see from Eq.~\eqref{eq:latency} then that the hybrid architecture in which $N$ can be reduced by increasing $m$ will, in most practical cases, also have significantly smaller latency than the architecture consisting solely of type-A repeaters (for which $N$ is large and $m=0$). Again for comparison, we expect the latency of the GKP repeater chain to be $t_l = \frac{L_{\text{tot}}}{c} +(m_{\text{GKP}} + 1) \tau_{\text{GKP}}$, where $m_{\text{GKP}}$ is the total number of repeaters in this architecture and the $+1$ term accounts for Alice's encoding. The latency of this scheme could clearly be shorter than for the schemes based on two levels of coding.

\end{document}